\documentclass[10pt,a4paper]{article}
\pdfoutput=1
\usepackage{curves,multicol}
\usepackage[export]{adjustbox}
\usepackage{subcaption}
\usepackage[makeroom]{cancel}
\usepackage{scalerel}
\usepackage{natbib}
\usepackage{graphicx}
\usepackage{hyperref}
\usepackage{authblk}
\usepackage{vmargin}
\usepackage{amssymb}
\usepackage{amsfonts}
\usepackage{multirow}
\usepackage{framed}
\usepackage{latexsym}
\usepackage{array}
\usepackage{tensor}
\usepackage{url}
\usepackage{xcolor}
\usepackage{graphicx}
\usepackage{epstopdf, epsfig}
\usepackage{booktabs}
\usepackage{commath}
\usepackage{rotating}
\usepackage{tabularx}
\usepackage{bibentry}
\usepackage{doi}

\providecommand{\keywords}[1]{\textbf{\textit{Keywords ---}} #1}
\newcommand{\fref}[1]{Fig.~\ref{#1}}
\newcommand{\tref}[1]{Tab.~\ref{#1}}
\newcommand{\eref}[1]{Eq.~\ref{#1}}
\newcommand{\sref}[1]{\S~\ref{#1}}
\newcommand{\aref}[1]{Appendix~\ref{#1}}

\newcolumntype{A}{ >{$} r <{$} @{} >{${}} l <{$} } 
\newcolumntype{L}[1]{>{\raggedright\let\newline\\\arraybackslash\hspace{0pt}}m{#1}}
\newcolumntype{C}[1]{>{\tiny\centering\let\newline\\\arraybackslash\hspace{0pt}}m{#1}}
\newcolumntype{R}[1]{>{\raggedleft\let\newline\\\arraybackslash\hspace{0pt}}m{#1}}
\newcolumntype{N}{@{}m{0pt}@{}}

\newcommand{\avg}[1]{\left\langle #1 \right\rangle}
\newcommand{\pars}[1]{\left(#1 \right)}

\newcommand{\p}{%
    \mathchoice%
        {\turnbox{12}{$\displaystyle\,'$}\;}%
        {\turnbox{12}{$\textstyle\,'$}\;}%
        {\turnbox{12}{$\scriptstyle\,'$}\;}%
        {\turnbox{12}{$\scriptscriptstyle\,'$}\;}%
}%

\newcommand{\vect}[1]{\boldsymbol{#1}}

\newcommand{\smallgtrsim}{\smallsym{\mathrel}{\gtrsim}}
\makeatletter
\newcommand{\smallsym}[2]{#1{\mathpalette\make@small@sym{#2}}}
\newcommand{\make@small@sym}[2]{%
  \vcenter{\hbox{$\m@th\downgrade@style#1#2$}}%
}
\newcommand{\downgrade@style}[1]{%
  \ifx#1\displaystyle\scriptstyle\else
    \ifx#1\textstyle\scriptstyle\else
      \scriptscriptstyle
  \fi\fi
}
\makeatother

\hypersetup{
	unicode=true, 
	pdftitle={Deep learning of the spanwise-averaged Navier--Stokes equations},
	pdfauthor={Bernat Font},
	pdfnewwindow=true, 
	colorlinks=true, 	
	pdfborder={0 0 0},	
	linkcolor=blue,
	linktoc=all, 		
	citecolor=blue,
	urlcolor=blue,
	breaklinks=false,
}
\setmarginsrb { 0.8in}  
              {0.25in}  
              { 0.8in}  
              { 0.5in}  
              {  20pt}  
              {0.25in}  
              {   9pt}  
              { 0.3in}  

\graphicspath{{img/}}

\title{\textbf{Deep learning of the spanwise-averaged Navier--Stokes equations}}
\author[1,2,\footnote{\href{mailto:bernat.font@bsc.es}{\texttt{bernat.font@bsc.es}}}]{Bernat Font}
\author[1,3]{Gabriel D. Weymouth}
\author[2]{Vinh-Tan Nguyen}
\author[1]{Owen R. Tutty}

\affil[1]{Faculty of Engineering and Physical Sciences, University of Southampton, SO17 1BJ Southampton, UK}
\affil[2]{Institute of High Performance Computing, Singapore Agency for Science, Technology and Research (A*STAR), 138632, Singapore}
\affil[3]{The Alan Turing Institute, NW1 2DB London, UK}
\date{\today}

\begin{document}

{\let\newpage\relax\maketitle}

\begin{abstract}
Simulations of turbulent fluid flow around long cylindrical structures are computationally expensive because of the vast range of length scales, requiring simplifications such as dimensional reduction.
Current dimensionality reduction techniques such as strip-theory and depth-averaged methods do not take into account the natural flow dissipation mechanism inherent in the small-scale three-dimensional (3-D) vortical structures.
We propose a novel flow decomposition based on a local spanwise average of the flow, yielding the spanwise-averaged Navier--Stokes (SANS) equations.
The SANS equations include closure terms accounting for the 3-D effects otherwise not considered in 2-D formulations.
A supervised machine-learning (ML) model based on a deep convolutional neural network  provides closure to the SANS system.
A-priori results show up to 92\% correlation between target and predicted closure terms; more than an order of magnitude better than the eddy viscosity model correlation.
The trained ML model is also assessed for different Reynolds regimes and body shapes to the training case where, despite some discrepancies in the shear-layer region, high correlation values are still observed.
The new SANS equations and ML closure model are also used for a-posteriori prediction.
While we find evidence of known stability issues with long time ML predictions for dynamical systems, the closed SANS simulations are still capable of predicting wake metrics and induced forces with errors from 1-10\%.
This results in approximately an order of magnitude improvement over standard 2-D simulations while reducing the computational cost of 3-D simulations by 99.5\%.
\end{abstract}


\section{Introduction}

The vast range of spatial and temporal scales inherent in turbulence renders direct numerical simulations of most engineering and environmental flows impossible.
This is because the required resolution is governed by the balance of the flow convective and viscous forces, quantified by the Reynolds number $Re=UL/\nu$, where $U$ and $L$ are the characteristic flow speed and length, and $\nu$ is the fluid kinematic viscosity. 
For illustration, turbulent engineering or atmospheric flows have $Re\smallgtrsim 10^6-10^{10}$ and directly simulating homogeneous isotropic turbulence in an $L^3$ box for one convection cycle would require $O(Re^3)$ operations \citep[p. 346]{Coleman2010,Pope2000}.
To avoid resolving the smallest length and time scales, turbulence models such as Reynolds-averaged Navier--Stokes (RANS) equations, large eddy simulation (LES), and others have been developed to account for the unresolved scales of motion on coarse discretizations.

However, in many cases the scale of the flow is so large that turbulence models do not sufficiently reduce the computational cost and further simplifications are required, such as reducing the flow from three spacial dimensions down to two.
In the case of very long slender bodies, strip theory can be used to place two-dimensional (2-D) strips along the structure span, reducing cost by orders of magnitude.
In offshore engineering, strip-theory methods have been used to simulate the vortex-induced vibrations of a marine riser \citep{Herfjord1999,Willden2001,Willden2004}.
Other slender-body problems tackled with strip theory are found in naval engineering to simulate ship hydrodynamics \citep{Newman1979}, or in aerospace engineering for wing fluttering prediction \citep{Yates1966}.
A different dimensionality reduction technique is the depth-averaged (DA) method, typically used in the simulation of shallow open-channel flow such as rivers \citep{Molls1995}, where the dimensions of the problem are simplified by averaging along the shear direction of the flow.

The main shortcoming of the dimensionality reduction methods above is that the fluid dynamics of 2-D systems are intrinsically different from their 3-D counterparts.
In particular, 3-D turbulence cannot develop when the spanwise dimension is neglected.
Instead, 2-D turbulence promotes coherent and high-intensity vortices via vortex-thinning and vortex-merging events \citep{Xiao2009} ultimately inducing larger forces to the cylinder \citep{Mittal1995, Norberg2003}.
Because of this, these methods require additional dissipation mechanisms, often in the form of LES Smagorinsky models.
These turbulence models applied to 2-D systems decrease the vortices intensity thus inducing weaker forces on solid structures which can resemble those found in 3-D cases.
However, the dissipation mechanism is arbitrary and we show that it cannot be used to simulate 3-D turbulence dynamics in a 2-D simulation.

To address this issue, \citet{Bao2016, Bao2019} proposed a \textit{thick} strip-theory method, as hinted in \citet{Herfjord1999}.
Providing spanwise thickness to the strips allows the flow to locally develop 3-D turbulence, thus recovering the natural dissipation mechanism.
However, the minimum strip thickness at which 3-D turbulence dominates the flow is in the order of the structure diameter \citep{Bao2016,Font2019}. Hence, the inclusion of 3-D effects comes at the expense of increased computational cost, which can limit the simulation of realistic spans or Reynolds regimes. 

\begin{figure}[!ht]
\centering
\includegraphics[width=0.8\linewidth]{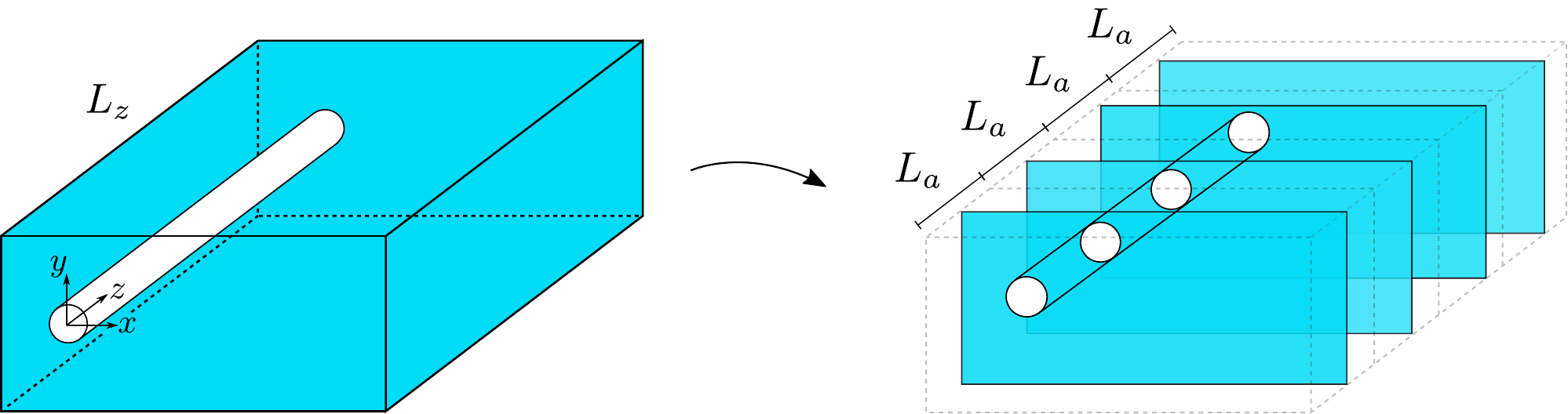}
\caption{Sketch of the spanwise-averaged strip-theory method.
The original $L_z$ span is decomposed into multiple $L_a$ spanwise segments for which the SANS equations are simultaneously solved, hence reducing the computational domain (colored region).
Solving the SANS equations for every spanwise-averaged strip allows to recover the local integrated forces induced to the cylinder.
The 2-D strips can be linked with an structural dynamic model similarly to \citet{Bao2016, Bao2019}.}
\label{fig:strips}
\vspace{-0.2cm}
\end{figure}

In this work, we propose a novel flow decomposition which separates the spanwise-averaged and the spanwise-fluctuating parts of the flow.
Instead of directly resolving the 3-D spatial scales, these are taken into account with an additional spanwise-fluctuating term plugged into the 2-D governing equations.
We call the resulting system of equations the spanwise-averaged Navier--Stokes equations (SANS).
In contrast to RANS or LES, the closure terms account for the 3-D effects enabling the simulation of unsteady wake flows in two dimensions.
Similarities can be drawn with DA methods.
Uniquely, the SANS equations do not assume hydrostatic pressure or approximately uniform velocity, instead being developed specifically to deal with fully 3-D wake turbulence.
And unlike RANS simulations which can only (at best) reproduce the time average forces, the SANS system is capable of reproducing the instantaneous spanwise-integrated forces on a locally cylindrical body. 
As such, a SANS strip theory is ideal for predicting the 3-D fluid-structure interactions with widely spaced 2-D simulation planes (see \fref{fig:strips}).

The SANS decomposition, like all turbulence equations, requires a model for the unresolved fluctuating terms, and we develop a data-driven model using machine learning (ML) for this purpose.
In recent years, ML methods have been successfully used in turbulence modelling for: 
pointwise classification and blending of appropriate turbulence closures \citep{Maulik2019}, 
sequential turbulent flow prediction \citep{Lee2019}, 
turbulent flow field super-resolution from coarse data \citep{Fukami2019,Liu2020}, 
among others (see \citet{Duraisamy2019} and \citet{Brunton2020} for extended reviews).
Specifically within turbulence closures, ML methods have been applied in RANS \citep{Weatheritt2016} and LES \citep{Gamahara2017} frameworks, establishing its utility for revealing hidden correlations between system variables for which physical laws are, a-priori, unknown \citep{Brenner2019}.
In particular, convolutional neural networks (CNNs) have been developed to efficiently learn spatial structures in data arrays by exploiting the translational invariance inherent in convolutions \citep{LeCun1995}.
This makes them a very attractive tool for modelling novel and spatially-complex fields such as the new SANS closures.

\section{The spanwise-averaged Navier--Stokes equations}\label{sec:SANS_eqs}

When a flow is essentially homogeneous in one direction this can be exploited to simplify modeling in a number of ways;
the turbulent flow statistics are invariant to translation along the homogeneous direction, there are no mean flow gradients in the homogeneous direction, and the component of mean velocity in the homogeneous direction can always be set 0 in an appropriate reference frame \citep[p. 76]{Pope2000}.
Therefore, a single homogeneous direction can be used to reduce a 3-D flow to a 2-D system of equations, potentially reducing the cost of simulation by many orders of magnitude.

\subsection{Mathematical description}

The SANS equations are based on a decomposition of the flow field into a spanwise-averaged and a spanwise-fluctuating part, i.e. $q=\avg{q}+q\p$, where $\avg{\cdot}$ and $\cdot\p$ indicate an averaging operator along the spanwise direction and its corresponding spanwise-fluctuating part, respectively.
The spanwise average, also noted as $\avg{q}=Q$, can be defined on a $z\in\left[a,b\right]$ interval as
\begin{equation}
Q=\avg{q}\pars{x,y,t} = \frac{1}{b-a}\int^b_a q\pars{x,y,z,t}\,\mathrm{d}z.
\label{eq:spanwise_average}
\end{equation}

Consider the non-dimensional 3-D incompressible Navier--Stokes equations which can be written as
\begin{equation}
\partial_t\vect{u}+\nabla \cdot \pars{\vect{u}\otimes\vect{u}}=-\nabla p+Re^{-1}\nabla^2\vect{u},\label{eq:N-S}
\end{equation}
where $\vect{u}\pars{\vect{x},t}=\pars{u,v,w}$ is the velocity vector field, $p\pars{\vect{x},t}$ is the scalar pressure field, $\vect{x}=\pars{x,y,z}$ is the spatial vector and $t$ is the time.
The Reynolds number is defined as $Re=UL/\nu$, where $U$ and $L$ are respectively the system scaling velocity and length, and $\nu$ is the fluid kinematic viscosity.
Velocity and pressure can be decomposed into a spanwise-averaged and a spanwise-fluctuating part for the (e.g.) $\vect{e}_x$ component as follows,
\begin{gather}
\partial_t\pars{U+u\p}+\partial_x\pars{UU+2Uu\p+u\p u\p}+\partial_y\pars{VU+Vu\p+v\p U+v\p u\p}+\partial_z\pars{WU+Wu\p+w\p U+w\p u\p}=\nonumber\\
=-\partial_x\pars{P+p\p}+Re^{-1}\left[\partial_{xx}\pars{U+u\p}+\partial_{yy}\pars{U+u\p}+\partial_{zz}\pars{U+u\p}\right].
\label{eq:N-S_decomp}
\end{gather}
The following relations arising from \eref{eq:spanwise_average} are considered for two arbitrary quantities $q$ and $s$
\begin{alignat}{3}
&\,\,q=Q+q\p,                      &&\qquad \,\,s=S+s\p,                       &&\qquad \avg{q+s}=Q+S,              \label{eq:rule1}\\
&\avg{Q}=Q,                  &&\qquad \avg{q\p}=0,                             &&\qquad \avg{Qs\p}=Q\avg{s\p}=0,      \label{eq:rule2} \\
&\avg{q\p q\p} \neq0,                    &&\qquad \avg{\partial_z q}\neq\partial_zQ, &&\qquad \,\,\partial_zQ=0,                \\
&\avg{\partial_t q} =\partial_t Q, &&\qquad \avg{\partial_x q} =\partial_x Q,  &&\qquad \avg{\partial_y q} =\partial_y Q.\label{eq:rule3}
\end{alignat}
Applying the averaging operator to the whole \eref{eq:N-S_decomp} and considering the relations above yields the governing equation of the $U$ spanwise-averaged velocity component
\begin{gather}
\partial_tU+\partial_x\pars{UU+\avg{u\p u\p}}+\partial_y\pars{VU+\avg{v\p u\p}}+\avg{\partial_z\pars{Wu\p+w\p U+w\p u\p}}=\nonumber \\
=-\partial_xP+Re^{-1}\pars{\partial_{xx}U+\partial_{yy}U+\avg{\partial_{zz} u\p}}, \\
\partial_tU+\partial_x\pars{UU}+\partial_y\pars{VU}+U\avg{\partial_z w\p}=-\partial_xP+Re^{-1}\pars{\partial_{xx}U+\partial_{yy}U}-\nonumber\\
-\partial_x\avg{u\p u\p}-\partial_y\avg{v\p u\p}-W\avg{\partial_z u\p}-\avg{\partial_z\pars{w\p u\p}}+Re^{-1}\avg{\partial_{zz} u\p}. \label{eq:N-S_decom_avg}
\end{gather}

In order to rewrite the equation above in its non-conservation form, the continuity equation is considered,
\begin{equation}
\nabla\cdot\vect{u}=0, \label{eq:cont}
\end{equation}
and applying the same decomposition and averaging procedure yields
\begin{equation}
\partial_xU+\partial_yV+\avg{\partial_z w\p}=0.\label{eq:cont_decomp_avg}
\end{equation}
By substituting \eref{eq:cont_decomp_avg} into \eref{eq:N-S_decom_avg}, the non-conservation form is obtained and the $U\avg{\partial_z w\p}$ term vanishes yielding
\begin{gather}
\partial_tU+U\partial_xU+V\partial_yU=-\partial_xP+Re^{-1}\pars{\partial_{xx}U+\partial_{yy}U}-\partial_x\avg{u\p u\p}-\partial_y\avg{v\p u\p}-\nonumber\\
-W\avg{\partial_z u\p}-\avg{\partial_z\pars{w\p u\p}}+Re^{-1}\avg{\partial_{zz} u\p}.\label{eq:z-avg_X}
\end{gather}

Analogously to the spanwise decomposition and averaging on the $\vect{e}_x$ component of the momentum equations, the same procedure can be applied to the $\vect{e}_y$ and $\vect{e}_z$ components which respectively leads to
\begin{gather}
\partial_tV+U\partial_xV+V\partial_yV=-\partial_yP+Re^{-1}\pars{\partial_{xx}V+\partial_{yy}V}-\partial_x\avg{u\p v\p}-\partial_y\avg{v\p v\p}-\nonumber\\
-W\avg{\partial_z v\p}-\avg{\partial_z\pars{w\p v\p}}+Re^{-1}\avg{\partial_{zz} v\p},\\
\partial_tW+U\partial_xW+V\partial_yW=-\avg{\partial_z p\p}+Re^{-1}\pars{\partial_{xx}W+\partial_{yy}W}-\partial_x\avg{u\p w\p}-\partial_y\avg{v\p w\p}-\nonumber\\
-W\avg{\partial_z w\p}-\avg{\partial_z\pars{w\p w\p}}+Re^{-1}\avg{\partial_{zz} w\p}.\label{eq:z-avg_Z}
\end{gather}
Note that Eqs. \ref{eq:z-avg_X} to \ref{eq:z-avg_Z} are the 2-D incompressible Navier--Stokes equations with additional spanwise stresses (first line) and terms arising from the finite averaging length (second line).

With this, the complete 2-D SANS momentum equations can be written in a more compact form as
\begin{equation}
\partial_t\vect{U}+\vect{U}\cdot\nabla\vect{U}=-\nabla P+Re^{-1}\nabla^2\vect{U}-\nabla\cdot\boldsymbol\tau_{ij}^R-W\avg{\partial_z \vect{u}\p}-\avg{\partial_z\pars{w\p \vect{u}\p}}+Re^{-1}\avg{\partial_{zz}\vect{u}\p},\label{eq:sans_full}
\end{equation}
where $\vect{U}=\pars{U,V}$, and the spanwise-stress residual (SSR) tensor $\boldsymbol\tau_{ij}^R$ is defined as
\begin{equation}
\boldsymbol\tau_{ij}^R= \avg{\vect{u}\otimes\vect{u}}-\avg{\vect{u}}\otimes\avg{\vect{u}} =
  \begin{pmatrix}
    \avg{u\p u\p} & \avg{u\p v\p} \\
    \avg{u\p v\p} & \avg{v\p v\p} 
  \end{pmatrix}.
\end{equation}

The SANS equations can be greatly simplified when averaging over a periodic interval since
\begin{equation}
\avg{\partial_z q} = \avg{\partial_z q\p}=\frac{1}{b-a}\int^b_a \frac{\partial q\p}{\partial z}\,\mathrm{d}z=\frac{q\p\left(b\right)-q\p\left(a\right)}{b-a}=0,\label{eq:periodic_assumption}
\end{equation}
because $q\p(a)=q\p(b)$ in a periodic interval.
This yields a simplified version of SANS which can be written as
\begin{gather}
\partial_tU+U\partial_xU+V\partial_yU=-\partial_xP+Re^{-1}\pars{\partial_{xx}U+\partial_{yy}U}-\partial_x\avg{u\p u\p}-\partial_y\avg{v\p u\p},\\
\partial_tV+U\partial_xV+V\partial_yV=-\partial_yP+Re^{-1}\pars{\partial_{xx}V+\partial_{yy}V}-\partial_x\avg{u\p v\p}-\partial_y\avg{v\p v\p},\\
\partial_tW+U\partial_xW+V\partial_yW=Re^{-1}\pars{\partial_{xx}W+\partial_{yy}W}-\partial_x\avg{u\p w\p}-\partial_y\avg{v\p w\p}.
\end{gather}
Note that the spatial dependency on the spanwise direction is lost during the process (there are no $z$ derivatives in the final form), and also the momentum equations for $U$ and $V$ are decoupled from $W$.
The momentum equation for $W$ becomes a convective--diffusive equation (with additional forcing terms) since the pressure term vanishes under the assumption of spanwise periodicity.
In a more compact form and dropping the $\vect{e}_z$ momentum equation, the simplified SANS equations can be written as
\begin{equation}
\partial_t\vect{U}+\vect{U}\cdot\nabla\vect{U}=-\nabla P+Re^{-1}\nabla^2\vect{U}-\nabla\cdot\boldsymbol\tau_{ij}^R.\label{eq:SANS}
\end{equation}
The resulting system of equations is equivalent to the 2-D Navier--Stokes equations with the addition of the SSR forcing term, which encapsulates the impact of the turbulent 3-D fluctuations on the SANS 2-D flow.
From here, the concept ``SANS'' assumes spanwise periodicity and refers to the simplified SANS equations (\eref{eq:SANS}).

\subsection{Fluid dynamics solver} \label{sec:solver}

The implementation of the SANS equations has been carried in Lotus, a 3-D finite-volume implicit large eddy simulation (iLES) in-house code.
The temporal evolution is integrated with a predictor-corrector algorithm, which is second-order accurate in time and uses an adaptive time step based on the local Péclet and Courant numbers.
Velocity and pressure are coupled with the pressure-Poisson equation and iteratively solved using a multigrid method, which also enforces the continuity condition in the velocity field.
The subgrid-scale structures are modelled by the numerical dissipation inherent in the spatial discretization scheme.
In this sense, we use the flux-limited quadratic upstream interpolation for convective kinematics (QUICK)  method, which is second-order accurate in space.
The effect of the implicit turbulence modelling, which can be quantified with
\begin{equation}
\nu_e=\frac{|\varepsilon|_{\scaleto{\Omega}{5pt}}}{|S_{ij}S_{ij}|_{\scaleto{\Omega}{5pt}}},
\end{equation}
where $\varepsilon$ is the turbulence kinetic energy dissipation rate, $\nu_e$ is the effective iLES viscosity, $S_{ij}$ is the velocity rate-of-strain tensor and $|\cdot|_{\scaleto{\Omega}{5pt}}$ denotes a volume average,
has been quantified for our solver in \citet{Hendrickson2019} by monitoring the grid-scaled effective viscosity $\nu_e\Delta^{-1}$ for the Taylor--Green vortex (TGV) case.
It was shown that the QUICK scheme linearly scales the effective viscosity with $\Delta$, the cell width.

\begin{figure}[!ht]
\centering
\includegraphics[width=0.35\linewidth]{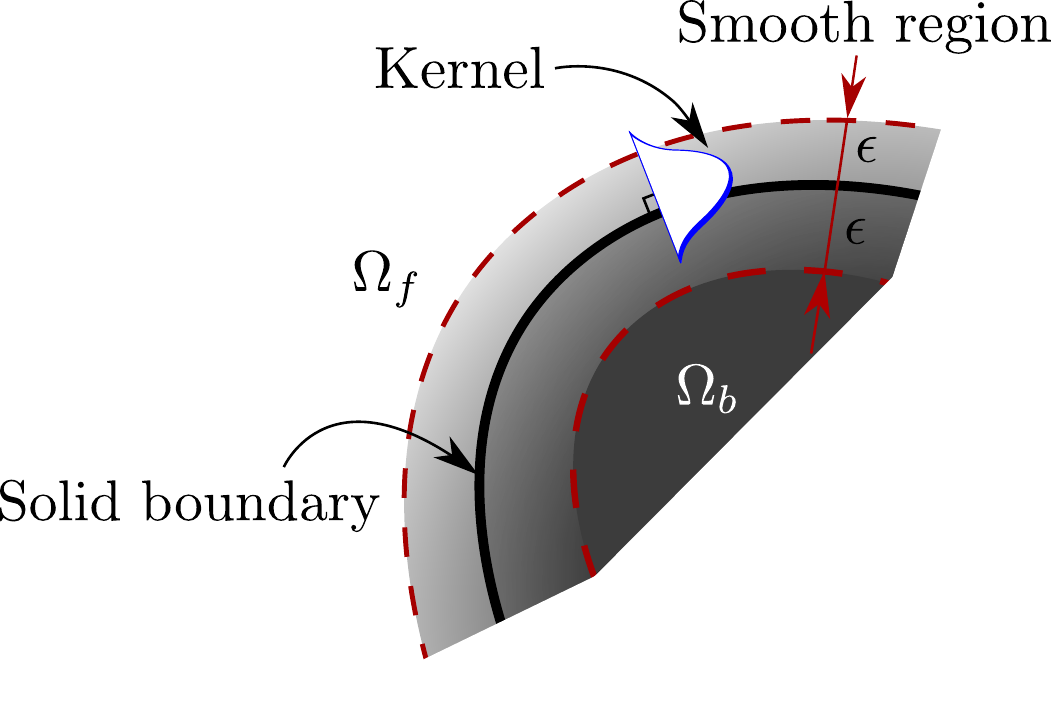}
\caption{BDIM sketch adapted from \citet{Maertens2015}. A convolutional kernel with radius $\epsilon$ smooths the interface between solid $(\Omega_b)$ and fluid $(\Omega_f)$ domains extending the governing equations applicability to the whole computational domain $\Omega$.}
\label{fig:bdim}
\end{figure}

The no-slip condition at the solid boundaries is enforced by the boundary data immersion method (BDIM) \citep{Weymouth2011}.
The BDIM is an immersed boundary method in which the fluid and solid governing equations are mapped together in a non-conforming rectilinear staggered grid.
A convolutional kernel provides a smooth  transition between the mediums (see \fref{fig:bdim}).
The BDIM has been tested in multiple applications and the reader is referred to \citet{Maertens2015} for a detailed mathematical description of the method.
Furthermore, our Louts in-house code has been validated for the simulation of bluff bodies in \citet{Schulmeister2017} and \citet{Font2019}, where the latter investigates flow past a circular cylinder at $Re=10^4$ with the same set-up as used in this work.

\subsection{Spanwise-averaged dynamics}\label{sec:spanwise-averaged_dynamics}

\begin{figure}[!ht]
\centering
\includegraphics[width=0.3\linewidth]{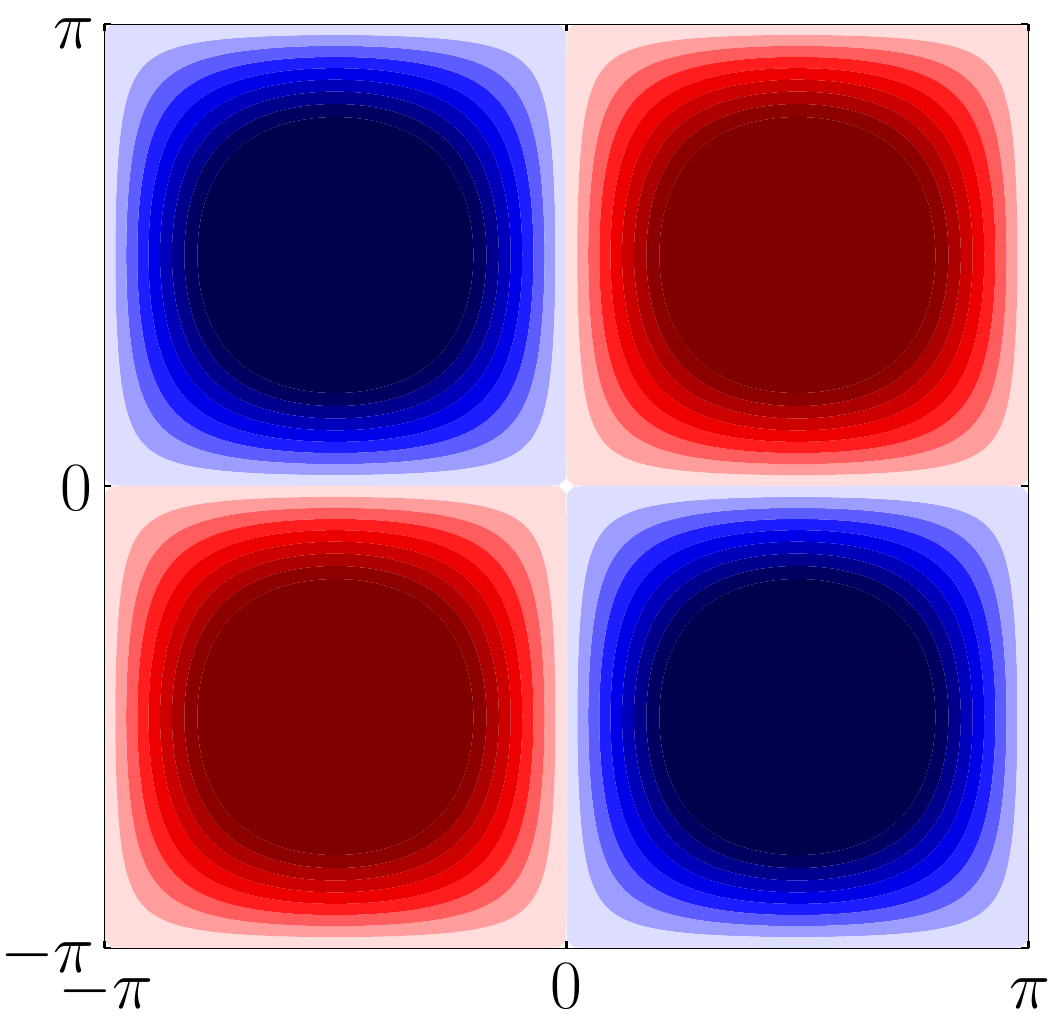}
\caption{Taylor-Green vortex initial vorticity field at $z=\pi/4$.}
\label{fig:t-g_0}
\end{figure}

To illustrate the effect of the SSR term, the TGV test case is considered. 
The TGV is broadly used in the computational fluid dynamics literature for the validation of numerical methods and turbulence models.
It consists of an energy decaying flow defined in a triple periodic box $(L\times L\times L)$ with initial flow field $\vect{u}_0$ given by
\begin{gather}
u_0(x,y,z) = U_0\sin(\kappa x)\cos(\kappa y)\cos(\kappa z), \nonumber \\
v_0(x,y,z) = -U_0\cos(\kappa x)\sin(\kappa y)\cos(\kappa z), \nonumber \\
w_0(x,y,z) = 0,
\end{gather}
where $\kappa=2\pi/L$, and we select $L=2\pi$ and $U_0=1$. 
The initial flow field (\fref{fig:t-g_0}) breaks down into smaller structures because of the vortex-stretching mechanism, yielding a phase of global enstrophy growth.
Once most of the large-scale structure have been destroyed, the inertial subrange and the dissipative structures cannot be sustained, which leads to an enstrophy decay because of viscous effects.
Since the purpose of this exercise is to emphasise the difference between 3-D, 3-D spanwise-averaged, and 2-D dynamics rather than fully resolving all the 3-D turbulence scales of motion, we set a resolution of 128 cells in all directions and $Re_L=1600$.

\begin{figure*}[!ht]
\centering
\setlength{\columnsep}{0cm}
\begin{multicols}{4}
    \includegraphics[width=\linewidth]{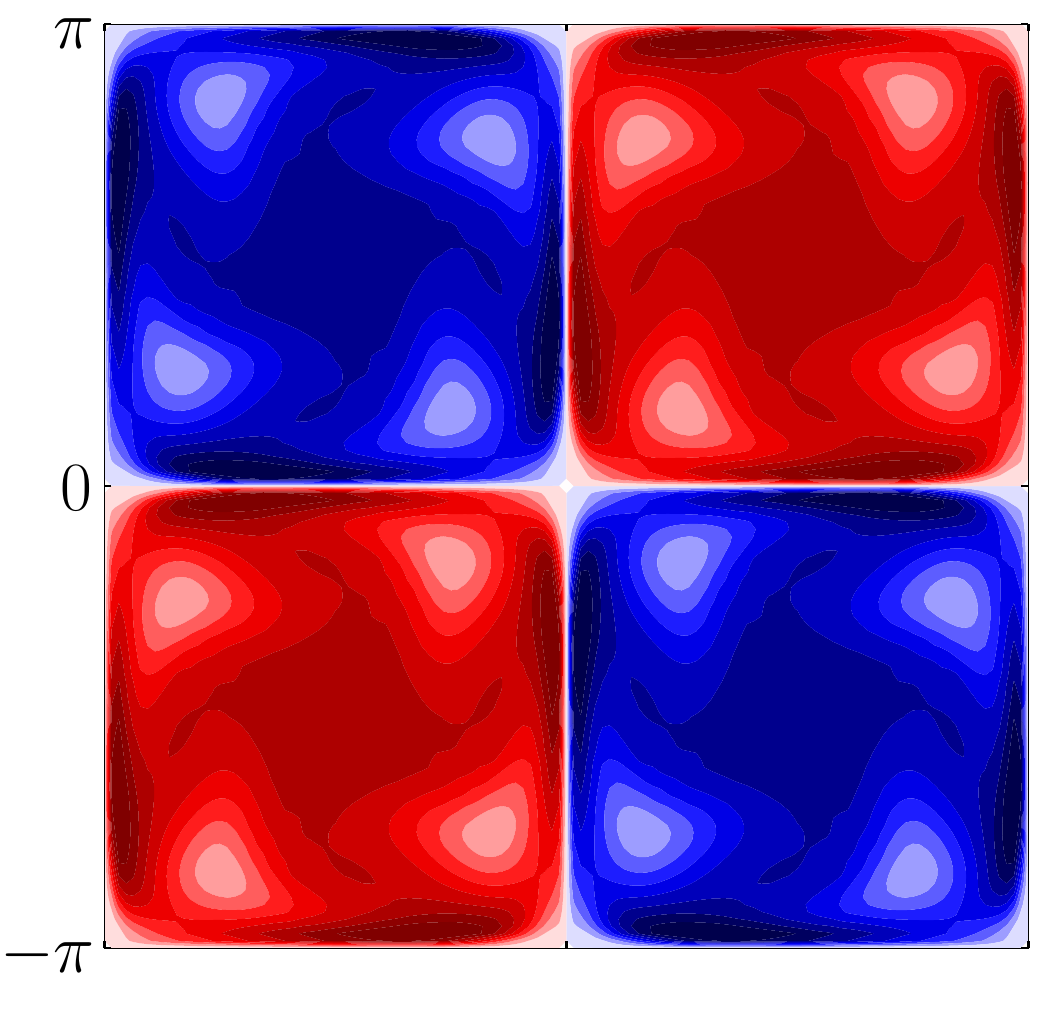}\par
    \includegraphics[width=\linewidth]{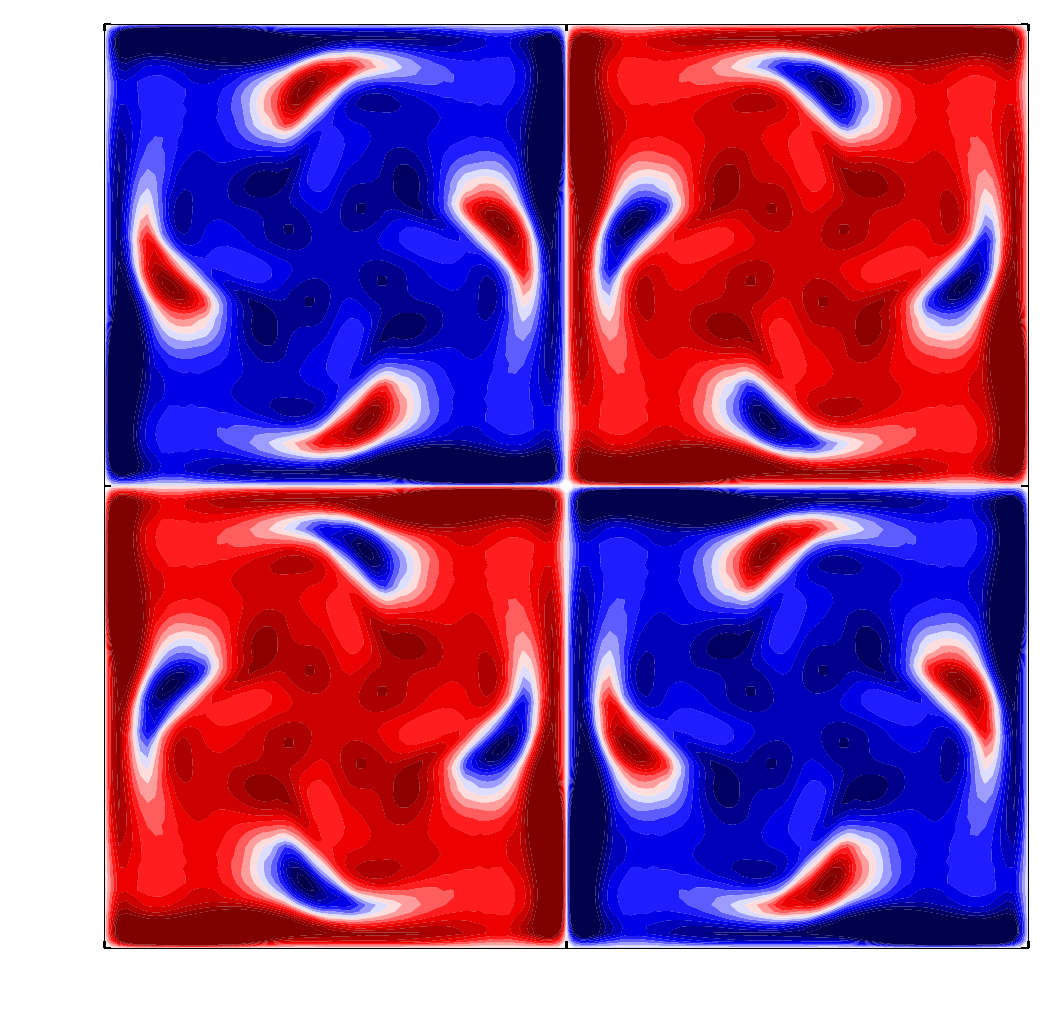}\par
    \includegraphics[width=\linewidth]{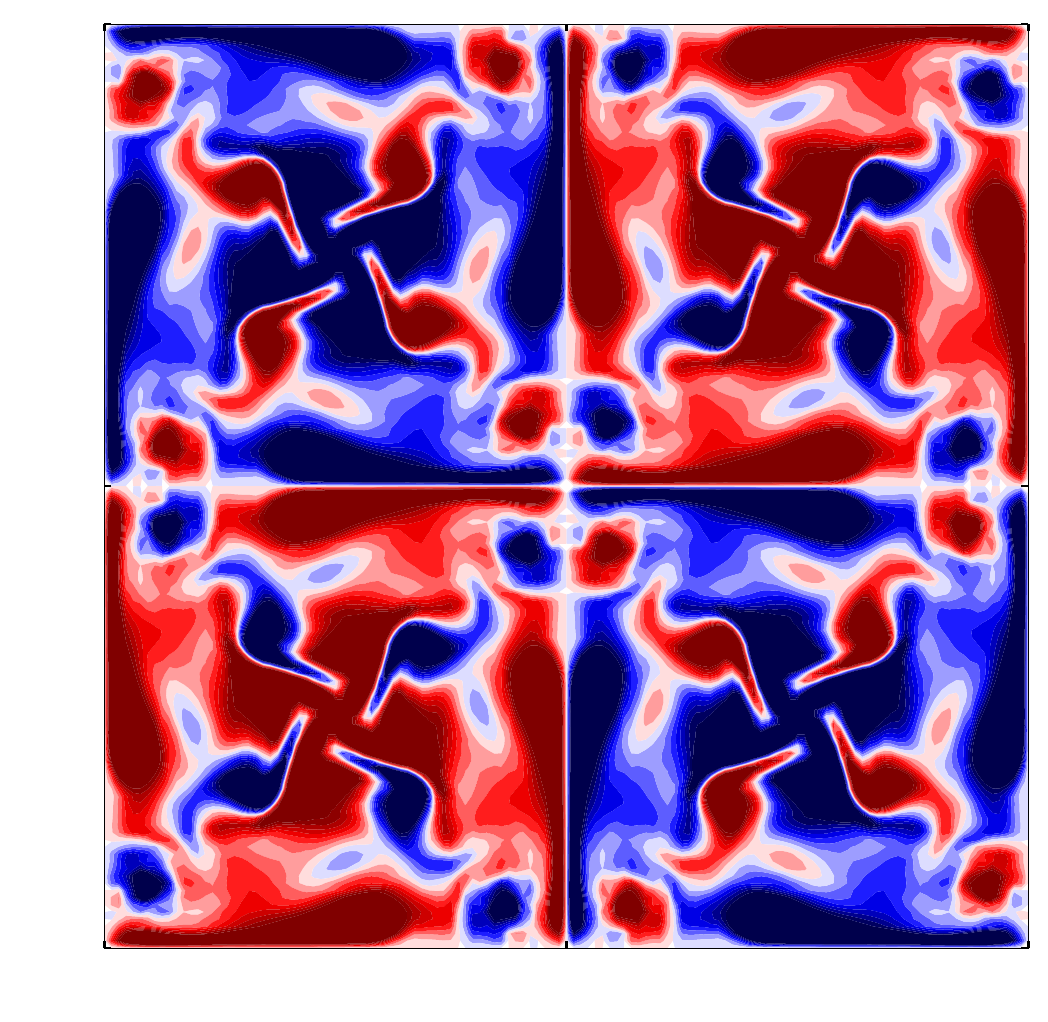}\par
    \includegraphics[width=\linewidth]{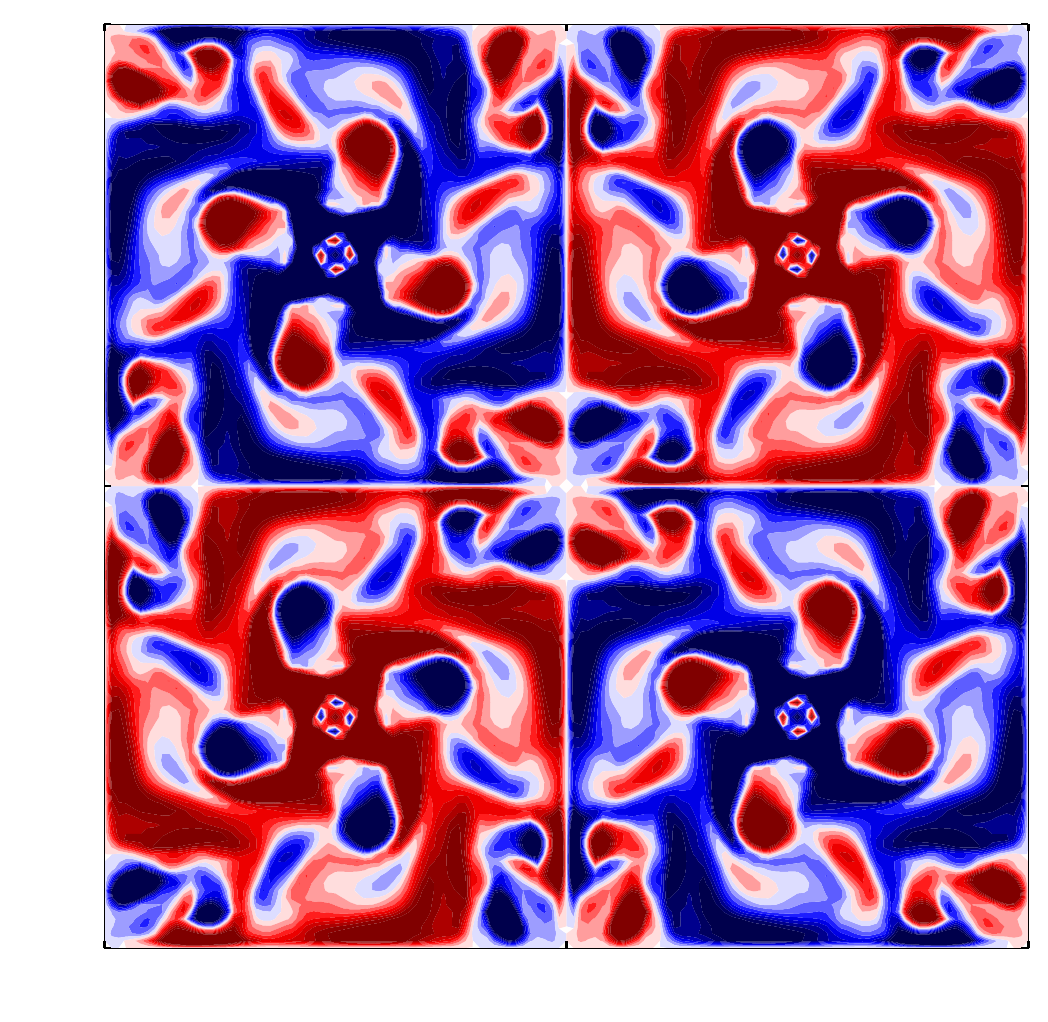}\par
\end{multicols}
\vspace{-0.9cm}
\begin{multicols}{4}
    \includegraphics[width=\linewidth]{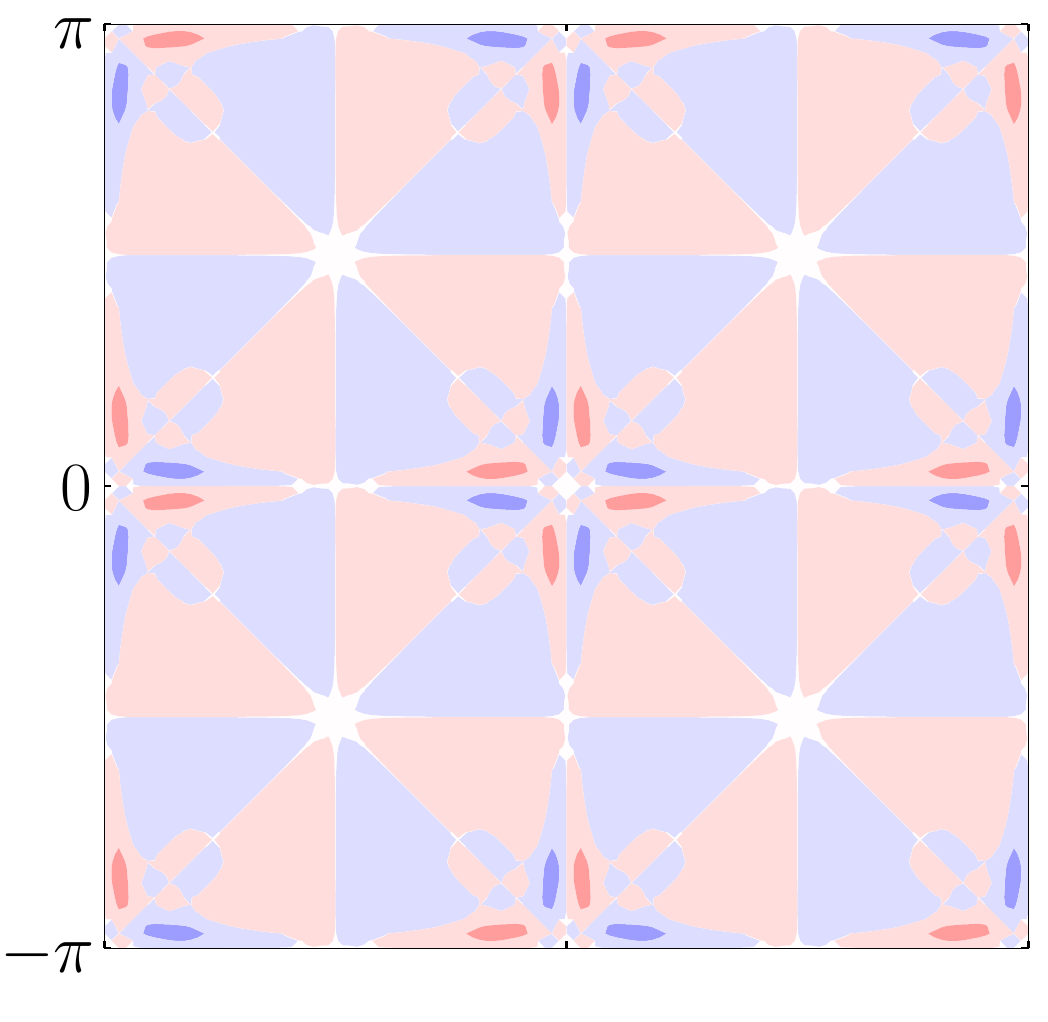}\par
    \includegraphics[width=\linewidth]{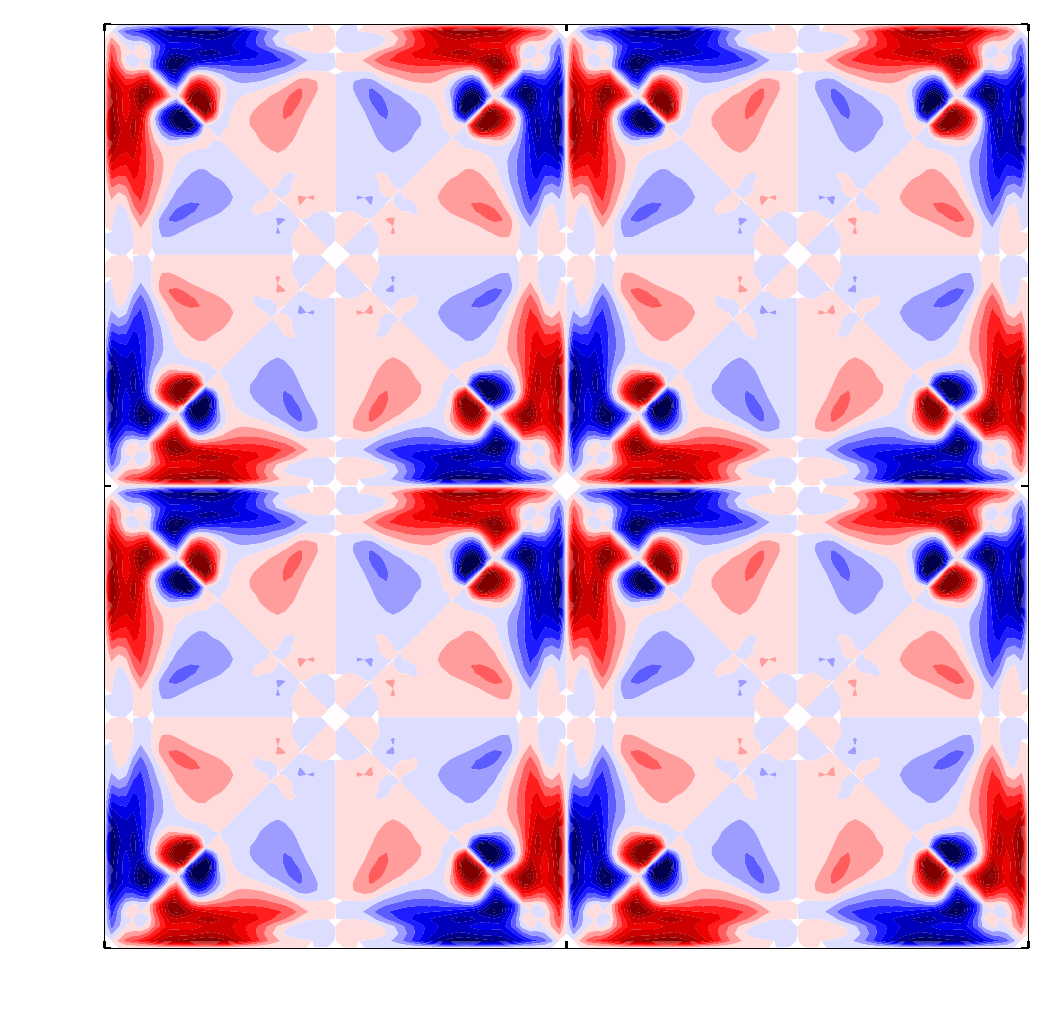}\par
    \includegraphics[width=\linewidth]{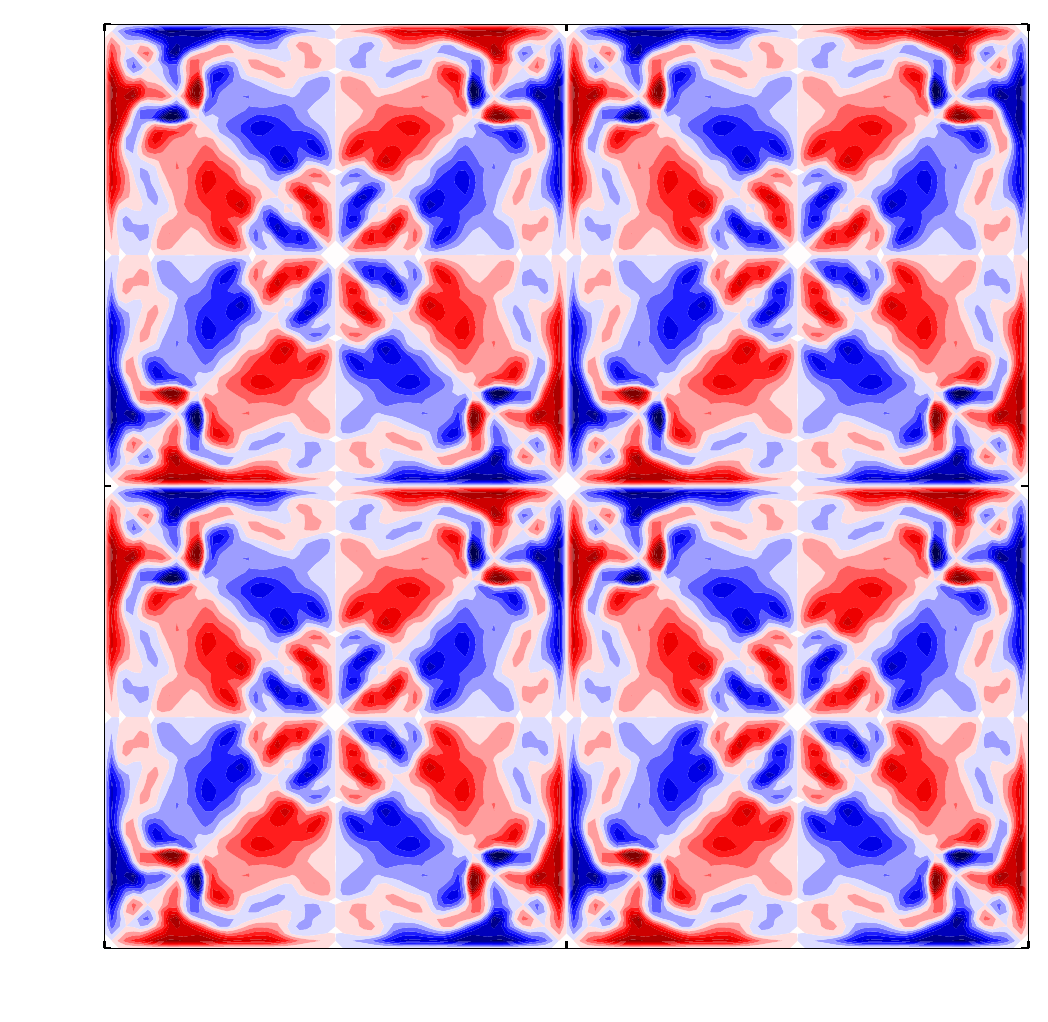}\par
    \includegraphics[width=\linewidth]{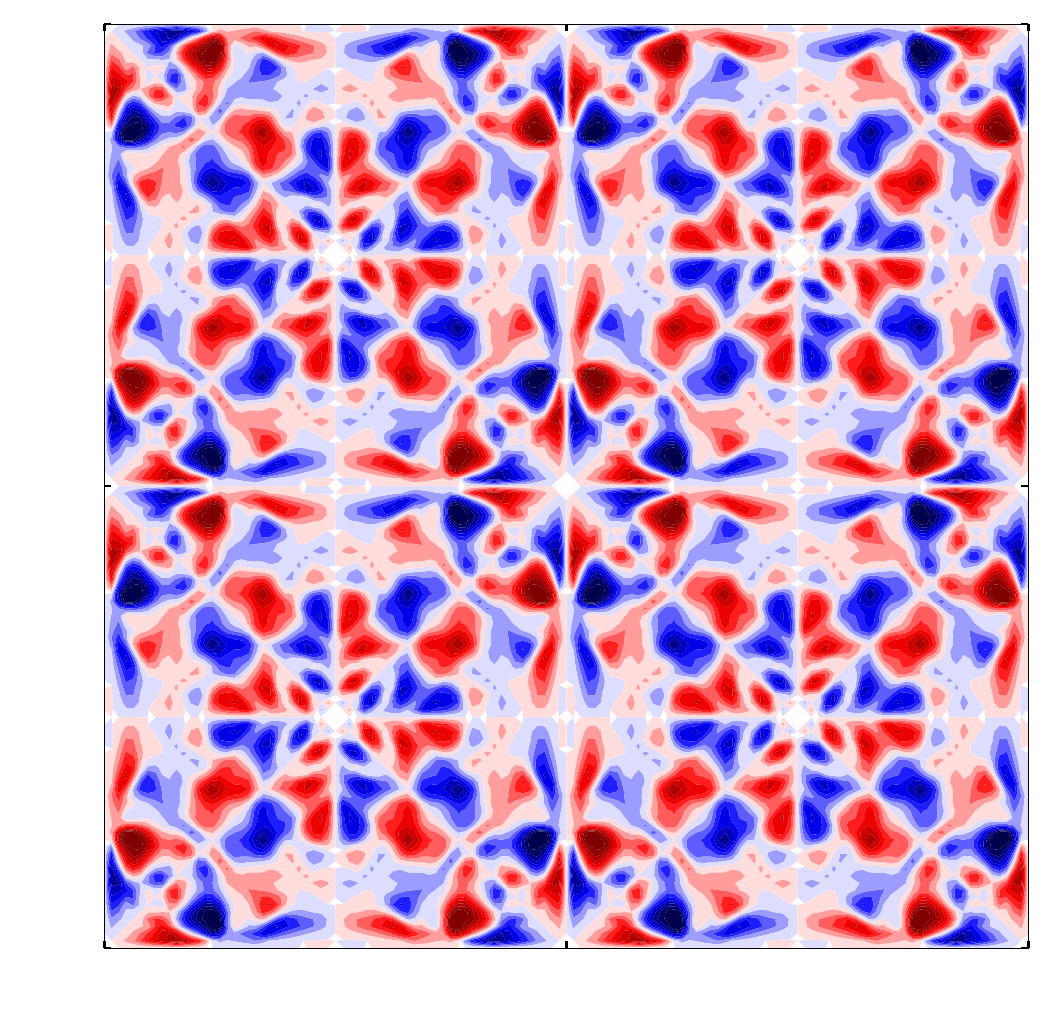}\par
\end{multicols}
\vspace{-0.9cm}
\begin{multicols}{4}
    \includegraphics[width=\linewidth]{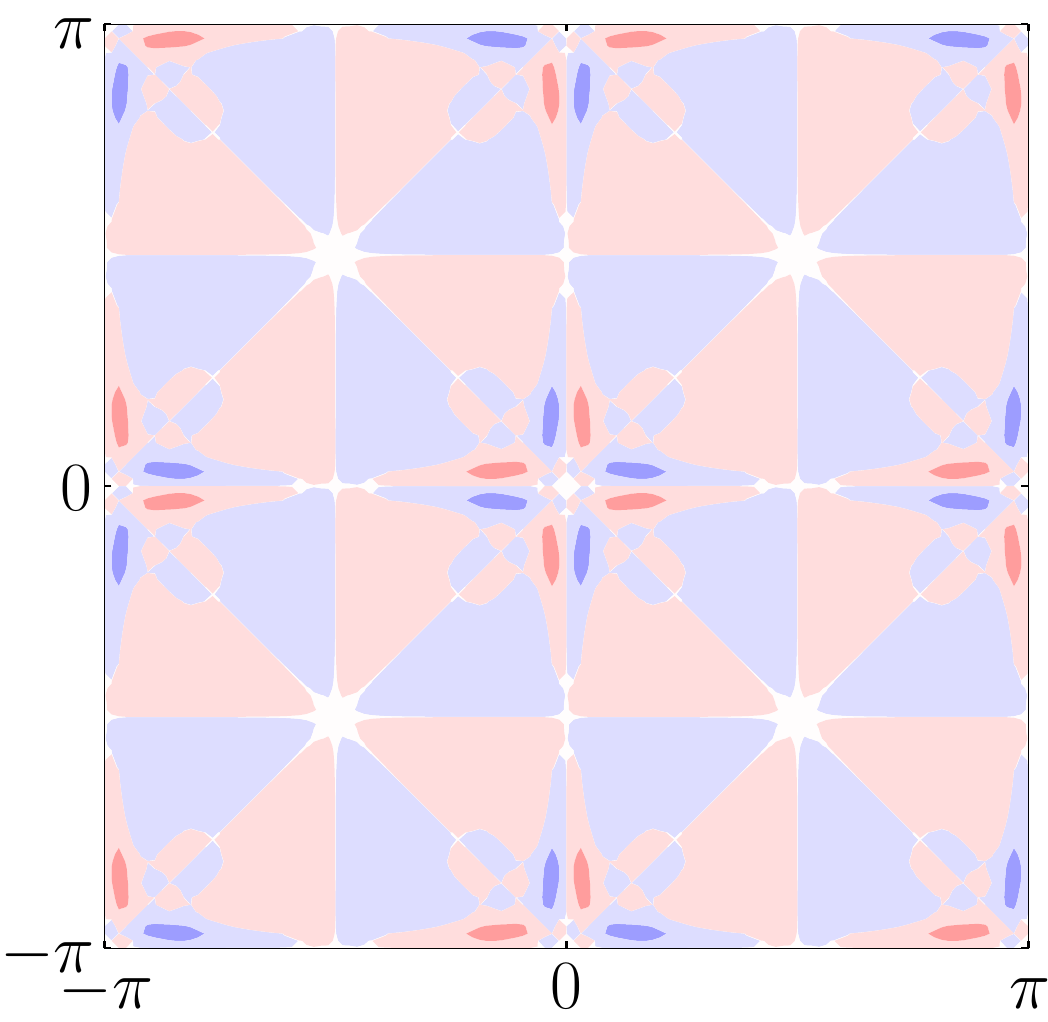}\par
    \includegraphics[width=\linewidth]{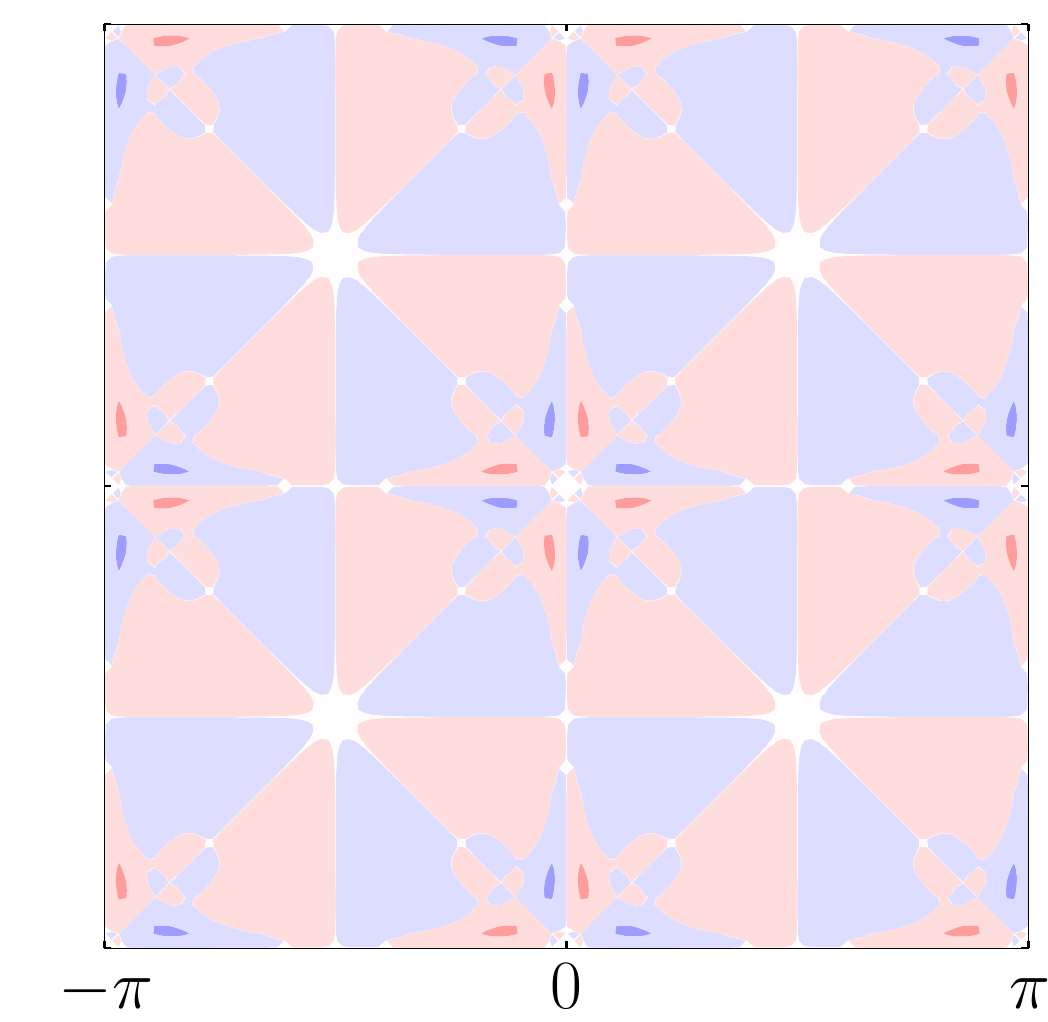}\par
    \includegraphics[width=\linewidth]{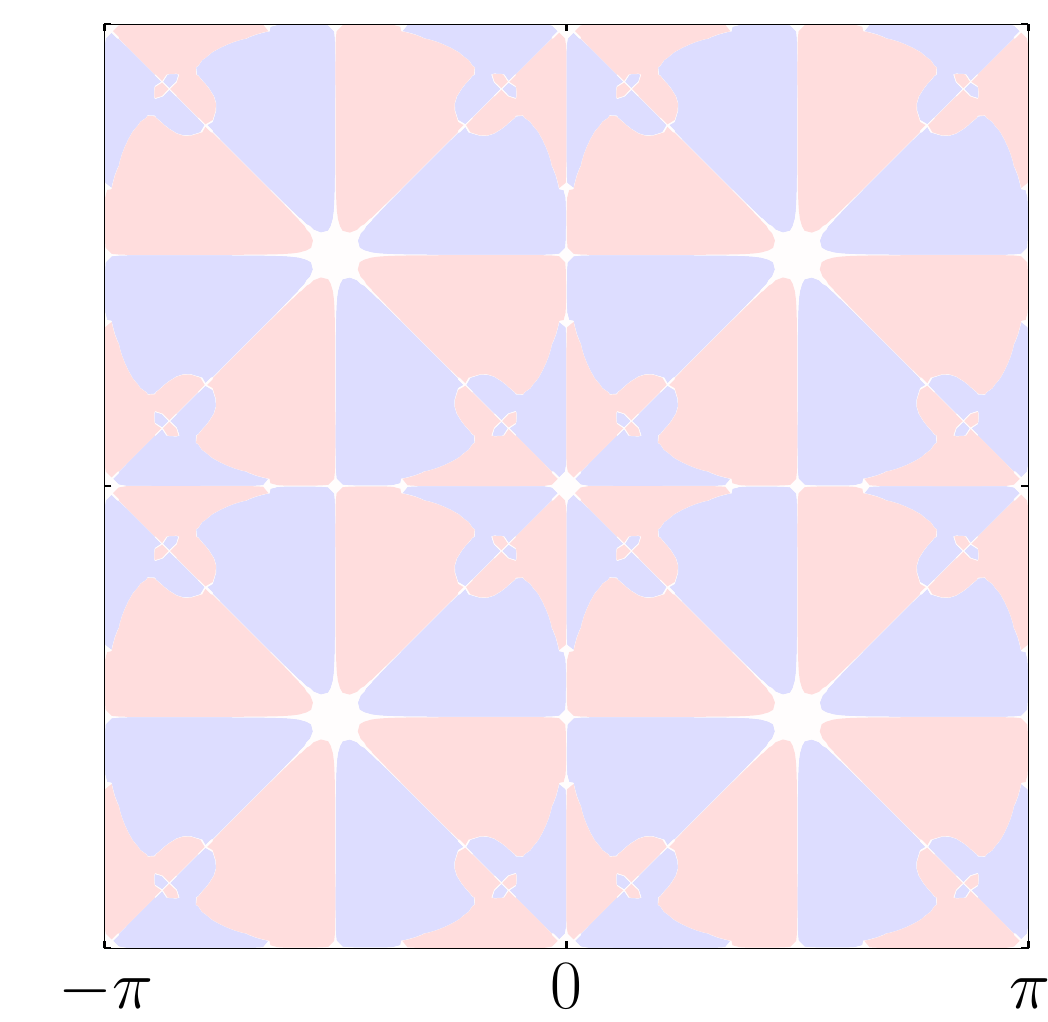}\par
    \includegraphics[width=\linewidth]{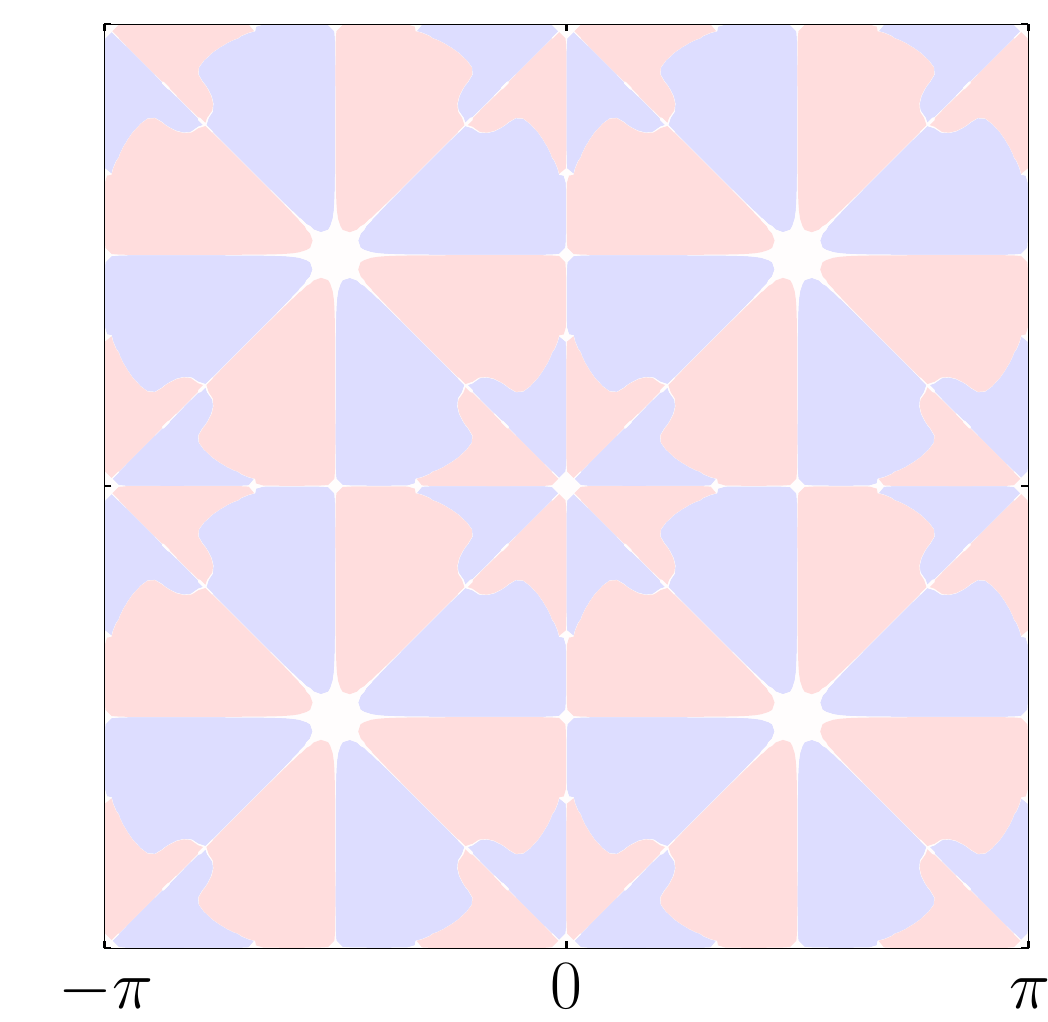}\par
\end{multicols}
\caption{Spanwise vorticity evolution of the Taylor-Green vortex at $Re=1600$ in a $2\pi$ periodic box from left to right: $t^*=4,6,8,10$, where $t^*$ is the non-dimensional time unit scaled with $t^*=t\kappa U_0$.
Top: 3-D slice at $z=\pi/4$.
Middle: 3-D spanwise-averaged.
Bottom: 2-D started from the 3-D spanwise-average snapshot at $t^*=4$, where transition to turbulence is expected on the 3-D system.
}
\label{fig:t-g_flow_field}
\end{figure*}

\fref{fig:t-g_flow_field} displays the flow evolution for the 3-D (top), spanwise-averaged (middle), and 2-D TGV cases.
The 2-D simulation is started from a spanwise-averaged snapshot of the 3-D system at $t^*=4$ (where $t^*=t\kappa U_0$).
It can be observed that the spanwise-averaged flow evolution is completely different from the purely 2-D turbulence dynamics.
Hence, the inclusion of the spanwise stresses into the 2-D Navier--Stokes solver (which can be computed in a 3-D simulation) allows to directly simulate spanwise-averaged dynamics, as we demonstrate in \S \ref{sec:perfect_closure}.

Mathematically, the kinetic energy $\pars{E=1/2\int \vect{u}^2\,\mathrm{d}\Omega}$ and enstrophy $\pars{Z=1/2\int \boldsymbol{\omega}^2\,\mathrm{d}\Omega}$ of 2-D flows are invariant in the incompressible inviscid limit \citep{Kraichnan1967}, i.e. $\mathrm{D}E/\mathrm{D}t=\mathrm{D}Z/\mathrm{D}t=0$, where $\boldsymbol\omega=\nabla\times\vect{u}=\pars{\omega_x,\omega_y,\omega_z}$ is the vorticity vector field.
This is completely different in the 3-D system, where the vortex-stretching term $\pars{\boldsymbol\omega\cdot\nabla\vect{u}}$ is responsible for the energy transfer from large-scales to small-scales of motion.
In this scenario, enstrophy is no longer conserved and helicity $\pars{H=1/2\int \vect{u}\cdot\boldsymbol{\omega}\,\mathrm{d}\Omega}$ is the quadratic invariant, together with the kinetic energy, conserved in the incompressible inviscid limit.
In 3-D systems, the rate of kinetic energy dissipation $\pars{\varepsilon}$ is related to enstrophy with $\varepsilon=-\mathrm{d}E/\mathrm{d}t=2\nu Z$.
Similarly to 3-D flows, the inclusion of the spanwise stresses to the 2-D equations locally removes or adds rotational energy to the system; hence enstrophy is no longer conserved.
Additionally, the spanwise stresses causes the kinetic energy to not be conserved in the incompressible inviscid limit either.

\section{Closure models}

The challenge of turbulence models is to adequately recover the physical information lost during the flow averaging process.
Closing the turbulent equations requires expressing the turbulent stresses in terms of the mean flow variables.
This section will explore a variety of closures for the SANS equations, including a new deep-learning data-driven closure.

\subsection{Perfect closure}\label{sec:perfect_closure}

The perfect closure $\pars{\mathcal{S}^R}$ is defined as the difference between the 2-D Navier--Stokes spatial operator $\tilde{\mathcal{S}}$ and the spanwise-averaged 3-D Navier--Stokes spatial operator $\avg{\mathcal{S}}$,
\begin{equation}
\mathcal{S}^R\pars{\vect{u\p}}\equiv \tilde{\mathcal{S}}-\avg{\mathcal{S}},
\end{equation}
where,
\begin{gather}
\tilde{\mathcal{S}}\pars{\avg{\vect{u}},\avg{p}}=\avg{\vect{u}}\cdot\nabla\avg{\vect{u}}+\nabla \avg{p}-Re^{-1}\nabla^2\avg{\vect{u}},\\
\avg{\mathcal{S}\left(\vect{u}, p\right)}=\avg{\vect{u}\cdot\nabla \vect{u} + \nabla p - Re^{-1} \nabla^2 \vect{u}}.
\end{gather}

With the previous definitions and using \eref{eq:SANS}, the $\nabla\cdot\boldsymbol\tau_{ij}^R$ closure term is redefined as the residual between the operators,
\begin{gather}
\partial_t\avg{\vect{u}}+\tilde{\mathcal{S}}=\tilde{\mathcal{S}}-\avg{\mathcal{S}}\equiv\mathcal{S}^R,\label{eq:perfect_SANS}\\
\partial_t\avg{\vect{u}}+\tilde{\mathcal{S}}\pars{\avg{\vect{u}},\avg{p}}=\mathcal{S}^R\pars{\vect{u\p}}.
\end{gather}

Note that, differently from $\nabla\cdot\boldsymbol\tau_{ij}^R$, the $\mathcal{S}^R$ residual term includes the spatial discretization error of the resolved spanwise-averaged quantities since $\tilde{\mathcal{S}}$ cancels out in \eref{eq:perfect_SANS} (see \citet{Beck2019} for an analogous explanation for LES closure terms).
Hence, the perfect spanwise-averaged solution can be recovered from the original definition of the SANS equations, i.e. $\avg{\partial_t\vect{u}+\mathcal{S}}=0$.
The SANS equations are numerically verified by including the perfect closure of the spanwise stresses in a 2-D simulation of the TGV case and flow past a circular cylinder at $Re=10^4$, as detailed next. 

\subsubsection*{Taylor--Green vortex}

The TGV as defined in \S \ref{sec:spanwise-averaged_dynamics} is used to verify the correct derivation of the SANS equations and the implementation of the perfect closure into the fluid dynamics solver.
Starting from a spanwise-averaged snapshot of the 3-D TGV case at $t^*=4$, a 2-D simulation and a SANS simulation are performed.
The SANS system includes the perfect closure previously computed in the 3-D case at every time step (which is kept constant so that the perfect closure is in sync with the resolved variables).

\begin{figure}[!ht]
\centering
\includegraphics[width=0.4\linewidth]{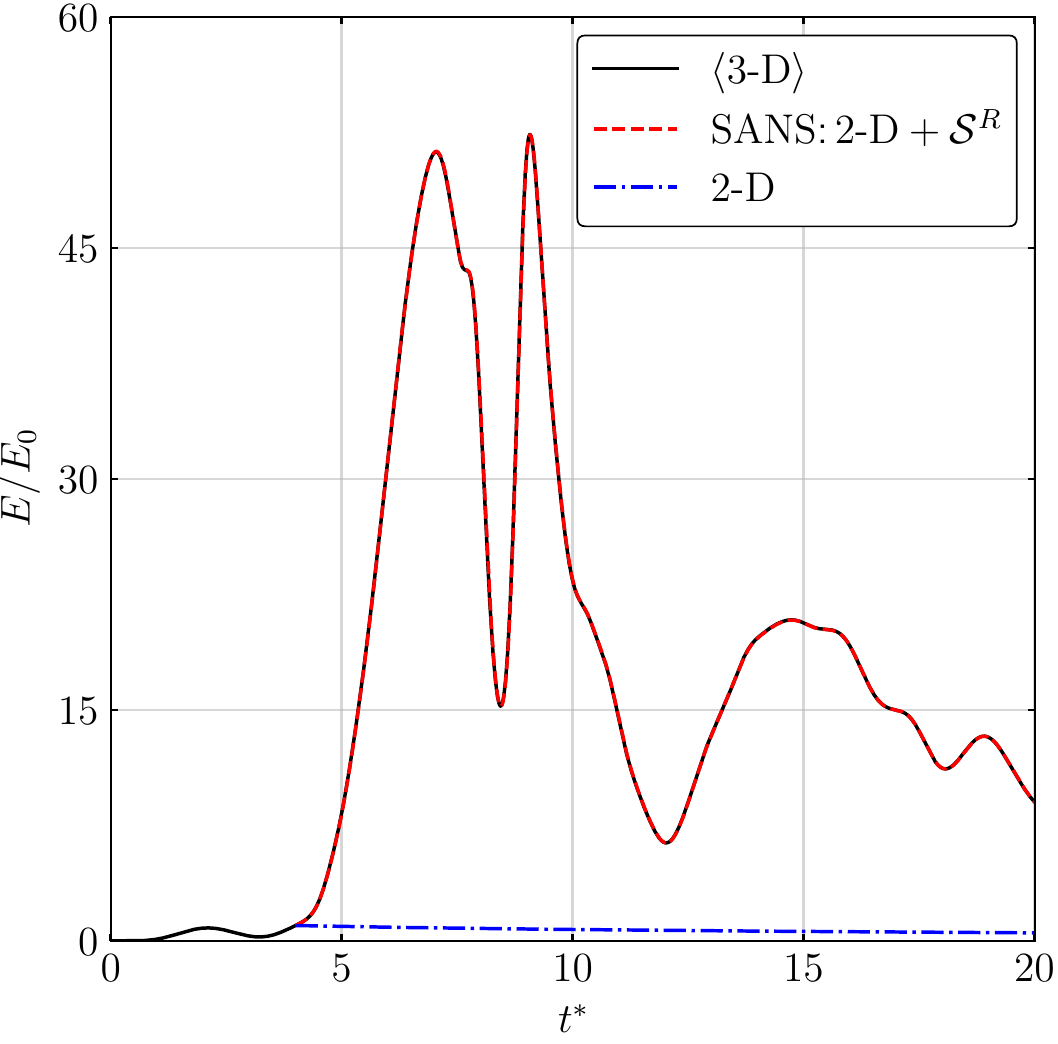}\hspace{0.5cm}
\includegraphics[width=0.4\linewidth]{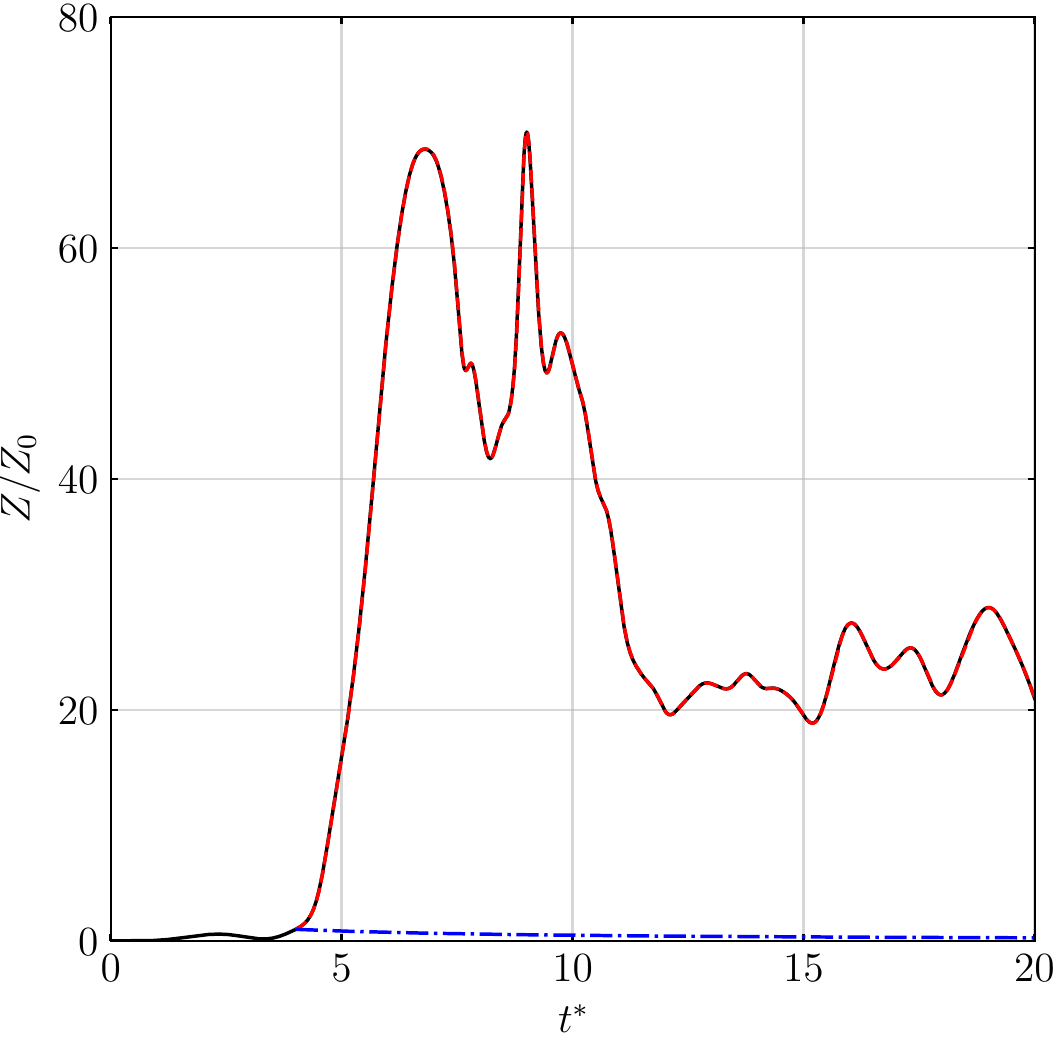}
\caption{Kinetic energy (left) and enstrophy (right) of the Taylor-Green vortex case.  
$E_0$ and $Z_0$ correspond to the energy and enstrophy at $t^*=4$. 
Energy is computed as $E=1/(2\Omega)\int \vect{U}^2\,\mathrm{d}\Omega$, where $\vect{U}=(U,V)$, and $\Omega$ is the 2-D volume.
Enstrophy is computed as $Z=1/(2\Omega)\int\Omega_z^2\,\mathrm{d}\Omega$, where $\Omega_z=\nabla\times\vect{U}$.}\label{fig:t-g_perfect}
\end{figure}

The evolution of the flow energy and enstrophy (\fref{fig:t-g_perfect}) display how the SANS system perfectly matches the 3-D spanwise-averaged flow when including the perfect closure (to machine precision).
It can also be observed that energy and enstrophy are not conserved.
Again, this is due to the inclusion of the perfect closure term $(\mathcal{S}^R)$, which changes the system dynamics from standard 2-D Navier--Stokes to SANS.
On the contrary, the 2-D case almost conserves energy and enstrophy, and the decrease of the metrics is due to viscous and numerical effects.

\subsubsection*{Flow past a circular cylinder at \texorpdfstring{$\mathbf{\boldsymbol{Re}=10^4}$}{Re=10000}} \label{sec:circular_cylinder}

\begin{figure}[!ht]
\centering
\includegraphics[width=0.6\linewidth]{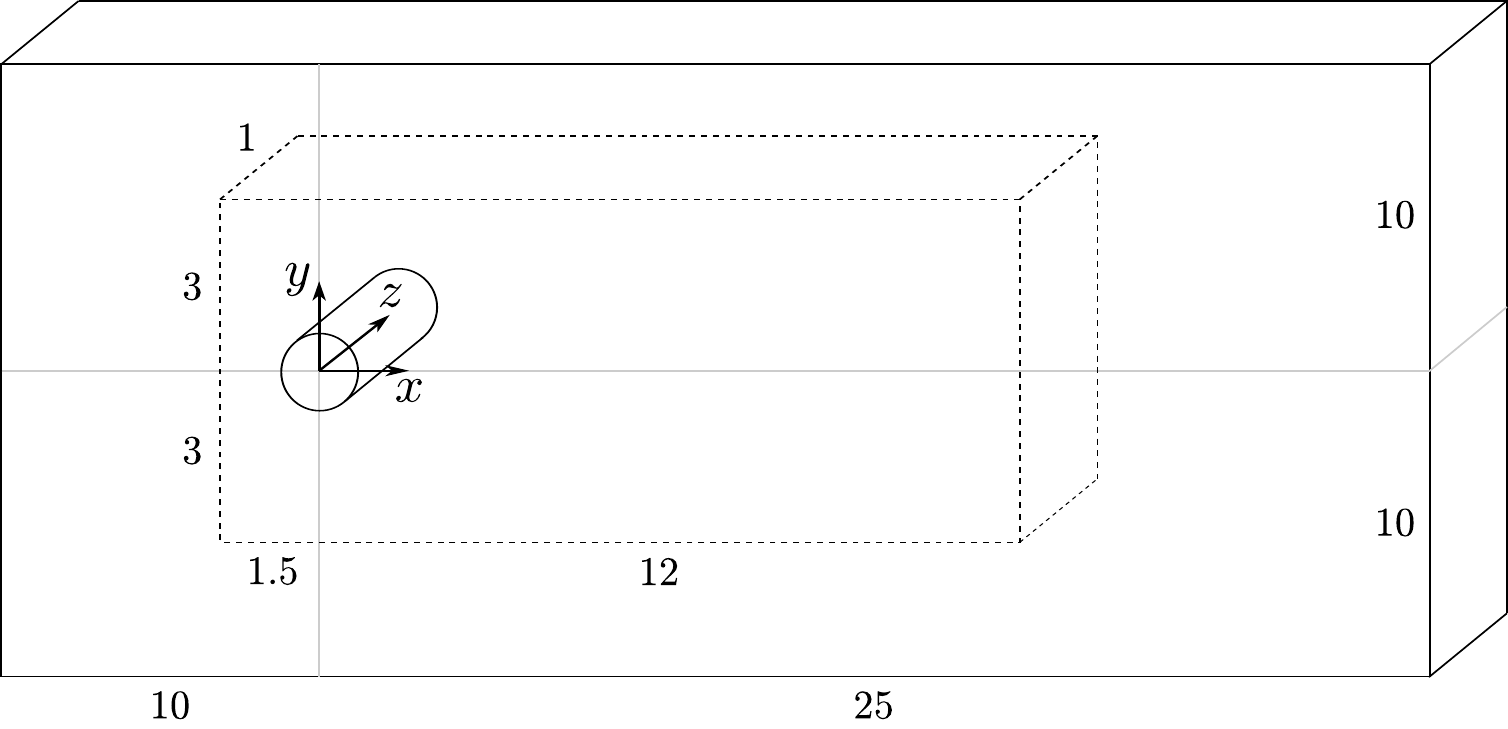}
\caption{Computational domain of the circular cylinder test case.
Dimensions are scaled with the cylinder diameter $D$. 
The dashed line defines the Cartesian domain. 
The solid line defines the boundaries of the computational domain.
A stretched grid transitions from the Cartesian grid to the domain boundaries.}
\label{fig:computational_details}
\end{figure}

Similarly to the TGV case, the SANS equations are verified with the perfect closure for flow past a circular cylinder at $Re=10^4$.
The 3-D computational domain is defined with a Cartesian grid for the wake region close to the cylinder and a stretched rectilinear grid for the region far from the cylinder (see Fig \ref{fig:computational_details}).
A resolution of 90 cells per diameter is selected for the Cartesian grid region, which proved sufficient to correctly capture the small-scale structures effect at this Reynolds regime in \citet{Font2019}.
The computational domain boundary conditions are: 
(a) uniform inlet, $\vect{u}=(1,0,0)$.
(b) natural convection outlet.
(c) no-penetration slip at the upper and lower boundaries.
(d) periodic condition in the spanwise constricting planes.
(e) no-slip condition on the cylinder boundary.

The 3-D simulation is started from a 2-D snapshot extruded along the spanwise direction to speed up the convergence to the statistically steady state.
All the data provided in this work belongs to $t^*>900$, where $t^*=tU/D$, equivalent to 180 wake cycles. 
We show in \citet{Font2019} that metrics after $t^*>200$ are already statistically steady, where a detailed verification of the solver is also provided for this exact test case.

\begin{figure}[!ht]
\centering
\includegraphics[width=0.415\linewidth]{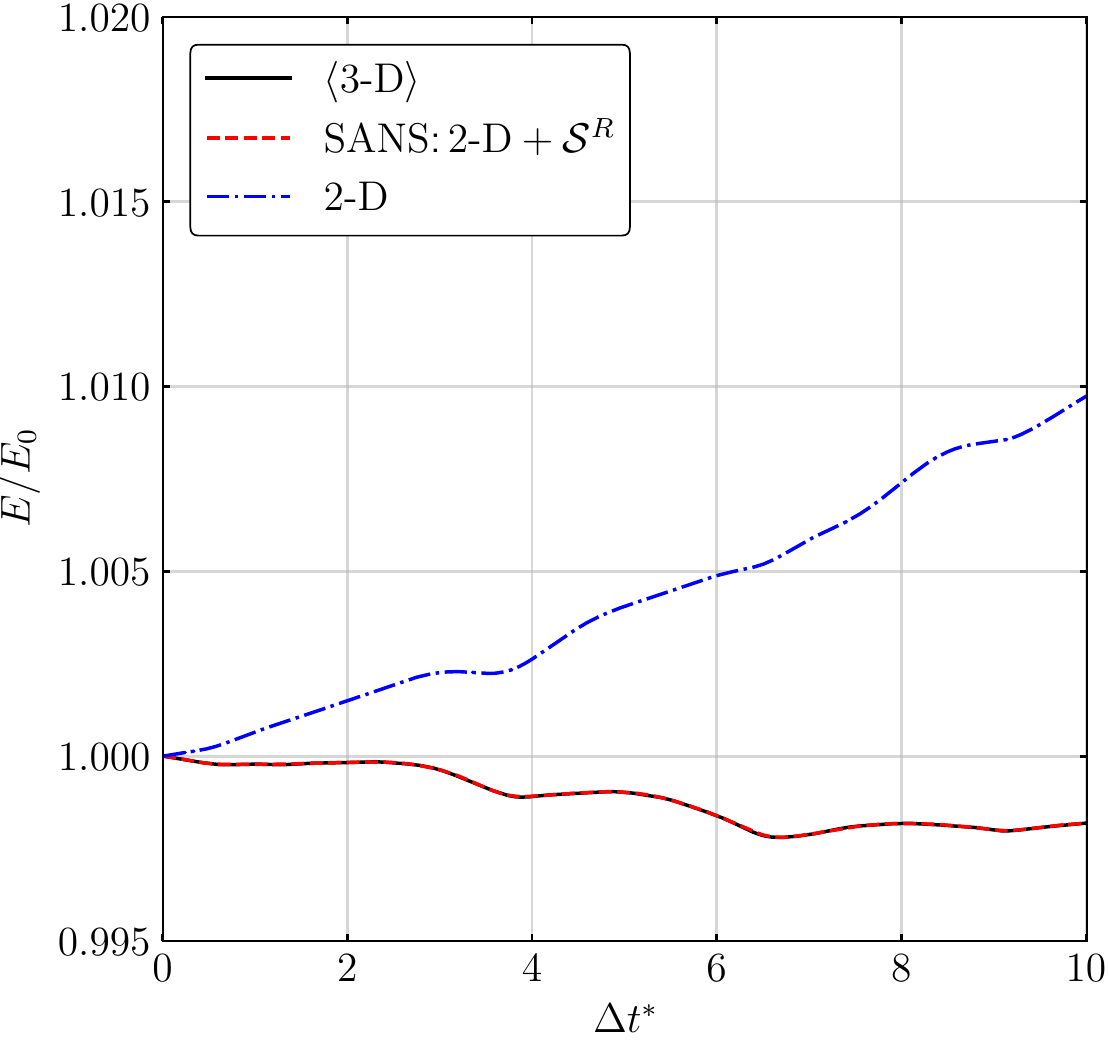}\hspace{0.5cm}
\includegraphics[width=0.4\linewidth]{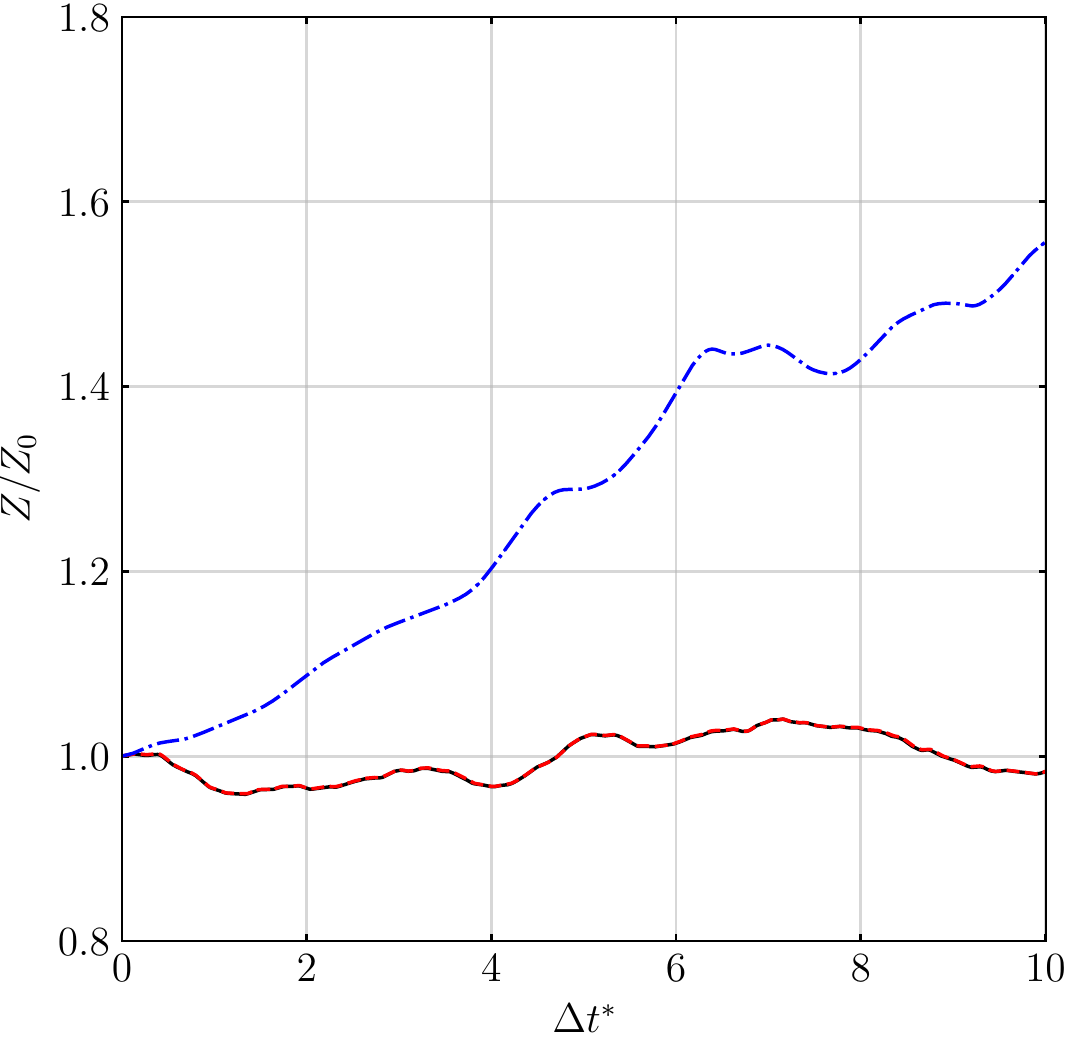}
\caption{Kinetic energy (left) and enstrophy (right) of the circular cylinder case.  
$E_0$ and $Z_0$ correspond to the energy and enstrophy at an arbitrary $t^*_0$. 
Energy and enstrophy are computed as detailed in \fref{fig:t-g_perfect}.}\label{fig:perfect_cc_EZ}
\end{figure}

\begin{figure}[!ht]
\centering
\includegraphics[width=0.4\linewidth]{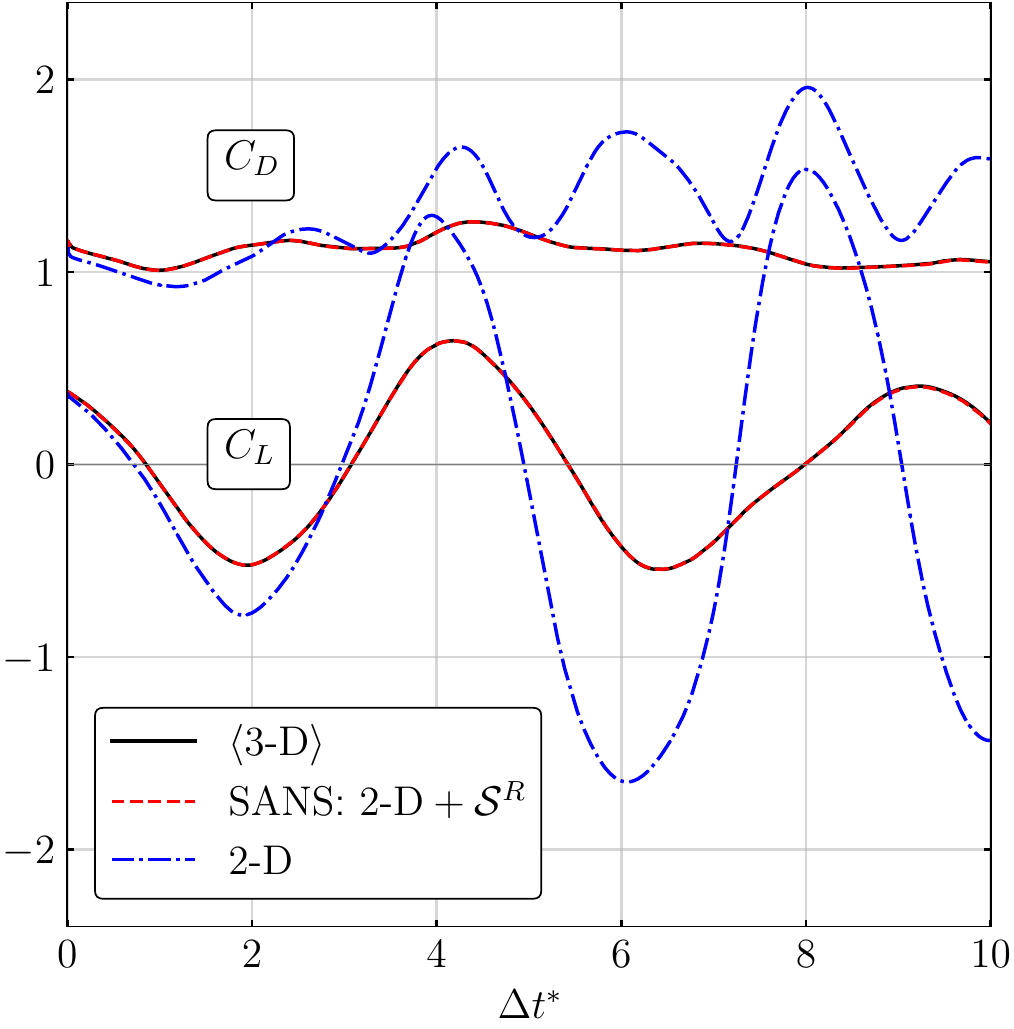}
\caption{Lift (bottom) and drag (top) force coefficients for the circular cylinder case at $Re=10^4$.
The force coefficients in the 3-D system have been calculated as $C_{D,L}=2F_{x,y}/(\rho U^2 D L_z)$, where $F_x$ and $F_y$ are respectively the horizontal and the vertical pressure forces, $\rho$ is the constant density, $U$ is the free-stream velocity, and $D$ is the cylinder diameter.
The force coefficients in the 2-D system are calculated analogously without factoring by $1/L_z$.
}
\label{fig:perfect_cc_forces}
\end{figure}

\newpage
Starting from a spanwise-averaged 3-D snapshot, the inclusion of the perfect closure in the 2-D solver at every time step allows to recover the unsteady spanwise-averaged flow.
Quantitatively, the difference between the SANS and the 2-D dynamics is emphasized in the kinetic energy and enstrophy evolution (\fref{fig:perfect_cc_EZ}), and the forces induced to the cylinder (\fref{fig:perfect_cc_forces}).
In the 2-D system, both kinetic energy and enstrophy rapidly increase as a result of its natural 2-D system attractor, and in turn larger forces are measured on the cylinder.
On the other hand, the spanwise-averaged dynamics dictate the evolution of the flow when the perfect closure is plugged into the 2-D solution, which is not altered by the presence of solid walls.

\begin{figure}[!ht]
    \centering
    \begin{subfigure}[!ht]{0.5\linewidth}        
        \centering
        \includegraphics[width=\linewidth]{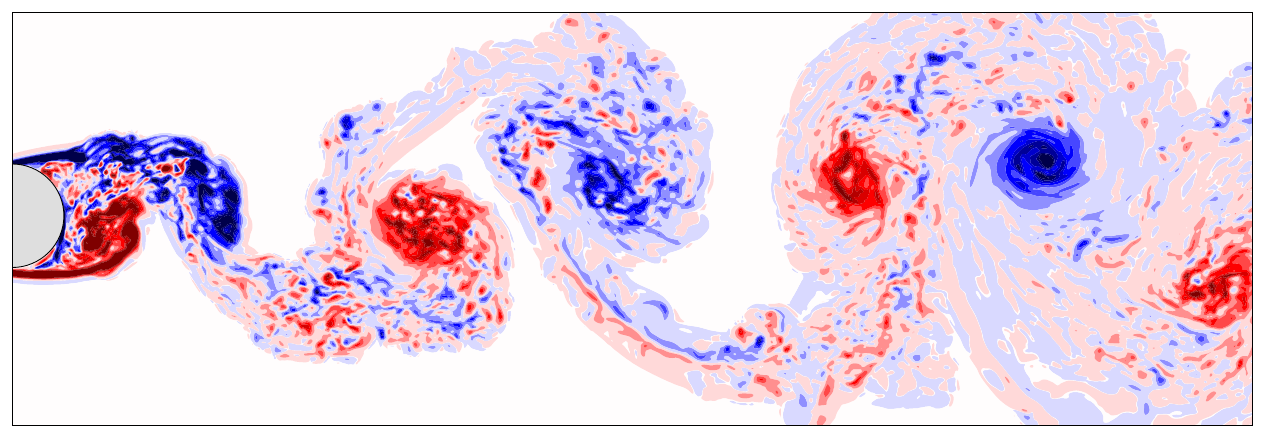}
        \caption{$\avg{3\text{-}\mathrm{D}}$, $t^*_0$.}\vspace{0.4cm}
    \end{subfigure}
    \begin{subfigure}[!ht]{0.5\linewidth}
        \centering
        \includegraphics[width=\linewidth]{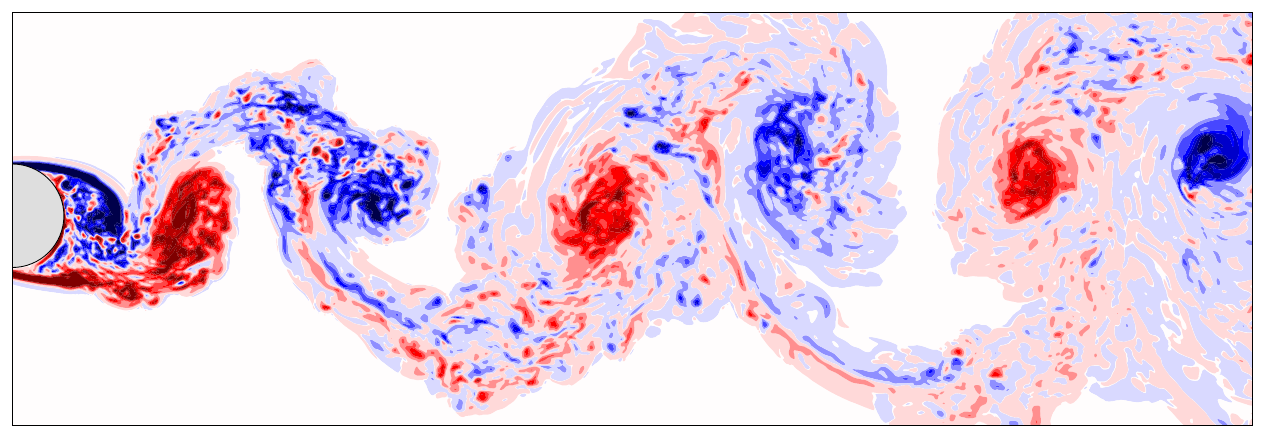}
        \caption{SANS: $2\text{-}\mathrm{D}+\mathcal{S}^{R}$, $\Delta t^*=2$.}\vspace{0.4cm}
    \end{subfigure}
    \begin{subfigure}[!ht]{0.5\linewidth}
        \centering
        \includegraphics[width=\linewidth]{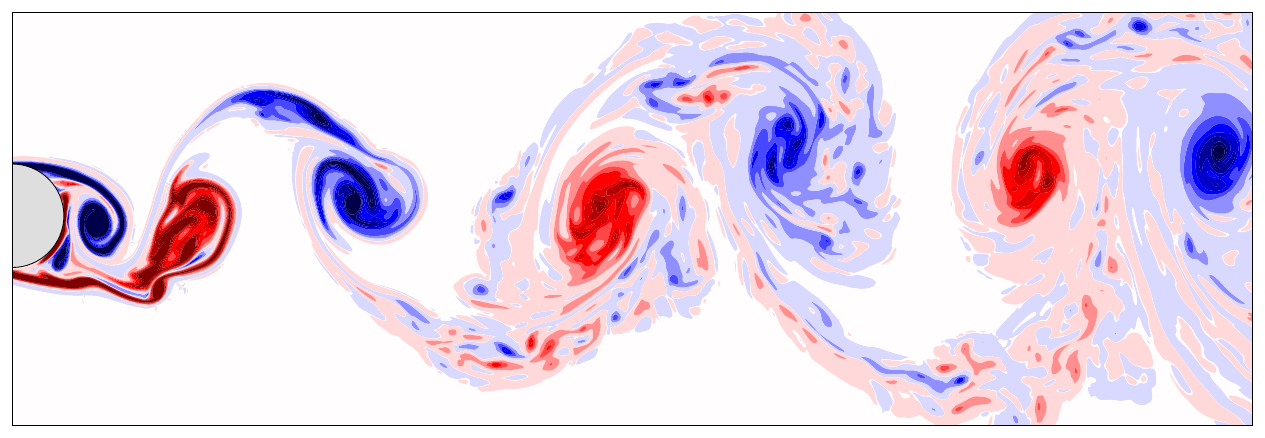}
        \caption{$2\text{-}\mathrm{D}$, $\Delta t^*=2$.}
    \end{subfigure}
    \captionsetup[subfigure]{width=\linewidth}
    \caption{Top: Vorticity of the spanwise-averaged 3-D flow used as initial condition for the SANS and the 2-D simulations.
Middle: Vorticity obtained in the SANS system after 2 convective time units.
Bottom: Vorticity obtained in the 2-D system after 2 convective time units.}
\label{fig:perfect_cc_vort}
\end{figure}

Qualitatively, it can be observed in Fig \ref{fig:perfect_cc_vort} that a rapid two-dimensionalisation of the wake develops when the perfect closure is not included in the 2-D solver.
In this scenario, vortical structures are highly coherent after just two convective time units (i.e. $\Delta t^*=t^*-t_0^*=2$).
This is reflected in the over-prediction of the forces induced to the cylinder as a result of the two-dimensionalisation mechanism yielding energized vortices in the near wake region.

The over-prediction of the hydrodynamic forces is the main short-coming of standard 2-D strip-theory methods, which arbitrarily include additional dissipation mechanisms such as 2-D turbulence models to tackle this issue.
Here we show that correctly modelling the SSR terms can bypass this problem by explicitly including 3-D effects into the governing equations while keeping the low computational cost of 2-D strip-theory methods.

\vspace{1cm}

\newpage
\subsection{Eddy-viscosity model} \label{sec:evm}

The most common model for the closure terms arising from time-averaged (RANS) or spatially-filtered (LES) Navier--Stokes equations is based on the eddy-viscosity (Boussinesq) hypothesis.
Under this hypothesis, it is assumed that the unresolved scales of motion can be linked to an effective (positive scalar) viscosity additional to the molecular viscosity of the fluid.
This implies that the closure terms can only have a dissipative effect in the governing equations.
Analogous to the definition of the stress tensor for Newtonian fluids, the eddy-viscosity hypothesis is defined as
\begin{equation}
\boldsymbol\tau^r_{ij}\equiv\avg{\vect{u\p}\otimes\vect{u\p}}-\frac{2}{3}k\delta_{ij}=-2\nu_t \avg{S_{ij}},
\label{eq:Boussinesq}
\end{equation}
where $\boldsymbol\tau^r_{ij}$ is the anisotropic turbulence stress tensor, $\avg{S_{ij}}=(\nabla\otimes\avg{\vect{u}}+\avg{\vect{u}}\otimes\nabla)/2$ is the mean rate-of-strain tensor, $\nu_t$ is the eddy viscosity, $k=\mathrm{tr}\avg{\vect{u\p}\otimes\vect{u\p}}/2$ is the turbulence kinetic energy and $\delta_{ij}$ is the Kronecker delta tensor.
The following relations arise for 2-D incompressible flow:
\begin{gather}
\avg{u\p u\p}-2k/3=-2\nu_t\partial_x U,\label{eq:t-v_h_1}\\
\avg{v\p v\p}-2k/3=-2\nu_t\partial_yV,\\
\avg{u\p v\p}=-\nu_t\pars{\partial_yU+\partial_xV}.\label{eq:t-v_h_3}
\end{gather}

The eddy viscosity field in \eref{eq:Boussinesq} has been a subject of research for decades and many different eddy-viscosity models (EVMs) have been developed for RANS and LES flow descriptions.
One of the most successful is the LES Smagorinsky model \citep{Smagorinsky1963} written as
\begin{equation}
\nu_t=\pars{C_s \Delta}^2\sqrt{2\avg{S_{ij}}\avg{S_{ij}}},
\end{equation}
where $\Delta$ is the averaging filter width, and $C_s$ is the Smagorinsky constant which tunes the amount of subgrid-scale kinetic energy to be removed.

\subsubsection*{EVM a-priori results}

To test the applicability of EVMs for SANS flow, we apply the Smagorinsky model to the circular cylinder flow field, manually selecting the value of $C_s$ to provide the best fit with the anisotropic part of the residual tensor.
As this a-priori assessment assumes knowledge of the best $C_s$ and does not require any modeling of the isotropic part of the stress tensor (i.e. the SSR kinetic energy, $k$), it represents the best possible results for the EVM.\footnotemark

\footnotetext{In an a-posteriori framework, turbulence modelling requires the full residual stress tensor to solve the averaged governing equations. Hence, EVMs usually incorporate the isotropic part $\pars{2k/3}$ in the total pressure defining a new pressure field, $p^*=p+2k/3$, on which a divergence-free velocity field is projected.
Alternatively, a transport equation for $k$ can also be considered.}

\begin{figure*}[!ht]
\begin{multicols}{2}
\begin{subfigure}{\linewidth}
    \includegraphics[width=\linewidth]{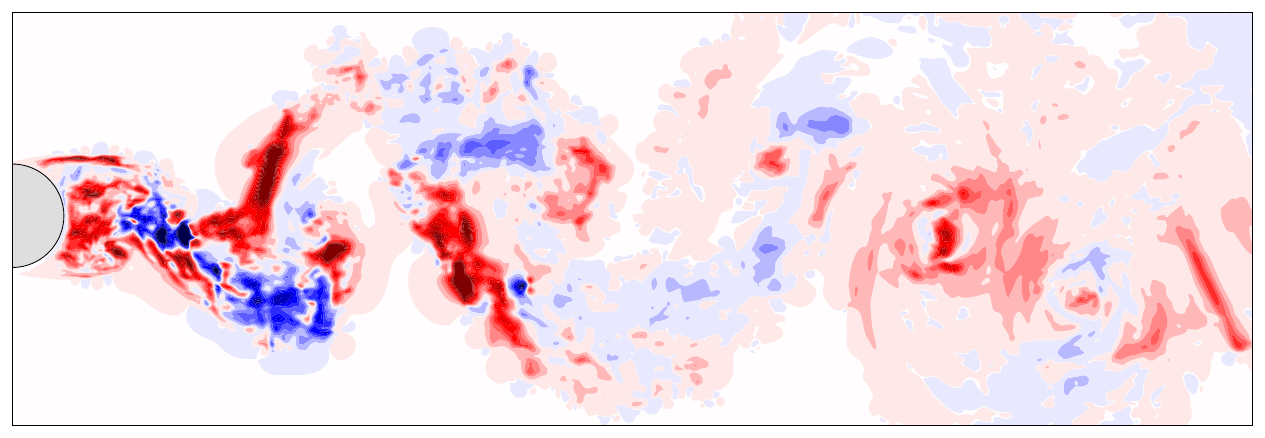}
    \caption{$\boldsymbol\tau^r_{11}$} 
\end{subfigure}
\begin{subfigure}{\linewidth}
    \includegraphics[width=\linewidth]{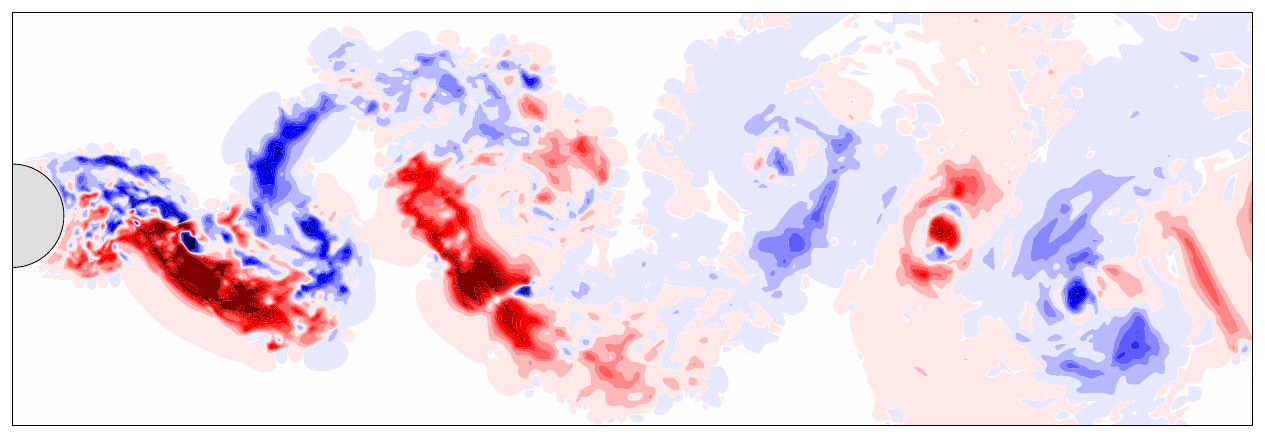}
    \caption{$\boldsymbol\tau^r_{12}$}
\end{subfigure}
\begin{subfigure}{\linewidth}
    \includegraphics[width=\linewidth]{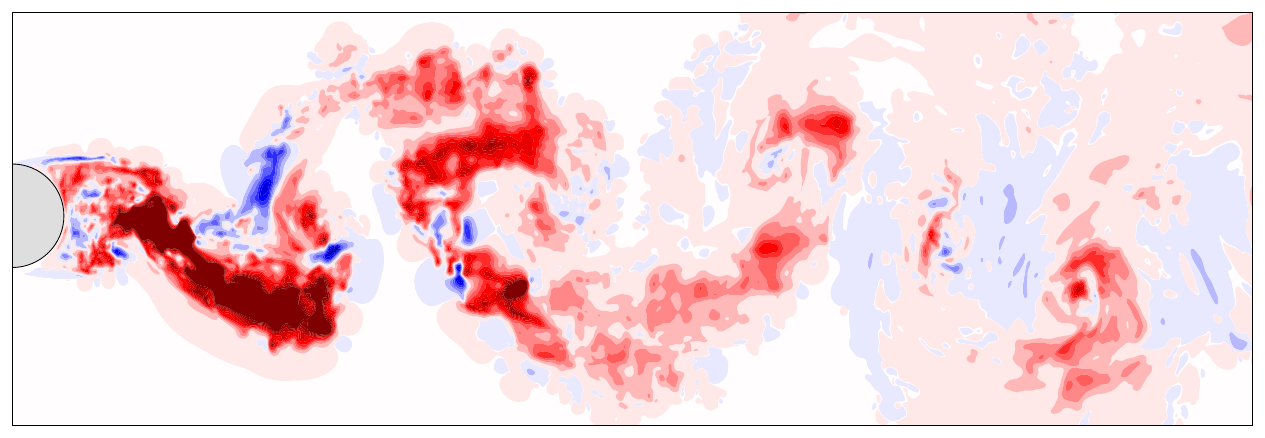}
    \caption{$\boldsymbol\tau^r_{22}$}
\end{subfigure}
\begin{subfigure}{\linewidth}
    \includegraphics[width=\linewidth]{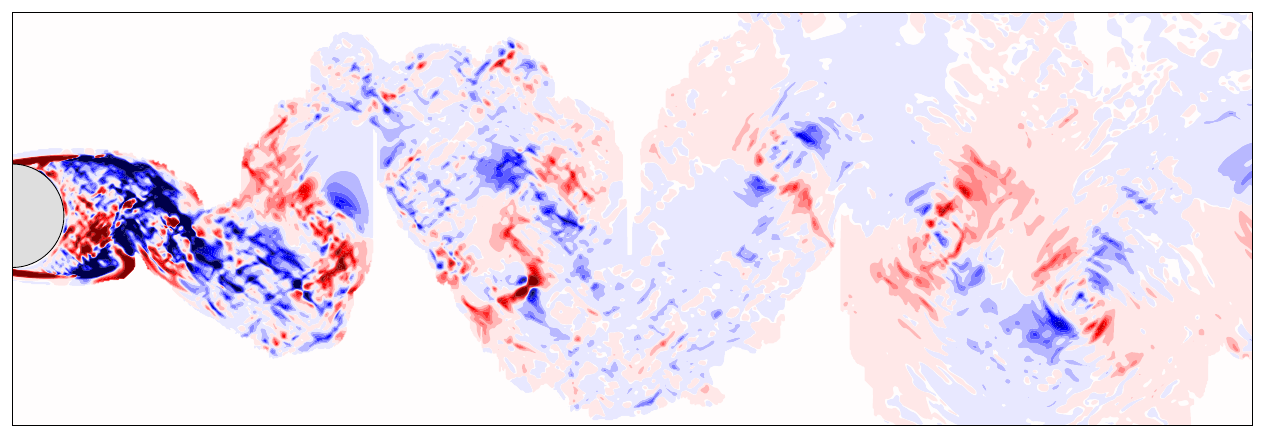}
    \caption{$\boldsymbol\tau^{r,\,\mathrm{EVM}}_{11}$}
\end{subfigure}
\begin{subfigure}{\linewidth}
    \includegraphics[width=\linewidth]{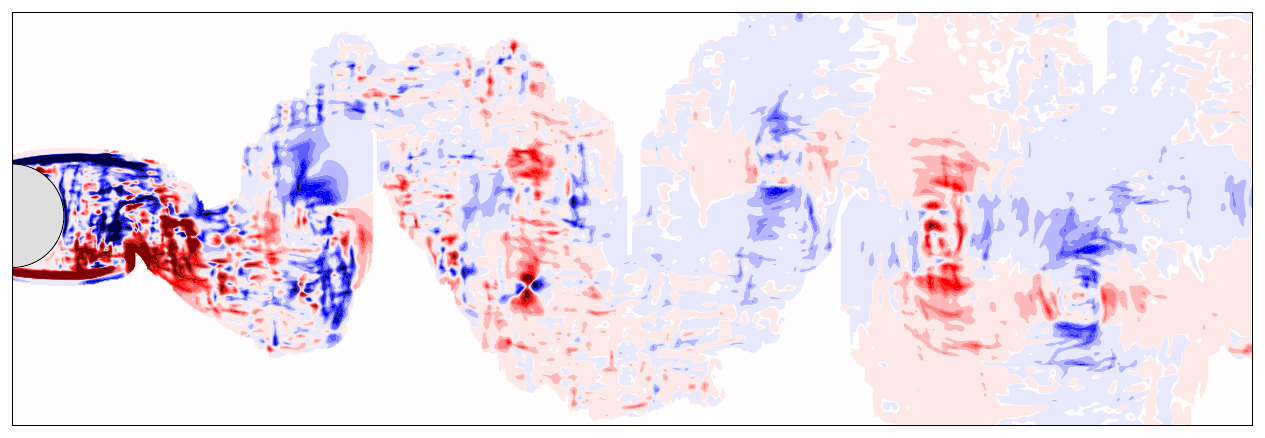}
    \caption{$\boldsymbol\tau^{r,\,\mathrm{EVM}}_{12}$}
\end{subfigure}
\begin{subfigure}{\linewidth}
    \includegraphics[width=\linewidth]{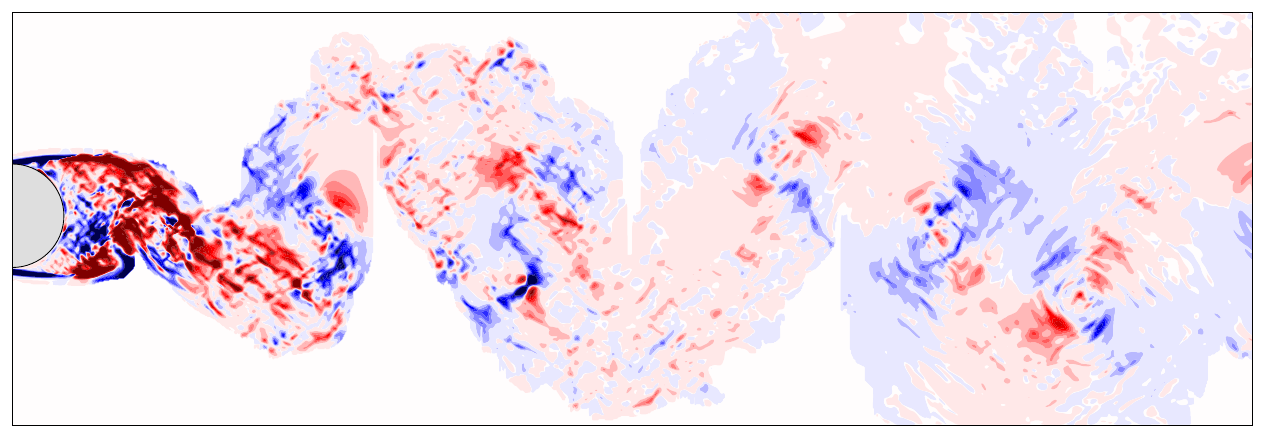}
    \caption{$\boldsymbol\tau^{r,\,\mathrm{EVM}}_{22}$}
\end{subfigure}
\end{multicols}
\caption{EVM predictions of the anisotropic SSR tensor components $(\boldsymbol\tau^{r,\,\mathrm{EVM}}_{ij})$ compared to reference data $(\boldsymbol\tau^{r}_{ij})$.
Note that a wake mask is applied to the EVM predictions as also used for the ML predictions (see \sref{sec:ML_model}).}
\label{fig:a-priori_EVM}
\end{figure*}

The a-priori results of the EVM are shown in \fref{fig:a-priori_EVM}.
Almost no qualitative similarity with the target components of the SSR anisotropic tensor can be observed, thus evidencing the poor performance of the EVM on capturing the spanwise fluctuations.
This poor performance is quantified in Tab. \ref{tab:a-priori_EVM} using the Pearson correlation coefficient
\begin{equation}
\mathcal{CC}\pars{\boldsymbol\tau^r_{ij},\boldsymbol\tau^{r,\,\mathrm{EVM}}_{ij}}=\frac{\mathrm{cov}\pars{\boldsymbol\tau^r_{ij},\boldsymbol\tau^{r,\,\mathrm{EVM}}_{ij}}}{\sqrt{\mathrm{cov}\Big(\boldsymbol\tau^r_{ij},\boldsymbol\tau^r_{ij}\Big)\mathrm{cov}\pars{\boldsymbol\tau^{r,\,\mathrm{EVM}}_{ij},\boldsymbol\tau^{r,\,\mathrm{EVM}}_{ij}}}},
\end{equation}
where the superscript $(\cdot)^{\mathrm{EVM}}$ denotes the EVM prediction.
The correlation coefficient is computed in the non-trivial wake region $x\in[0,12],y\in[-2,-2]$ (the region displayed in \fref{fig:a-priori_EVM}).
The low correlation values provide clear evidence that the EVM is not suited for the prediction of the spanwise stresses.

These poor results are ultimately expected since EVMs only account for the dissipative effects of turbulent fluctuations.
In the spanwise-averaged system, the spanwise stresses can have both a dissipative and energizing physical effect, and this is a fundamental difference which no EVM can reproduce.

\begin{table}[t]
\centering
\caption{Correlation coefficients between components of the target anisotropic SSR tensor and predictions provided by the EVM.}
\begin{tabular}{lrrr}
\toprule
$\mathcal{CC}$ & $\boldsymbol\tau^r_{11}$ & $\boldsymbol\tau^r_{12}$ & $\boldsymbol\tau^r_{22}$ \\
\midrule
EVM &  0.06& 0.06 & 0.14\\
\bottomrule
\label{tab:a-priori_EVM}
\end{tabular}

{\footnotesize The correlation coefficients are calculated for 500 snapshots and the average values are provided. \par}
\end{table}

\subsection{Machine-learning model} \label{sec:ML_model}

Data-driven closures are particularly helpful when physical information of the system to be modelled is not known a-priori, and this is certainly the case for the novel spanwise stresses developed in this work.
In this section we describe the use of machine learning to model the spanwise stresses, thus providing closure to the SANS equations.

A CNN is proposed to map resolved quantities, $\mathrm{X}_n=\left\lbrace U,V,P\right\rbrace$, to closure terms, $\mathrm{Y}_n$.
The need of a high-dimensional tunable function that can reveal hidden correlations in spatial data motivates the CNN choice \citep{LeCun1989}.
CNNs have been successfully used in the prediction of turbulent fields \citep{Beck2019, Lapeyre2019, Kim2020a}, as well as for the generation of unsteady flow fields \citep{Lee2019}.
Moreover, our immersed-boundary solver provides the image-type data structure that CNNs require so, differently from complex body-conforming grids, no interpolation errors are introduced in the framework.
The closure terms of the a-priori analysis are the SSR tensor components (similarly to \sref{sec:evm}), $\mathrm{Y}_n=\left\lbrace \avg{u\p u\p},\avg{u\p v\p},\avg{v\p v\p}\right\rbrace$.
On the other hand, the perfect closure is targeted as the output of the a-posteriori analysis, $\mathrm{Y}_n=\lbrace \mathcal{S}^R_x,\mathcal{S}^R_y\rbrace$, in order to improve the model stability as will be discussed in the ML a-posteriori results section.

\begin{figure}[!ht]
\centering
\includegraphics[width=0.9\linewidth]{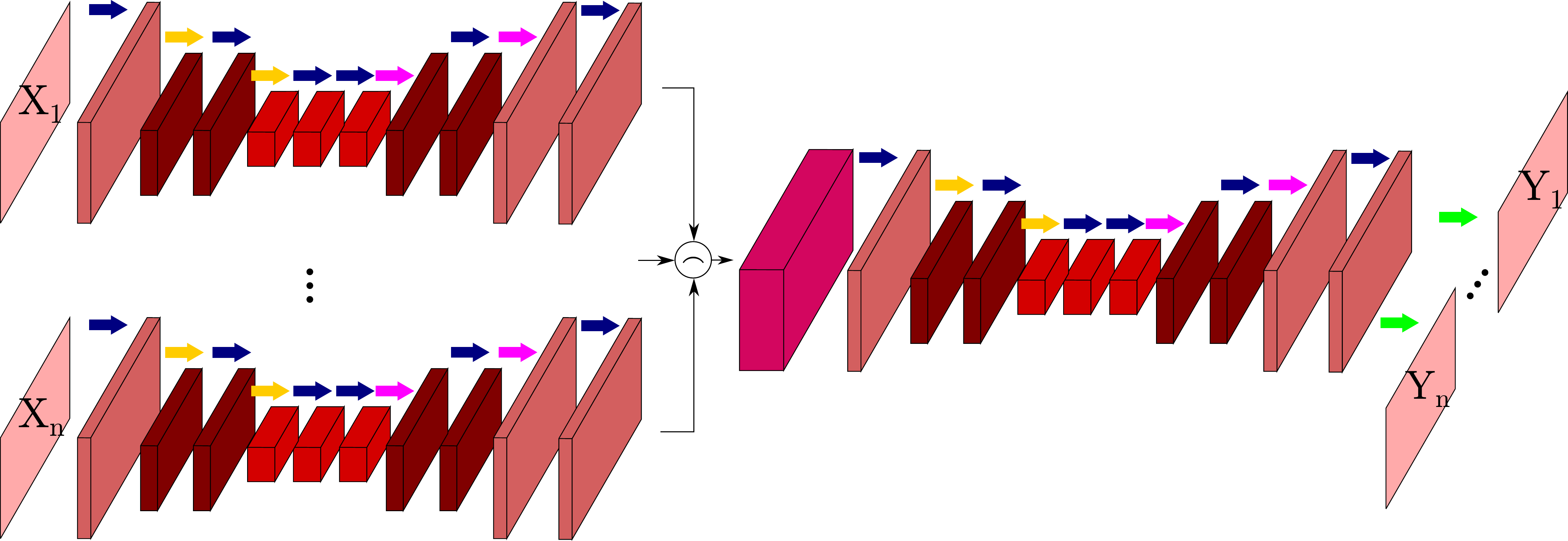}
\caption{CNN architecture.
An encoding-decoding branch is considered for each input field.
The input branches are then concatenated $(\frown)$ and another encoding-decoding branch follows.
The closure terms are separately extracted on the last layer.
Arrows legend: Dark blue: Conv2D 5x5 $+$ batch normalisation $+$ activation function.
Yellow: MaxPooling 2x2.
Magenta: UpSampling 2x2.
Green: Conv2D 1x1 $+$ wake mask $+$ Gaussian filter.
Block thickness indicates the number of filters, which double/halve at each convolutional operation in the encoding/decoding parts.}
\label{fig:cnn2}
\end{figure}

The CNN model architecture (depicted in Fig. \ref{fig:cnn2}) is based on a multiple-input multiple-output CNN (MIMO-CNN). 
The same model architecture is used for both a-priori and a-posteriori analysis, where only the number of outputs changes from three to two, respectively.
Each input branch is composed of an encoding part and a decoding part.
The number of filters doubles at every convolutional layer in the encoding part, reaching a maximum of 64 filters, and halves in the decoding part.
Each convolutional layer consists of a 2-D convolution operation with a $5\times5$ filter (convolution kernel), a batch normalisation operation, and a rectified linear unit (ReLU) activation function.
The total number of trainable parameters is 944,739.
The batch normalisation helps to speed up the neural network learning rate and improves its stability with respect to the random weights initialisation \citep{Ioffe2015}. 
The ReLU function allows higher learning rates and handles vanishing-gradient problems better than other standard activation functions such as the Sigmoid or the $\tanh$ functions \citep{Nair2010}.
All input branches are then concatenated into a single branch, for which another encoding-decoding branch follows.
The last layer consists of a wake mask function and a Gaussian filter.

\begin{figure*}[!ht]
\begin{multicols}{2}
\begin{subfigure}{\linewidth}
    \includegraphics[width=\linewidth]{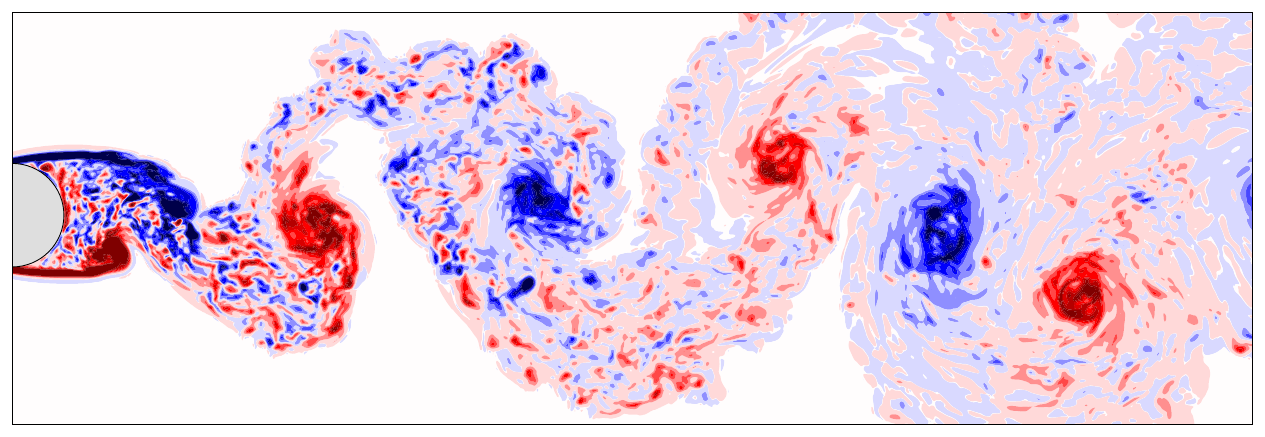}
 	\caption{$\Omega_z$} 
\end{subfigure}
\begin{subfigure}{\linewidth}
    \includegraphics[width=\linewidth]{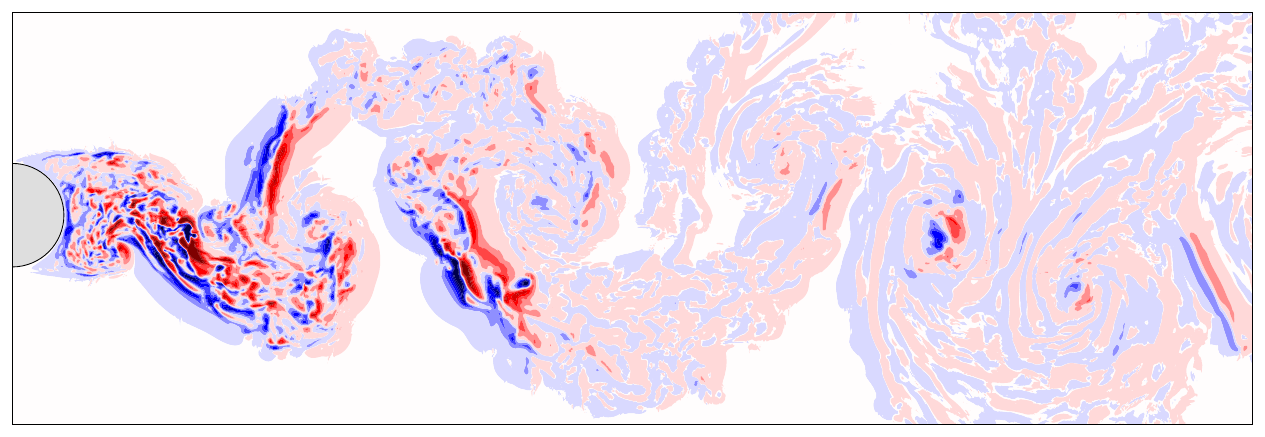}
	\caption{$\mathcal{S}^R_x$}
\end{subfigure}
\begin{subfigure}{\linewidth}
    \includegraphics[width=\linewidth]{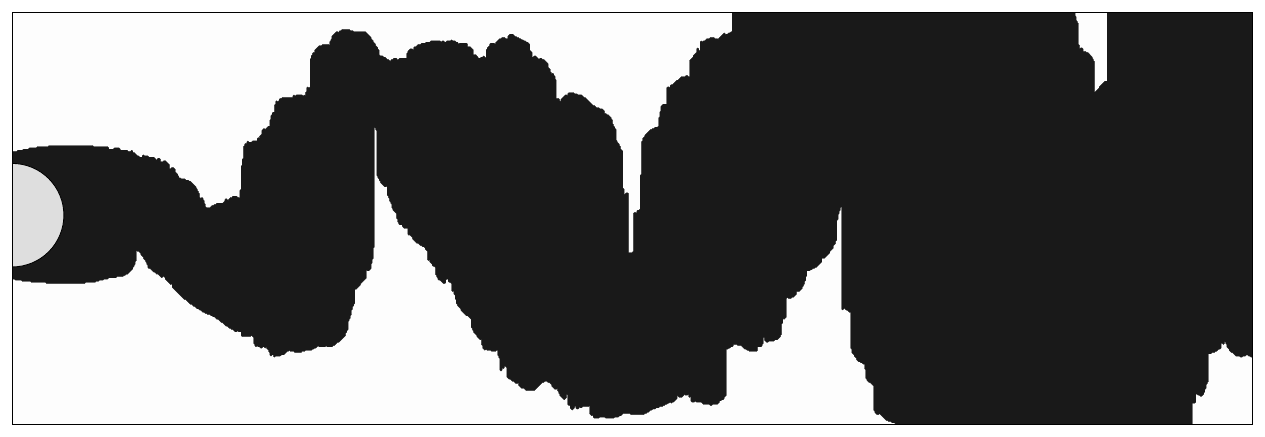}
	\caption{Wake mask used in the a-priori analysis.}\label{fig:wake1}
\end{subfigure}
\begin{subfigure}{\linewidth}
    \includegraphics[width=\linewidth]{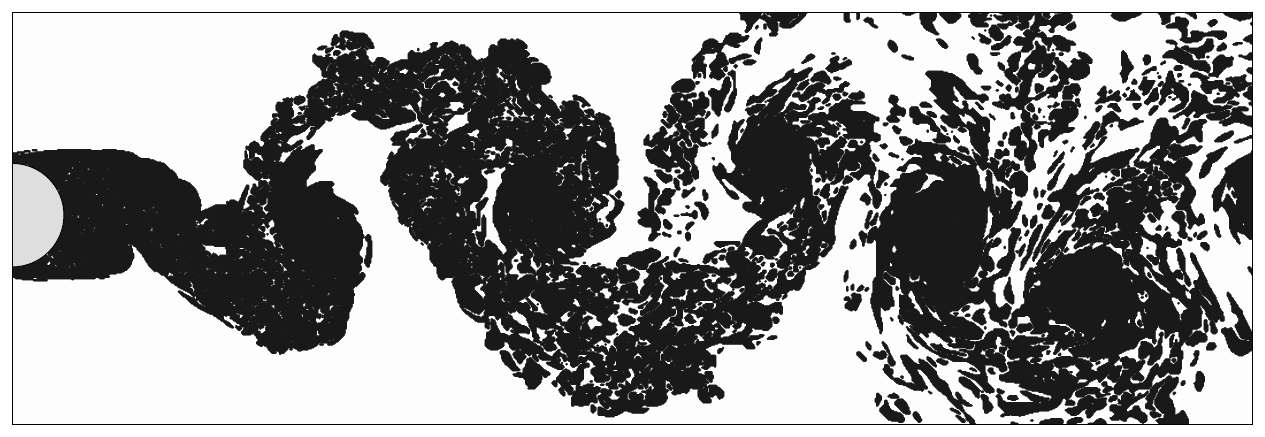} 
	\caption{Wake mask used in the a-posteriori analysis.}\label{fig:wake2}
\end{subfigure}
\end{multicols}
\caption{A wake mask from the vorticity field is generated to focus the ML model outputs on the non-trivial wake region alone.
The aim is to match the region where the target output $(\mathcal{S}^R_x)$ is activated.
In the a-posteriori analysis, the far-wake region is not fully activated in order to aid the closure stability.}
\end{figure*}

The wake mask helps providing a clean output in the regions of the domain where the closure terms are negligible.
The wake mask is based in the vorticity field for which the condition $|\Omega_z|>\epsilon$ is applied.
Hence, a field of 1s (and 0s) is generated when the condition is met (or not).
The wake mask for the a-priori stage considers all the points within the wake limit.
The wake limit is activated using the $\epsilon=1\cdot10^{-3}$ threshold (Fig. \ref{fig:wake1}).
With the use of such wake mask applied before the loss function calculation, the model can only optimize the non-trivial wake region which also encapsulates areas where the closure term is not active.
The wake mask for the a-posteriori stage applies the criterion in a pointwise manner, hence points within the wake region might not be activated, particularity in the far wake region (Fig \ref{fig:wake2}).
Also, the criterion is more strict by setting $\epsilon=3.5\cdot10^{-3}$.
This is enforced to improve the stability of the a-posteriori analysis so longer time dynamics can be evaluated.
Finally a Gaussian filter of $\sigma=0.5$ is applied to the output of the wake detection layer.

A detailed hyper-parametric study of the ML model has been conducted for the circular cylinder case in \cite{Font2020}.
In short, the present model architecture was compared against two different MIMO-CNNs, and this architecture offered the best accuracy among the tested models.
Also, it was found that the larger the CNN (in terms of trainable parameters), the faster high correlation coefficients were obtained for a similar number of training epochs (where epoch refers to an optimisation iteration on the training dataset).
The number of trainable parameters was modified by changing the number of filters per layer.
The sum of the squared error as a loss function provided the best performance when compared to the sum of the absolute error.
The use of primitive variables (velocity and pressure) did not yield significant differences when compared to the velocity gradient tensor components set. 
Both these input sets showed a better performance than a third input set containing the vorticity field and the distance function to the cylinder.

Unsteady turbulent flow past a circular cylinder with a 1 diameter span $(L_z=1)$ at $Re=10^4$ (as previously considered) has been used to generate a dataset containing a total of 7000 samples of $1216\times540$ size (i.e. the Cartesian subdomain in \fref{fig:computational_details}) during $\Delta t^*=1650$ and using a sampling rate of $\delta t^*=0.25$.
The input fields are normalised with the standard score, i.e. $\hat{x} = (x-\overline{x})/\sigma_x$.
As in \cite{Beck2019, Kim2020a}, a single $Re$ regime is used in the dataset generation, which is deemed sufficient because of the turbulent nature and unsteadiness of the flow.
The constant energy transfer rate across different scales in the inertial subrange of turbulence implies that closures trained on one $Re$ should generalise reasonably well to unseen $Re$ as long as the mean flow characteristics are similar.

The dataset is split with 6000 samples for optimisation (training), 500 samples for validation during training, and 500 for evaluation (testing) purposes.
To ensure the closure's ability to generalise and avoid overfitting, the time windows for testing, optimisation and validation do not overlap, i.e. the closure is assessed on a continuous window of previously unseen flow.
A mini-batch stochastic gradient descent method based on the Adam optimiser \citep{Kingma2014} is used to update the network weights during training.
The sum of the squared error (sse) function is used for optimisation
\begin{equation}
\mathrm{Loss}=\sum^m_i\sum^n_j \pars{\mathrm{Y}_{i,j}-\mathrm{Y}_{i,j}^{\mathrm{ML}}}^2,
\end{equation}
where $m$ is the size (number of samples) of the mini-batch, $n$ is the output index, and the superscript $(\cdot)^{\mathrm{ML}}$ denotes the ML model prediction.

The CNN training history for the a-priori output set, $\mathrm{Y}_n=\left\lbrace \avg{u\p u\p},\avg{u\p v\p},\avg{v\p v\p}\right\rbrace$, and the a-posteriori output set, $\mathrm{Y}_n=\lbrace \mathcal{S}^R_x,\mathcal{S}^R_y\rbrace$, is provided in \aref{app:ML_model_details}, where it can be appreciated that high correlation values are obtained when the sse converges to $\mathcal O(10^{-4}-10^{-6})$.
Additional details of the hardware and software used for the SANS simulation are given as well.

\subsubsection*{ML model a-priori results}\label{sec:ML_a-priori}

To allow a comparison with the EVM, the anisotropic part of the SSR tensor is recovered from the ML model predictions of the full SSR tensor.
The fact that the ML model yields the full SSR tensor already provides an advantage with respect to the EVM since no further modifications of the governing equations are required, nor an additional transport equation for the SSR kinetic energy.

\begin{figure*}[!ht]
\begin{multicols}{2}
\begin{subfigure}{\linewidth}
    \includegraphics[width=\linewidth]{a-priori/tau_r_11_lowres}
    \caption{$\boldsymbol\tau^r_{11}$} 
\end{subfigure}
\begin{subfigure}{\linewidth}
    \includegraphics[width=\linewidth]{a-priori/tau_r_12_lowres}
    \caption{$\boldsymbol\tau^r_{12}$}
\end{subfigure}
\begin{subfigure}{\linewidth}
    \includegraphics[width=\linewidth]{a-priori/tau_r_22_lowres}
    \caption{$\boldsymbol\tau^r_{22}$}
\end{subfigure}
\begin{subfigure}{\linewidth}
    \includegraphics[width=\linewidth]{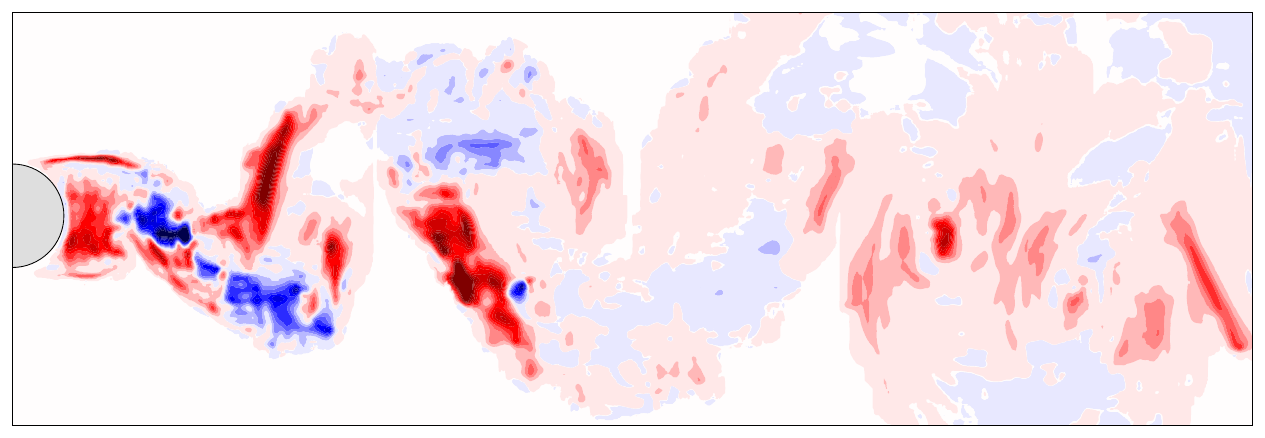}
    \caption{$\boldsymbol\tau^{r,\,\mathrm{ML}}_{11}$}
\end{subfigure}
\begin{subfigure}{\linewidth}
    \includegraphics[width=\linewidth]{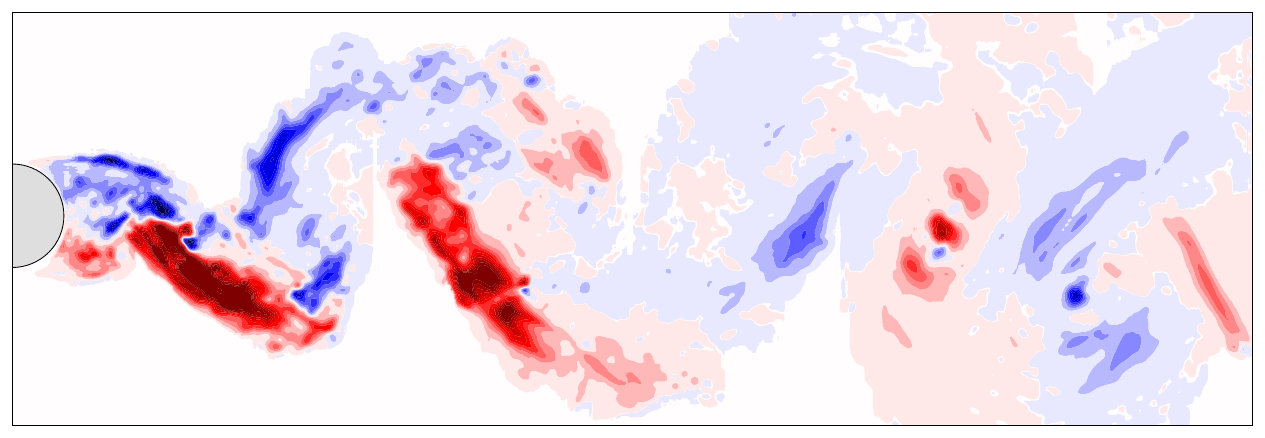}
    \caption{$\boldsymbol\tau^{r,\,\mathrm{ML}}_{12}$}
\end{subfigure}
\begin{subfigure}{\linewidth}
    \includegraphics[width=\linewidth]{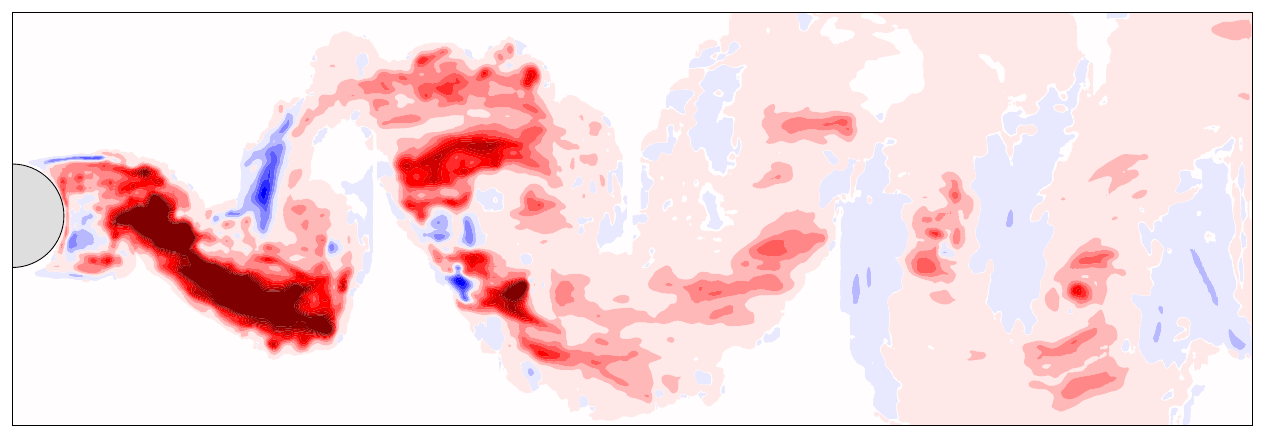}
    \caption{$\boldsymbol\tau^{r,\,\mathrm{ML}}_{22}$}
\end{subfigure}
\end{multicols}
\caption{ML model prediction of the anisotropic SSR tensor components compared to reference data.}
\label{fig:ML_a-priori}
\end{figure*}

\begin{figure*}[!ht]
\begin{multicols}{2}
\begin{subfigure}{\linewidth}
    \includegraphics[width=\linewidth]{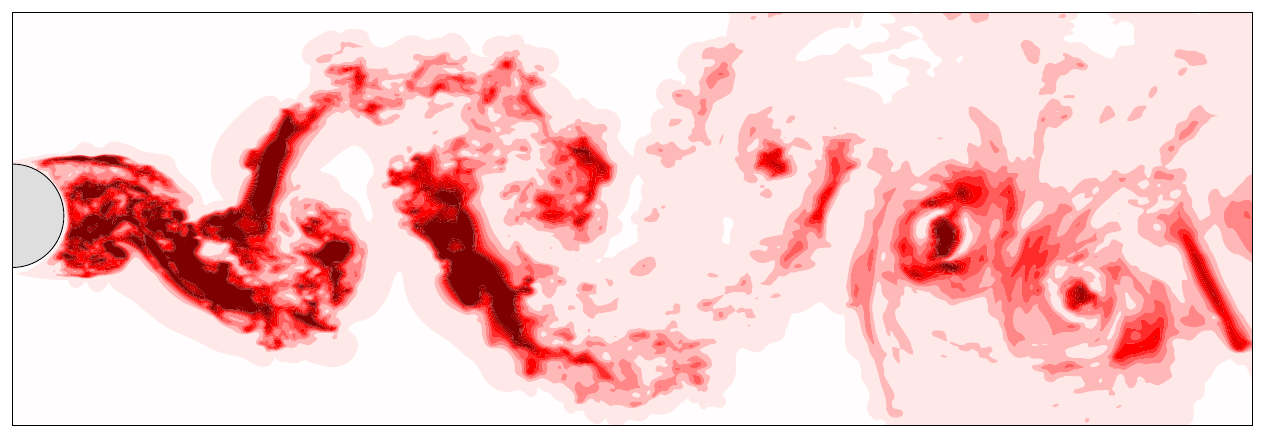}
    \caption{$\avg{u\p u\p}$} 
\end{subfigure}
\begin{subfigure}{\linewidth}
    \includegraphics[width=\linewidth]{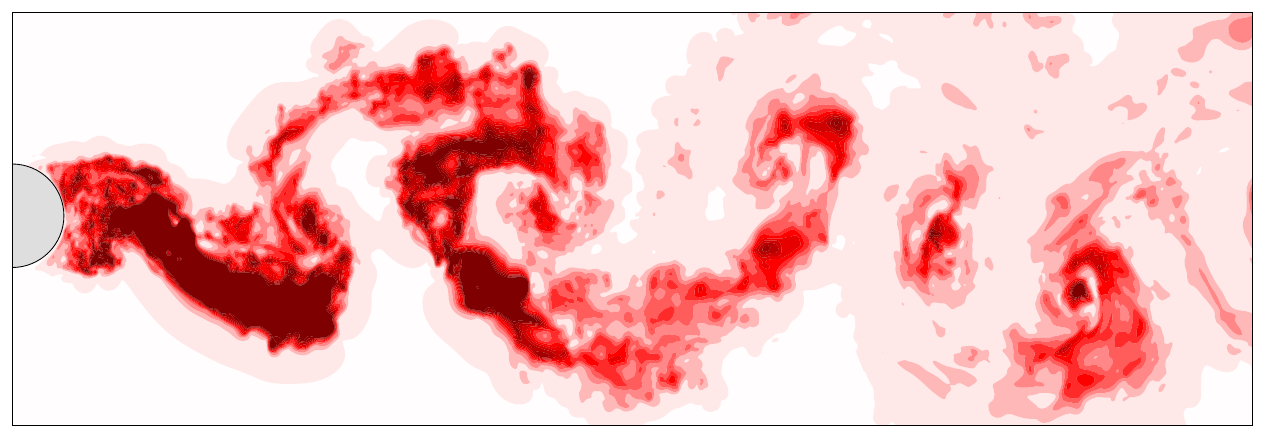}
    \caption{$\avg{v\p v\p}$}
\end{subfigure}
\begin{subfigure}{\linewidth}
    \includegraphics[width=\linewidth]{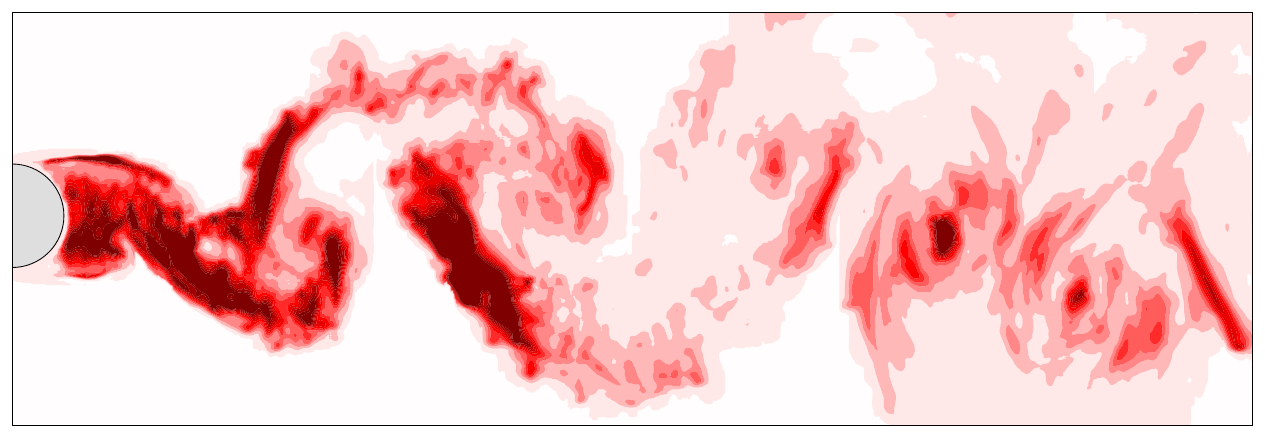}
    \caption{$\avg{u\p u\p}^{\mathrm{ML}}$}
\end{subfigure}
\begin{subfigure}{\linewidth}
    \includegraphics[width=\linewidth]{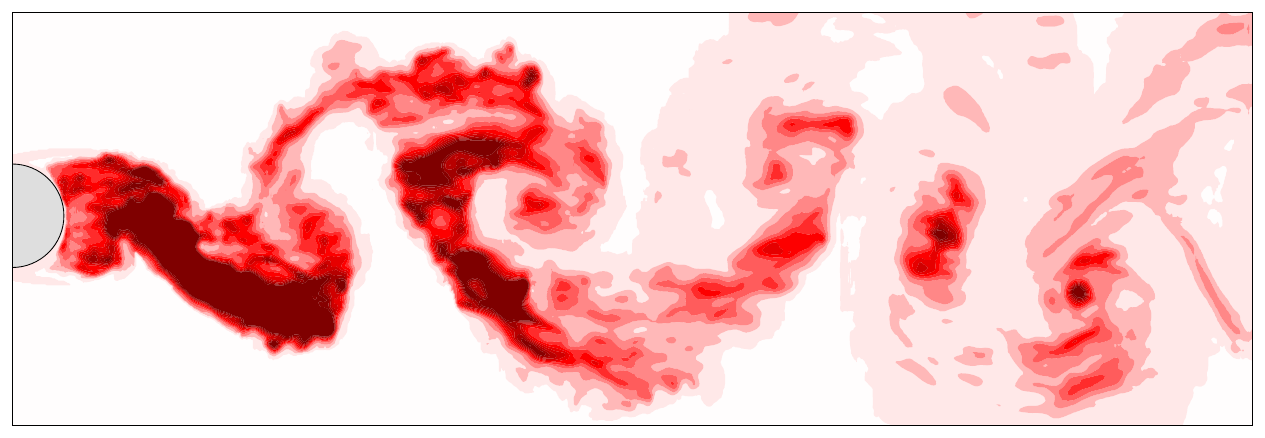}
    \caption{$\avg{v\p v\p}^{\mathrm{ML}}$}
\end{subfigure}
\end{multicols}
\caption{ML model prediction of the full SSR tensor components compared to reference data (the SSR tensor shear component is equivalent of the anisotropic SSR shear component of \fref{fig:ML_a-priori} and hence not displayed here).}
\label{fig:ML_ab_a-priori}
\end{figure*}

The predicted anisotropic SSR tensor components $(\boldsymbol\tau^{r,\,\mathrm{ML}}_{ij})$ are depicted in \fref{fig:ML_a-priori}, and the predicted components of the full SSR tensor $(\boldsymbol\tau^{R,\,\mathrm{ML}}_{ij})$ are depicted in \fref{fig:ML_ab_a-priori}.
It can be appreciated that large-scale structures are overall correctly captured.
On the other hand, small-scale structures, such as those encountered in near-wake region, are more difficult to predict.
Similar issues have been found for other CNN-based data-driven models \citep{Lee2019}.
The Gaussian filtering operation together with the wake detection feature in the CNN output layer aids to provide less noisy predictions focused on the wake region alone.
This helps mitigating error propagation in an a-posteriori framework (see next section), however, it can have an impact on the small-scale structures as well.

\begin{table}[!ht]
\centering
\caption{Correlation coefficients between components of the anisotropic SSR tensor and predictions provided by the EVM (repeated here from \tref{tab:a-priori_EVM}) and the ML model.}
\begin{tabular}{lrrr}
\toprule
$\mathcal{CC}$ & $\boldsymbol\tau^r_{11}$ & $\boldsymbol\tau^r_{12}$ & $\boldsymbol\tau^r_{22}$ \\
\midrule
EVM &  0.06& 0.06 & 0.14\\
ML &  0.90   & 0.92 & 0.92\\
\bottomrule \label{tab:ML_a-priori}
\end{tabular}

{\footnotesize The correlation coefficients are calculated for the 500 snapshots of the test dataset and the average values are provided. \par}
\end{table}

\begin{table}[!ht]
\centering
\caption{Correlation coefficients between target SSR tensor and perfect closure components and ML model predictions.}
\begin{tabular}{lrrrrr}
\toprule
$\mathcal{CC}$ & $\avg{u\p u\p}$ & $\avg{u\p v\p}$ &$\avg{v\p v\p}$ & $\mathcal{S}^R_x$ & $\mathcal{S}^R_y$ \\
\midrule
ML & 0.95 & 0.92 & 0.96 & 0.89 & 0.90\\
\bottomrule \label{tab:ML_ab_a-priori}
\end{tabular}

{\footnotesize The correlation coefficients are calculated for the 500 snapshots of the test dataset and the average values are provided. \par}
\end{table}

A significant improvement is qualitatively observed when comparing the predictions against the EVM, where the latter fails to capture almost all flow structures except for the largest ones (\fref{fig:a-priori_EVM}).
Furthermore, correlation values above $90\%$ have been achieved for all the anisotropic SSR tensor components (Tab. \ref{tab:ML_a-priori}), as well as the full SSR tensor components (Tab. \ref{tab:ML_ab_a-priori}).
It can be concluded that the ML model provides a significantly better prediction of the closure terms compared to the EVM, which is critical for SANS to correctly model turbulence wake dynamics.

\newpage

\subsubsection*{ML model generalization}\label{sec:ML_generalisation}

So far, it has been demonstrated that the trained ML model can correctly predict the SANS stresses of previously unseen snapshots of the cylinder case, i.e. the model has correctly generalized in time.
Here, the generalization of the ML model to other flow configurations such as body shapes and Reynolds regimes is assessed.
This analysis is important because, first, it provides information on whether training the ML model in a fixed set-up can still be useful for the prediction of the SANS stresses in a different case.
Second, it is a good test to check that the ML model has not over-fitted the training data even with the use of the early-stopping criterion.

Four different cases are studied:
Flow past a circular cylinder at $Re=1000$.
Flow past a circular cylinder at $Re=3900$.
Flow past an ellipse with axis $(x,y)=(2D,D)$ (ellipse 1) at $Re=10^4$.
Flow past an ellipse with axis $(x,y)=(0.5D,D)$ (ellipse 2) at $Re=10^4$.
Similarly to the circular cylinder case at $Re=10^4$, a test dataset of 500 snapshots has been generated during $\Delta t^*=125$ time units after reaching a statistically stationary state of the wake for all cases.
Both output target sets, i.e. the full SSR tensor components and the perfect closure components, are analyzed.

\begin{figure}[!ht]
\centering
\includegraphics[width=0.6\linewidth]{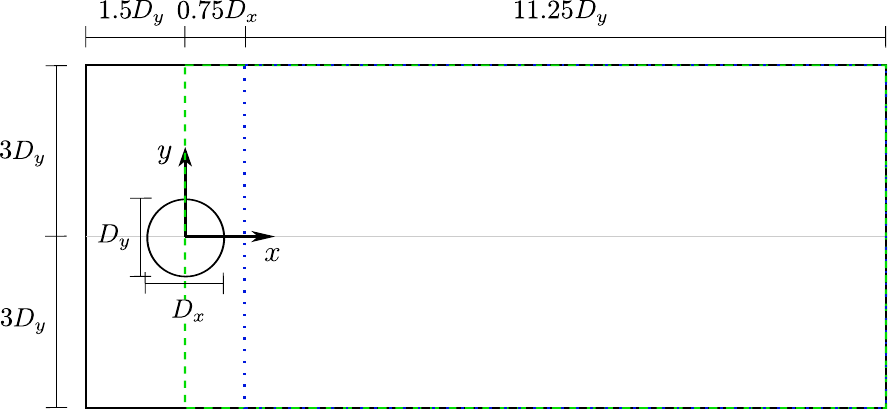}
\caption{Regions used for the ML model analysis.
Black: Input and output region of the ML model. 
Green dashed: Region used to compute the correlation coefficient between ML predictions and target data (full wake).
Blue dotted: Region used to compute the correlation coefficient between ML predictions and target data omitting the near-body domain (downstream wake).}
\label{fig:cc_regions}
\end{figure}

\begin{figure*}[!ht]
\begin{multicols}{2}
\begin{subfigure}{\linewidth}
    \includegraphics[width=\linewidth]{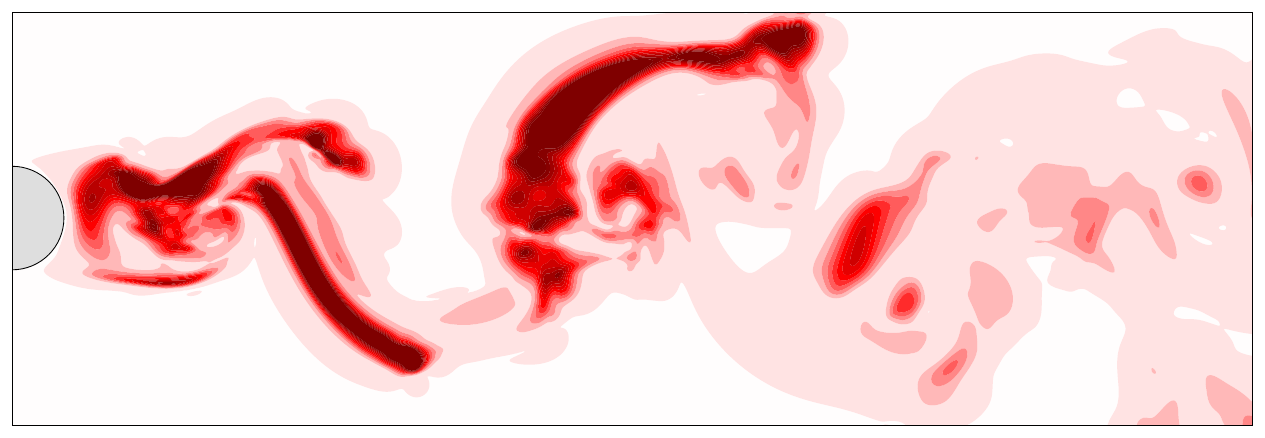}
    \caption{$\avg{u\p u\p}$} 
\end{subfigure}
\begin{subfigure}{\linewidth}
    \includegraphics[width=\linewidth]{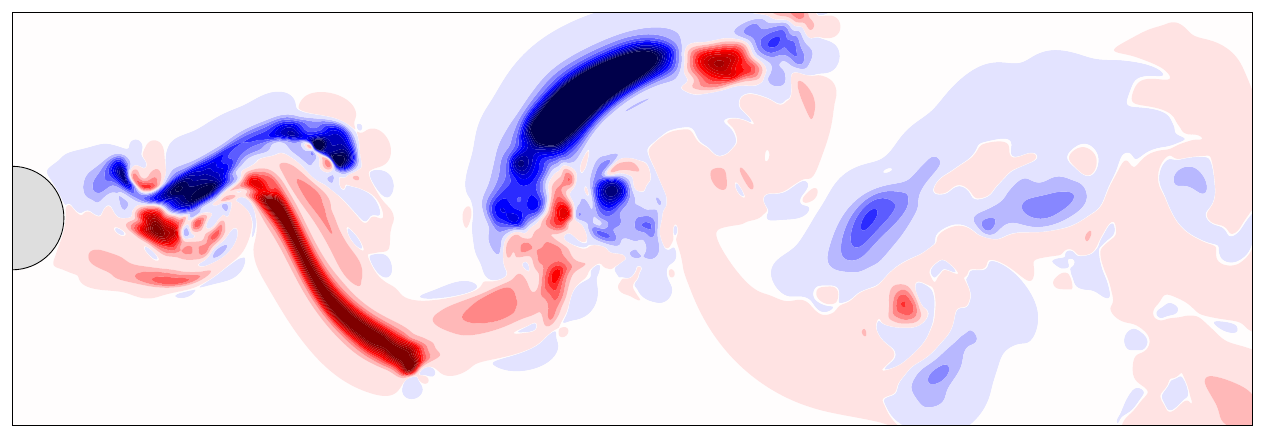}
    \caption{$\avg{u\p v\p}$}
\end{subfigure}
\begin{subfigure}{\linewidth}
    \includegraphics[width=\linewidth]{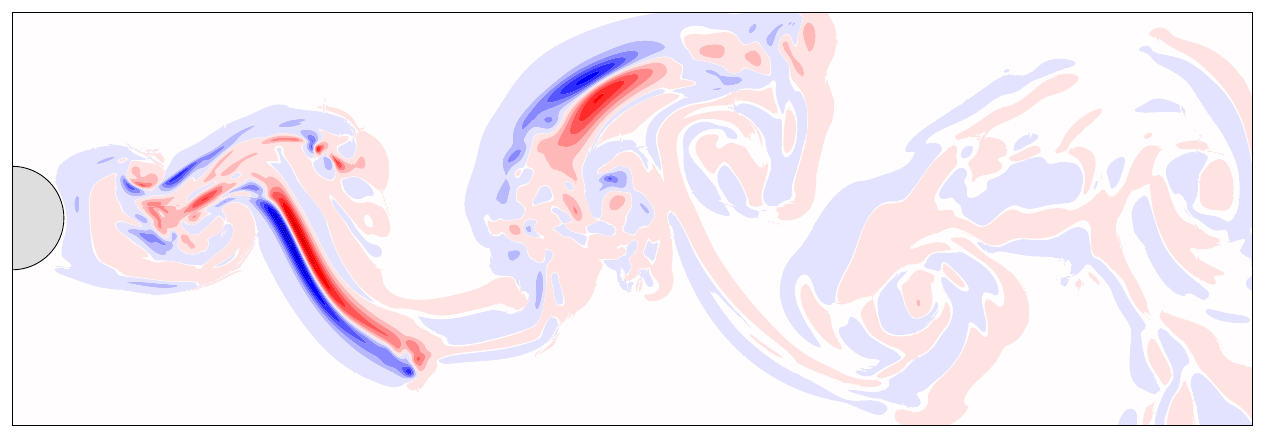}
    \caption{$\mathcal{S}^R_x$}
\end{subfigure}
\begin{subfigure}{\linewidth}
    \includegraphics[width=\linewidth]{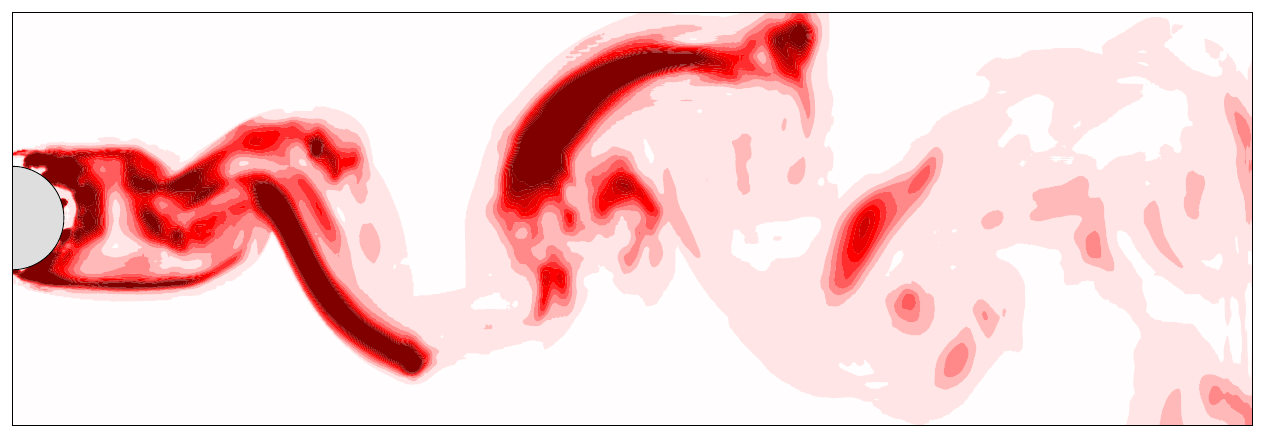}
    \caption{$\avg{u\p u\p}^{\mathrm{ML}}$} 
\end{subfigure}
\begin{subfigure}{\linewidth}
    \includegraphics[width=\linewidth]{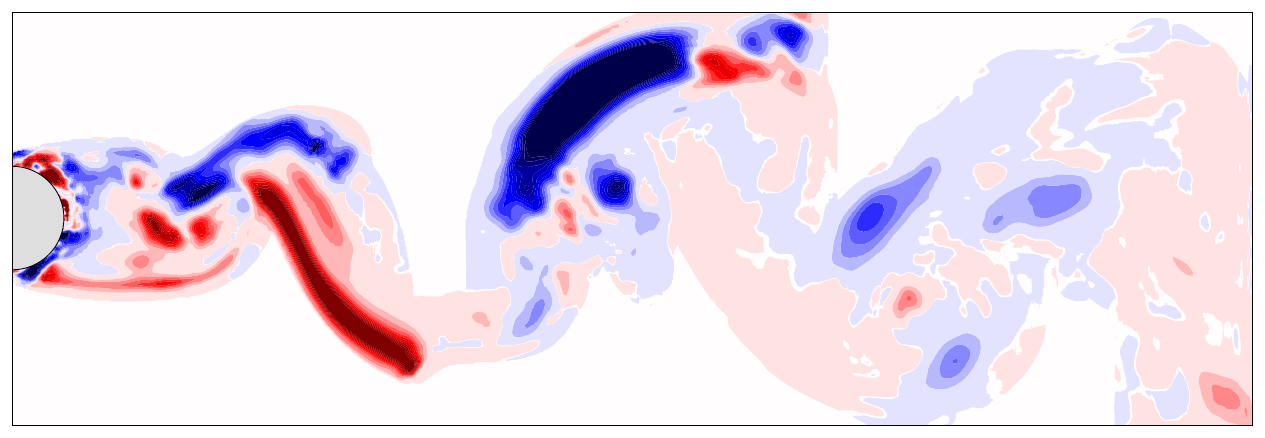}
    \caption{$\avg{u\p v\p}^{\mathrm{ML}}$}
\end{subfigure}
\begin{subfigure}{\linewidth}
    \includegraphics[width=\linewidth]{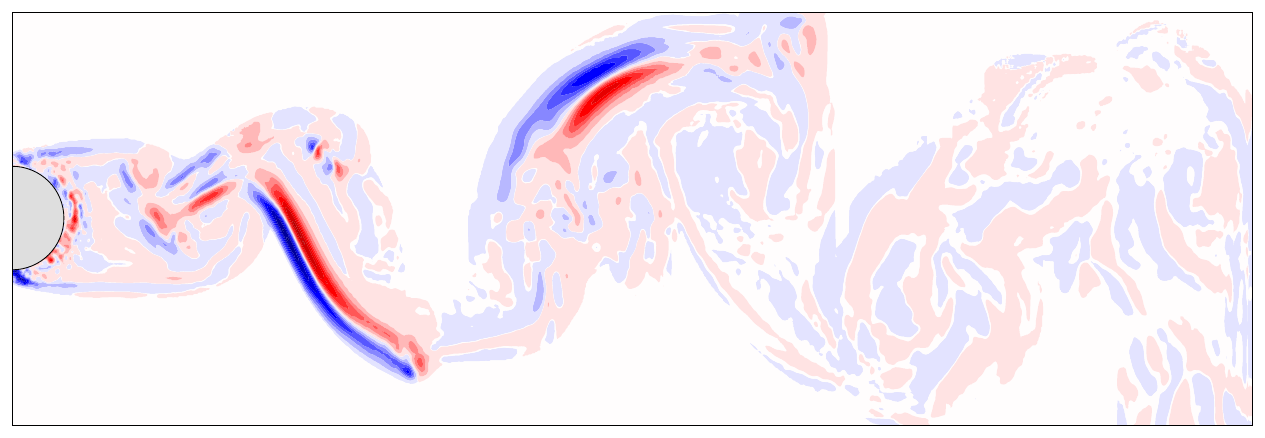}
    \caption{$\mathcal{S}_x^{R,\,\mathrm{ML}}$}
\end{subfigure}
\end{multicols}
\caption{Cylinder, $Re=1000$ case: ML model predictions of components of the SSR tensor and the perfect closure compared to reference data}
\label{fig:ML_generalisation_R1}
\end{figure*}

\begin{figure*}[!ht]
\begin{multicols}{2}
\begin{subfigure}{\linewidth}
    \includegraphics[width=\linewidth]{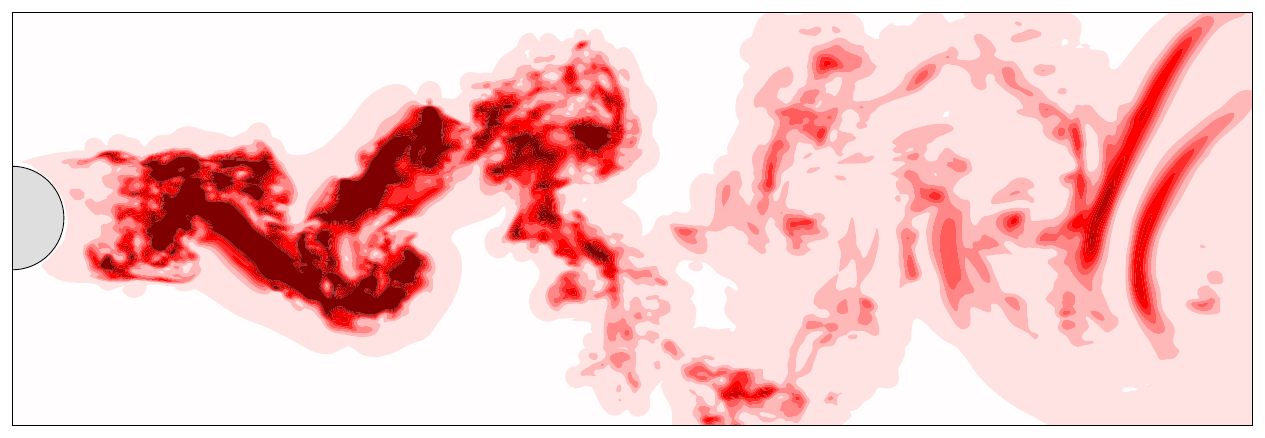}
    \caption{$\avg{u\p u\p}$} 
\end{subfigure}
\begin{subfigure}{\linewidth}
    \includegraphics[width=\linewidth]{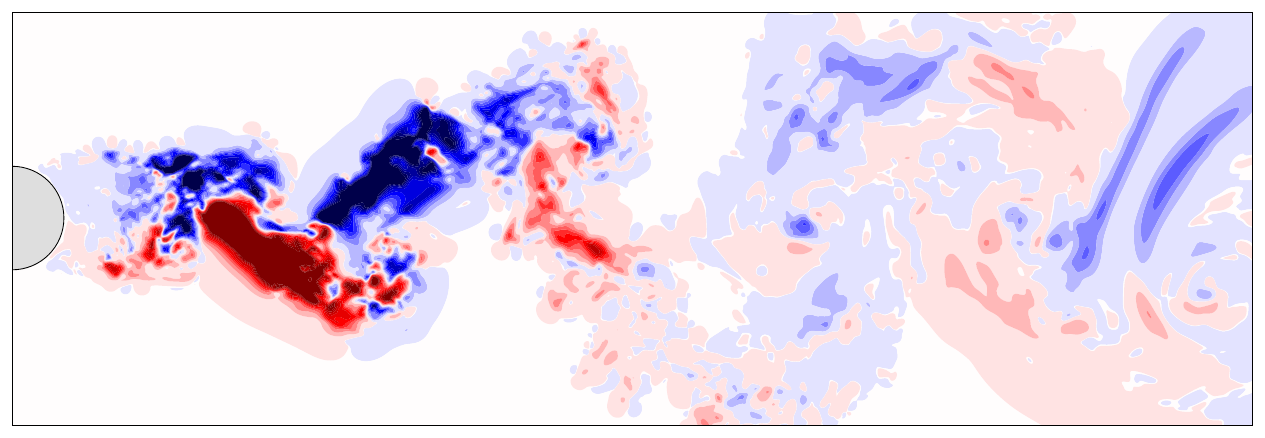}
    \caption{$\avg{u\p v\p}$}
\end{subfigure}
\begin{subfigure}{\linewidth}
    \includegraphics[width=\linewidth]{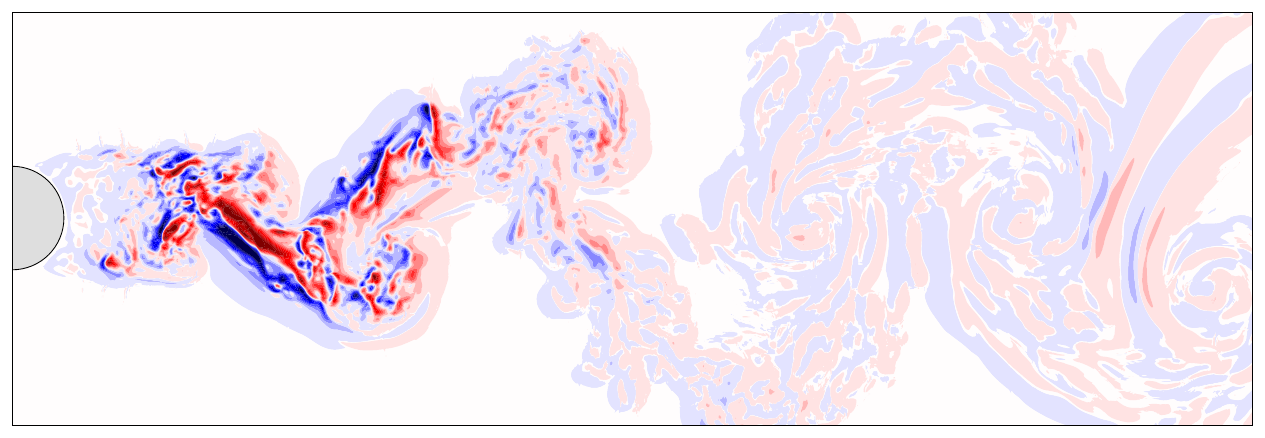}
    \caption{$\mathcal{S}^R_x$}
\end{subfigure}
\begin{subfigure}{\linewidth}
    \includegraphics[width=\linewidth]{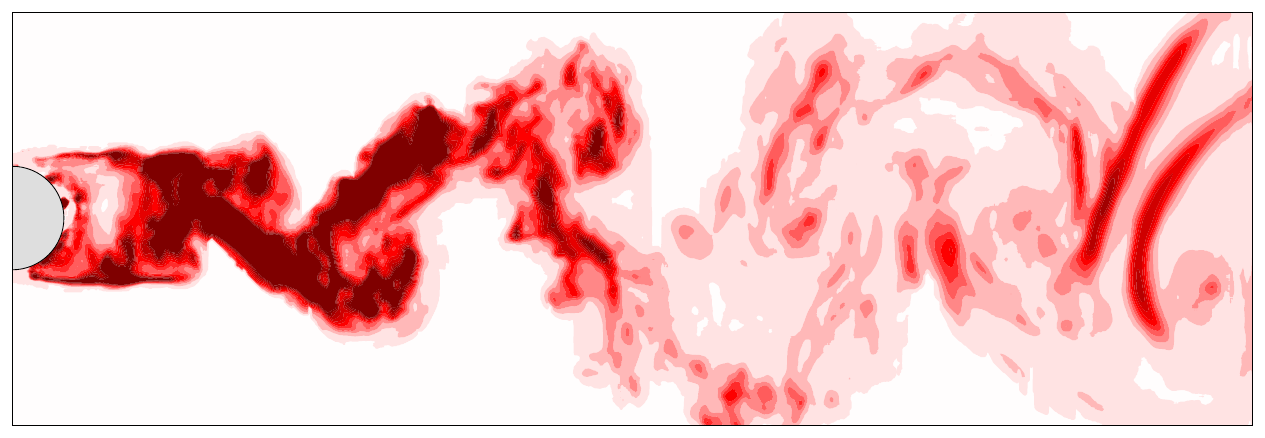}
    \caption{$\avg{u\p u\p}^{\mathrm{ML}}$} 
\end{subfigure}
\begin{subfigure}{\linewidth}
    \includegraphics[width=\linewidth]{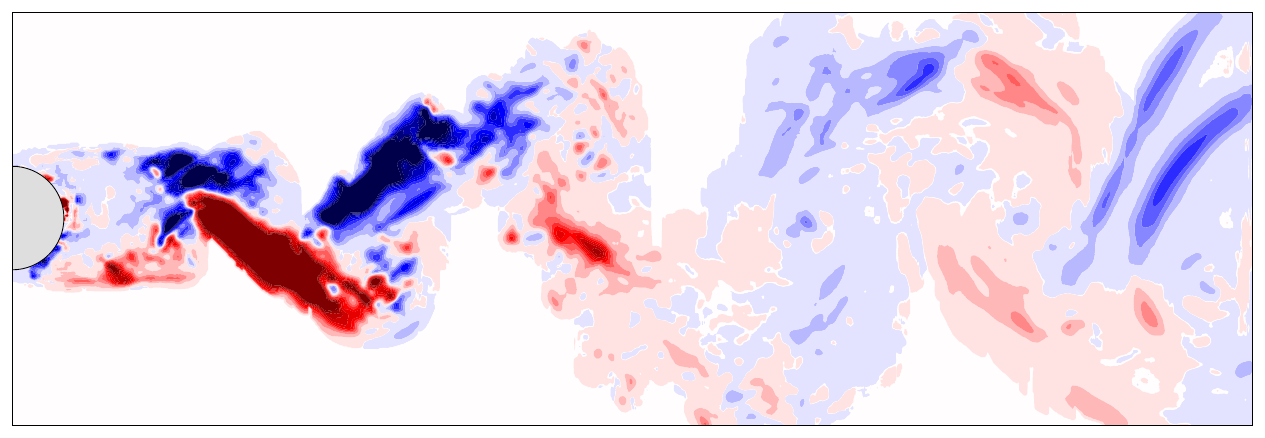}
    \caption{$\avg{u\p v\p}^{\mathrm{ML}}$}
\end{subfigure}
\begin{subfigure}{\linewidth}
    \includegraphics[width=\linewidth]{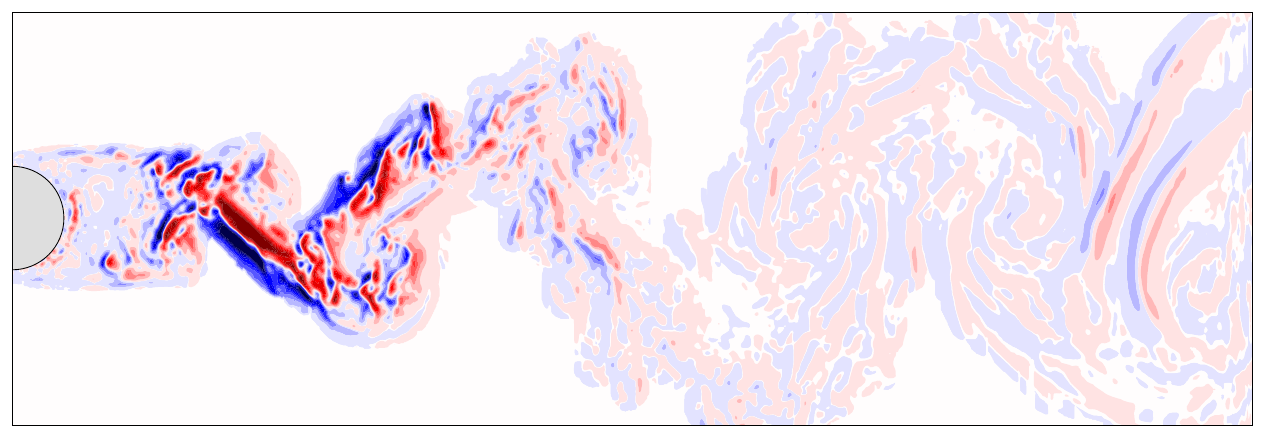}
    \caption{$\mathcal{S}_x^{R,\,\mathrm{ML}}$}
\end{subfigure}
\end{multicols}
\caption{Cylinder, $Re=3900$ case: ML model predictions of components of the SSR tensor and the perfect closure compared to reference data.}
\label{fig:ML_generalisation_R2}
\end{figure*}

\tref{tab:ML_generalisation} quantifies the accuracy of the ML model in terms of mean correlation coefficient to target data. 
Note that correlation values are provided for two wake regions, and these can be visualized in \fref{fig:cc_regions}:
one involving the non-trivial wake region (same as before, noted in green dashed lines) and one discarding the region surrounding the body (noted in blue dashed lines), i.e. $x\in[0.75D_x,12D_y]$ where $D_x$ and $D_y$ are the streamwise and crossflow body lengths, respectively.
With respect to the model generalization in different Reynolds regimes, its prediction ability on the $Re=3900$ case is similar to the training case ($Re=10^4$) since the correlation values provided in both cases are very close.
For the $Re=1000$ case, similar results are also obtained to the training case, specially when the near-body region is cropped out of the correlation analysis.
Qualitatively, the ML model prediction of the spanwise stresses is displayed in \fref{fig:ML_generalisation_R1} and \fref{fig:ML_generalisation_R2} for the $Re=1000$ and $Re=3900$ cases, respectively.
Again, a similar performance to the training case is observed;
mid- and large-scale structures are correctly captured whereas small-scale structures are not reconstructed as in the target data.
For the $Re=1000$ case, it can also be observed how the ML model struggles to predict the spanwise stresses in the region near to the body.

\begin{table}[!ht]
\centering
\caption{Correlation coefficients between target data and ML model predictions for the different generalization cases. FW and DW refer to the correlation coefficient for the green-dashed region and blue-dotted region of \fref{fig:cc_regions}, respectively.}
        \begin{tabular}{ccccccc}
            \toprule
            \multicolumn{1}{c}{Case}& $\mathcal{CC}$ & $\avg{u\p u\p}$ & $\avg{u\p v\p}$ &$\avg{v\p v\p}$ & $\mathcal{S}^R_x$ & $\mathcal{S}^R_y$ \\
            \midrule
            \multirow{2}{*}{Cylinder, $Re=1000$} & \multicolumn{1}{c}{FW} & \multicolumn{1}{c}{0.86} & \multicolumn{1}{c}{0.90} & \multicolumn{1}{c}{0.92} & \multicolumn{1}{c}{0.86} & \multicolumn{1}{c}{0.91} \\
                                & \multicolumn{1}{c}{DW} & \multicolumn{1}{c}{0.94} & \multicolumn{1}{c}{0.94} & \multicolumn{1}{c}{0.96} & \multicolumn{1}{c}{0.93} & \multicolumn{1}{c}{0.95} \\ 
            \midrule
            \multirow{2}{*}{Cylinder, $Re=3900$} & \multicolumn{1}{c}{FW} & \multicolumn{1}{c}{0.94} & \multicolumn{1}{c}{0.91} & \multicolumn{1}{c}{0.95} & \multicolumn{1}{c}{0.89} & \multicolumn{1}{c}{0.92} \\
                                & \multicolumn{1}{c}{DW} & \multicolumn{1}{c}{0.96} & \multicolumn{1}{c}{0.93} & \multicolumn{1}{c}{0.96} & \multicolumn{1}{c}{0.91} & \multicolumn{1}{c}{0.93} \\ 
            \midrule
            \multirow{2}{*}{Ellipse 1, $Re=10^4$} & \multicolumn{1}{c}{FW} & \multicolumn{1}{c}{0.61} & \multicolumn{1}{c}{0.60} & \multicolumn{1}{c}{0.79} & \multicolumn{1}{c}{0.67} & \multicolumn{1}{c}{0.69} \\
                                & \multicolumn{1}{c}{DW} & \multicolumn{1}{c}{0.93} & \multicolumn{1}{c}{0.90} & \multicolumn{1}{c}{0.95} & \multicolumn{1}{c}{0.86} & \multicolumn{1}{c}{0.91} \\ 
            \midrule
            \multirow{2}{*}{Ellipse 2, $Re=10^4$} & \multicolumn{1}{c}{FW} & \multicolumn{1}{c}{0.91} & \multicolumn{1}{c}{0.21} & \multicolumn{1}{c}{0.92} & \multicolumn{1}{c}{0.07} & \multicolumn{1}{c}{0.06} \\
                                & \multicolumn{1}{c}{DW} & \multicolumn{1}{c}{0.94} & \multicolumn{1}{c}{0.91} & \multicolumn{1}{c}{0.94} & \multicolumn{1}{c}{0.89} & 0.89 \\ 
            \bottomrule
        \end{tabular}\label{tab:ML_generalisation}

\vspace{0.5cm}
{\footnotesize The correlation coefficients are calculated for the 500 snapshots of the test dataset and the average values are provided. \par}
\end{table}

\begin{figure*}[!ht]
\begin{multicols}{2}
\begin{subfigure}{\linewidth}
    \includegraphics[width=\linewidth]{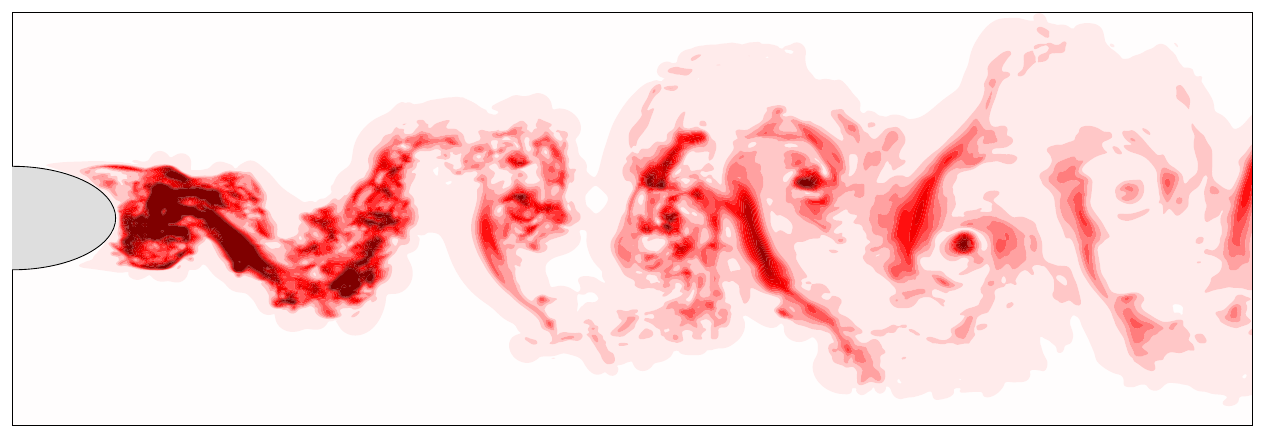}
    \caption{$\avg{u\p u\p}$} 
\end{subfigure}
\begin{subfigure}{\linewidth}
    \includegraphics[width=\linewidth]{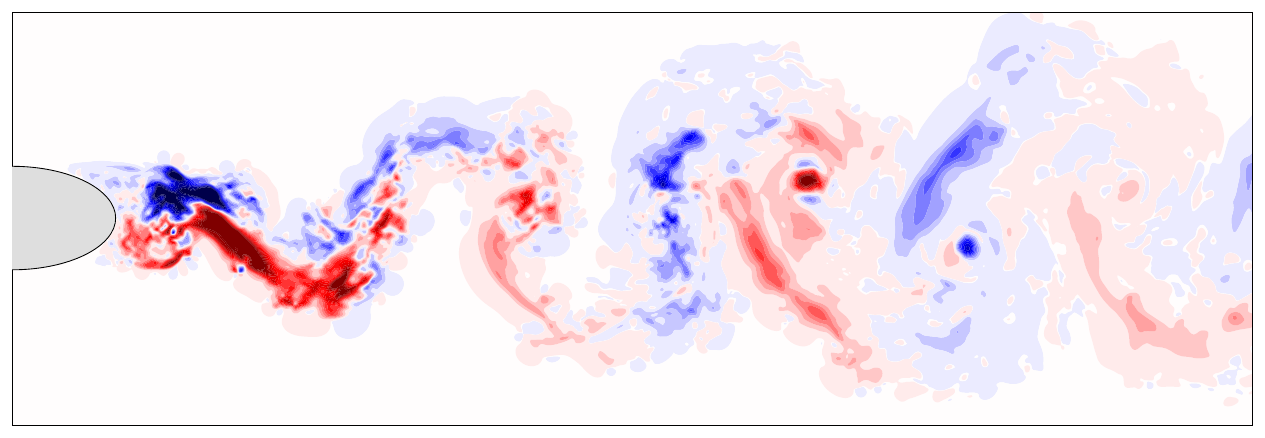}
    \caption{$\avg{u\p v\p}$}
\end{subfigure}
\begin{subfigure}{\linewidth}
    \includegraphics[width=\linewidth]{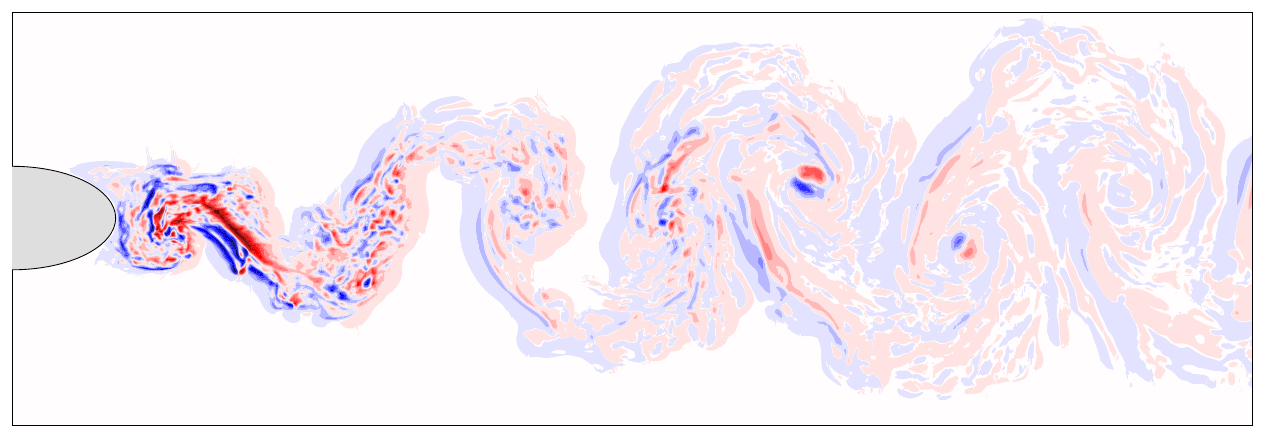}
    \caption{$\mathcal{S}^R_x$}
\end{subfigure}
\begin{subfigure}{\linewidth}
    \includegraphics[width=\linewidth]{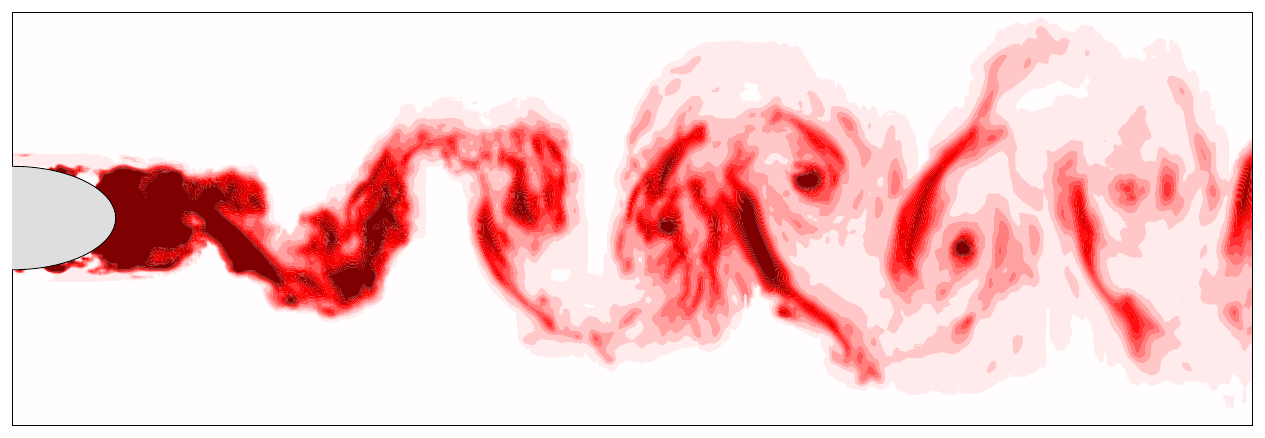}
    \caption{$\avg{u\p u\p}^{\mathrm{ML}}$} 
\end{subfigure}
\begin{subfigure}{\linewidth}
    \includegraphics[width=\linewidth]{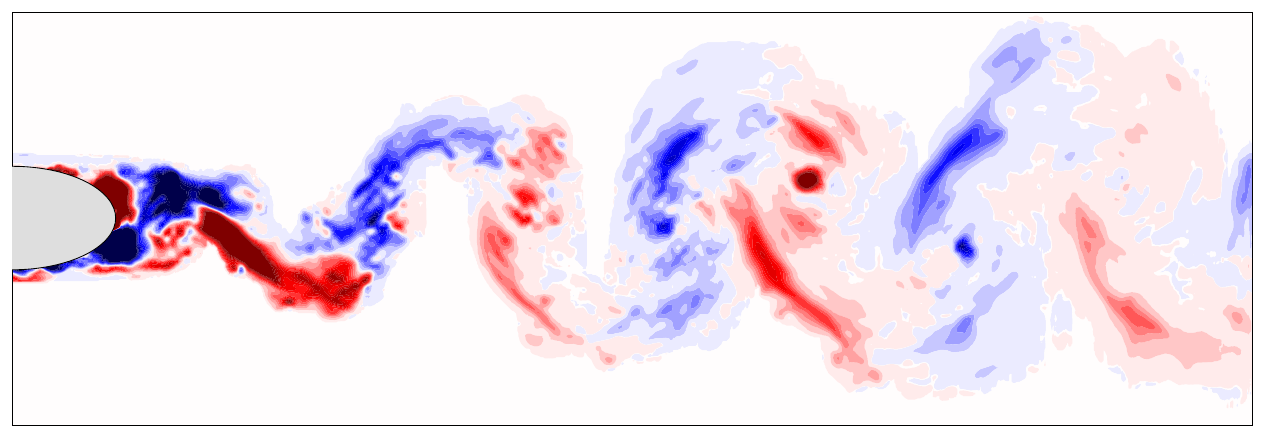}
    \caption{$\avg{u\p v\p}^{\mathrm{ML}}$} \label{fig:ML_generalisation_E1_uv}
\end{subfigure}
\begin{subfigure}{\linewidth}
    \includegraphics[width=\linewidth]{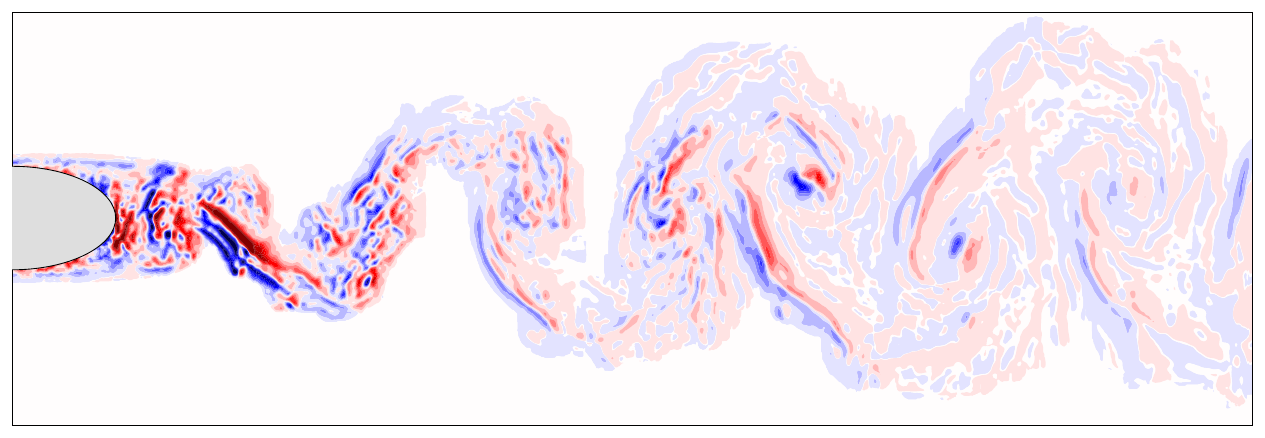}
    \caption{$\mathcal{S}_x^{R,\,\mathrm{ML}}$}
\end{subfigure}
\end{multicols}
\caption{Ellipse 1, $Re=10^4$ case: ML model predictions of components of the SSR tensor and the perfect closure compared to reference data.}
\label{fig:ML_generalisation_E1}
\end{figure*}

\begin{figure*}[!ht]
\begin{multicols}{2}
\begin{subfigure}{\linewidth}
    \includegraphics[width=\linewidth]{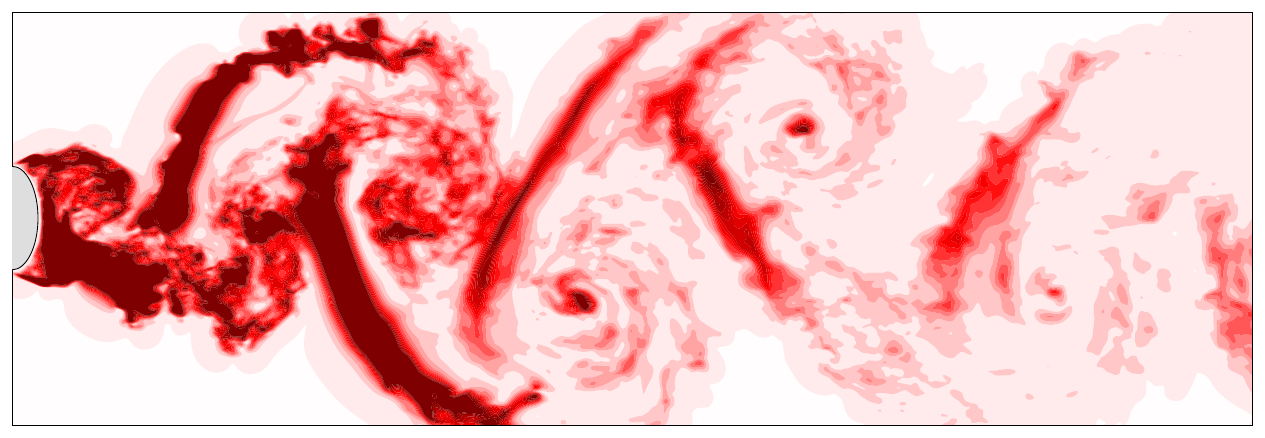}
    \caption{$\avg{u\p u\p}$} 
\end{subfigure}
\begin{subfigure}{\linewidth}
    \includegraphics[width=\linewidth]{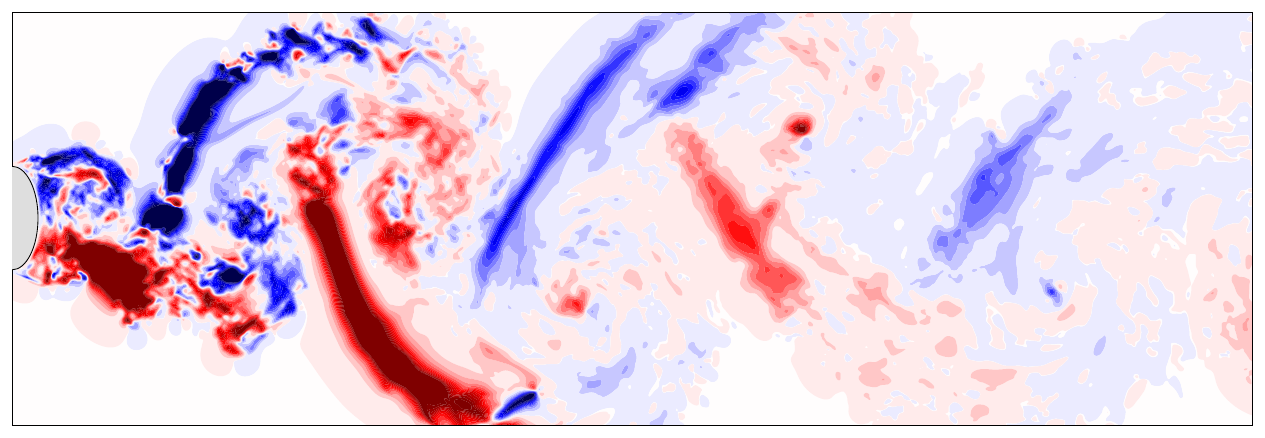}
    \caption{$\avg{u\p v\p}$}
\end{subfigure}
\begin{subfigure}{\linewidth}
    \includegraphics[width=\linewidth]{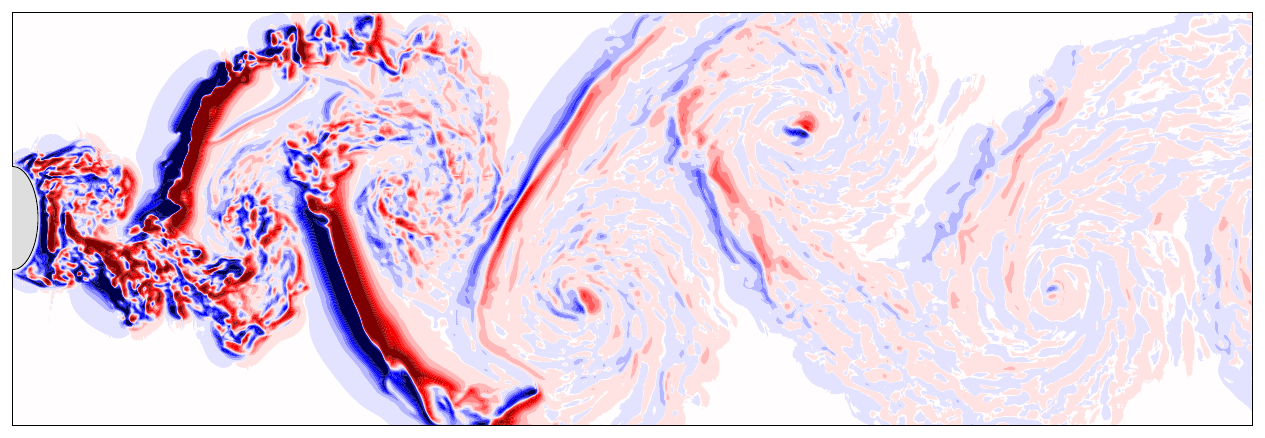}
    \caption{$\mathcal{S}^R_x$}
\end{subfigure}
\begin{subfigure}{\linewidth}
    \includegraphics[width=\linewidth]{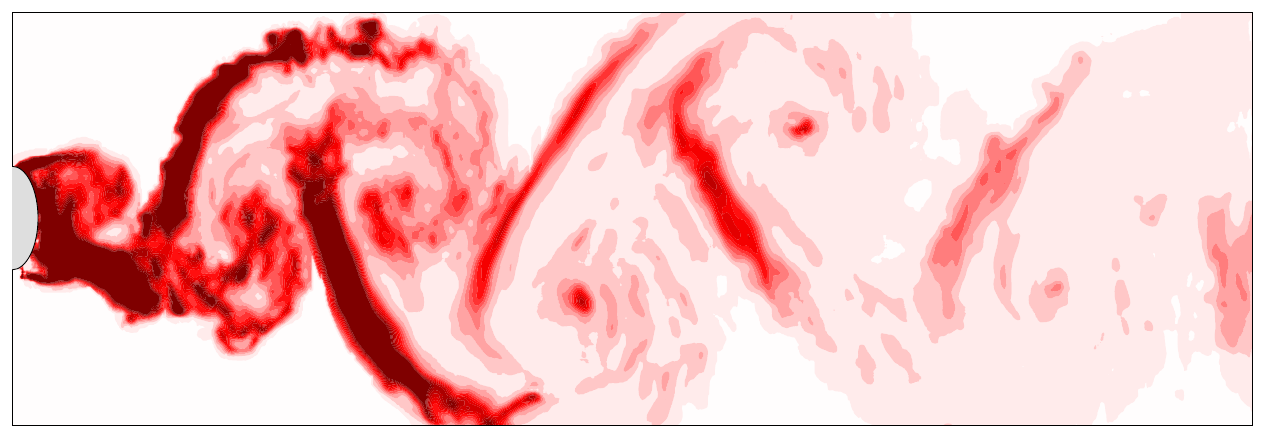}
    \caption{$\avg{u\p u\p}^{\mathrm{ML}}$} 
\end{subfigure}
\begin{subfigure}{\linewidth}
    \includegraphics[width=\linewidth]{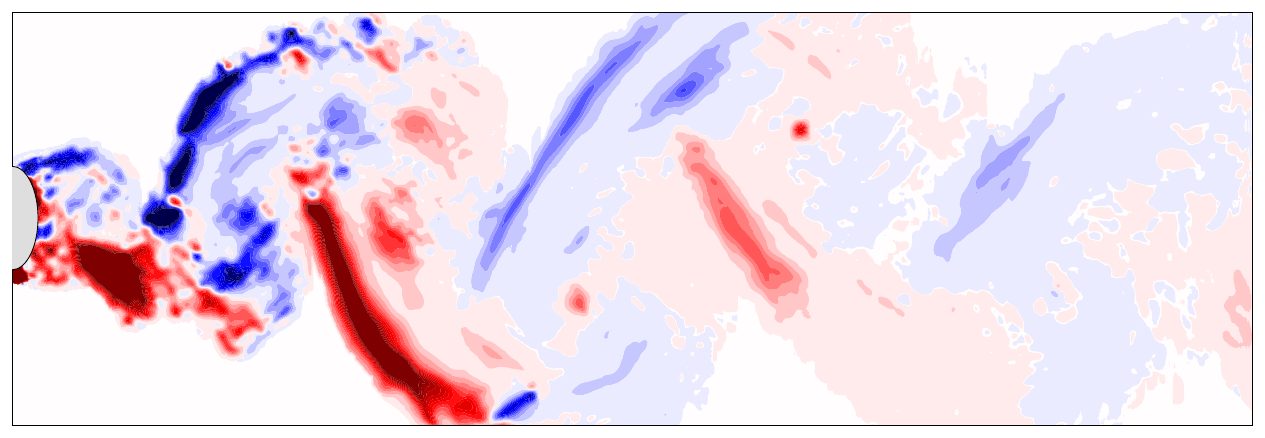}
    \caption{$\avg{u\p v\p}^{\mathrm{ML}}$}
\end{subfigure}
\begin{subfigure}{\linewidth}
    \includegraphics[width=\linewidth]{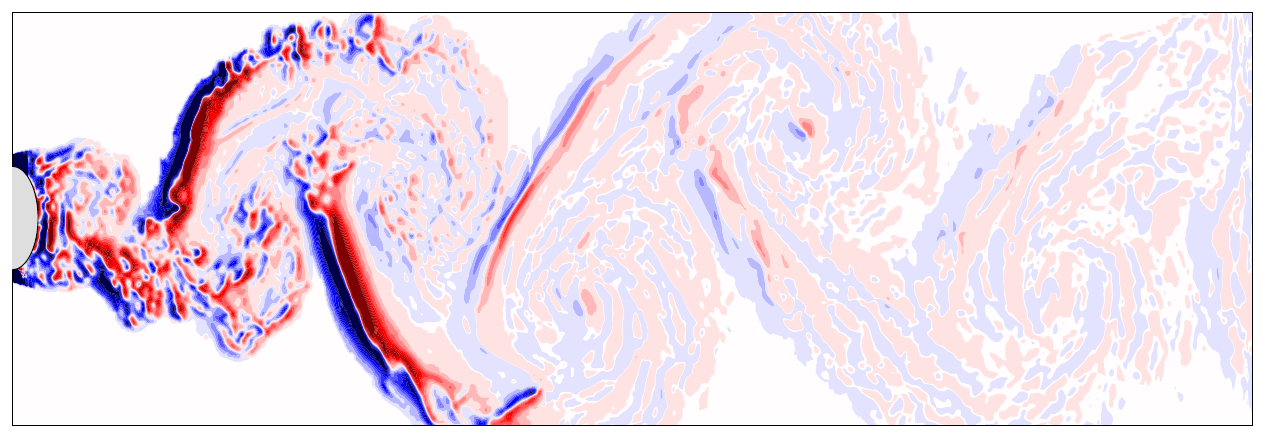}
    \caption{$\mathcal{S}_x^{R,\,\mathrm{ML}}$}
\end{subfigure}
\end{multicols}
\caption{Ellipse 2, $Re=10^4$ case: ML model predictions of components of the SSR tensor and the perfect closure compared to reference data.}
\label{fig:ML_generalisation_E2}
\end{figure*}

Regarding the ellipse cases at $Re=10^4$, high correlation values are observed when the near-body region is cropped out of the correlation analysis as displayed \tref{tab:ML_generalisation}.
On the other hand, the ML model fails to correctly predict the spanwise stresses in the shear-layer region.
This behaviour can be appreciated in \fref{fig:ML_generalisation_E1} and \fref{fig:ML_generalisation_E2}, where patches of high-intensity spanwise stresses arise at the vicinity of the body.
This can also be observed to a certain extent on the $Re=1000$ and $Re=3900$ cases, but is more pronounced in the ellipse cases.
It indicates that training the ML model in a single flow configuration generalises well for lower Reynolds regimes and body shape across the wake except for the shear-layer region, which performs better when it resembles the training data.
It can be the case that the shear layer information contained in the training case is limited so the spatial patterns learnt by the CNN related to this region cannot be recombined for previously unseen cases.
This limitation has been also reported in \citet{Lee2019}, where flow past a circular cylinder at $Re=300$ and $Re=500$ is used for training ML model then tested at $Re=150,400,3900$, and results show a better agreement for the test cases within the training Reynolds regime.

Another issue that arises is the over- or under-prediction of the spanwise stresses intensity (not reflected in the correlation coefficients), noting that the ML model predictions and the reference data are plotted using the same scale. 
This can be observed for example in \fref{fig:ML_generalisation_E1_uv}.
The current input data normalization, based in each snapshot standard score, is a factor that might result in this difference.

\newpage

Overall, the ML model has shown an excellent performance on the generalization across lower $Re$ and different body shapes with results close to the training case.
Additionally, this indicates that the training data is not over-fitted.
Two limitations on the generalization have been observed: the prediction of the spanwise-stresses in the shear-layer region and the over- or under-prediction of the spanwise stresses intensity in some cases.

\newpage

\subsubsection*{ML model a-posteriori results}\label{sec:ML_a-post}

The use of the ML model in the a-posteriori framework is assessed next, i.e. the trained closure is embedded directly into the simulation and no knowledge of the correct flow is assumed.
As before, the goal is to recover the unsteady spanwise-averaged solution of the original 3-D flow using only a 2-D system.
Results are compared against a pure 2-D simulation and spanwise-average reference data (extracted from a 3-D simulation).
This comparison demonstrates that the SANS simulation can yield results close to the 3-D reference data while keeping the computational cost order of the 2-D case.
In the a-posteriori setup, the SANS simulation is started with a spanwise-averaged initial condition and the ML model is incorporated into the 2-D solver providing closure to the SANS equations at every time step.

The most important aspect of the a-posteriori analysis revolves around the system stability.
Ideally, a stable closure would be provided at every time step even when noisy inputs are given to the ML model.
For the system to be stable, the spanwise-average attractor should dictate the temporal evolution of the flow and prevent the 2-D flow dynamics to dominate, as well as the growth of non-physical effects.

For our test case, it has been found that directly incorporating the raw output of the SSR tensor ML model prediction $(\boldsymbol\tau_{ij}^{R,\,\mathrm{ML}})$ into the fluid solver is rather unstable.
The noise in the predicted fields is amplified by the divergence operator, which in turn significantly contaminates the resolved quantities at every time step in a feedback loop.
To mitigate this issue, the ML model has been trained to predict the perfect closure $(\mathcal{S}^{R})$ instead of the SSR tensor.
Hence, the target outputs of the CNN in the a-posteriori framework are $\mathrm{Y}_n=\lbrace \mathcal{S}^R_x,\mathcal{S}^R_y\rbrace$.
Issues related to data-driven closures stability are discussed in detail in \citet{Cruz2019} under a RANS formulation.
It is suggested that learning the divergence of the residual tensor instead of its components helps reducing the error accumulation on the resolved quantities.

\begin{figure*}[!ht]
\begin{multicols}{2}
\begin{subfigure}{\linewidth}
    \includegraphics[width=\linewidth]{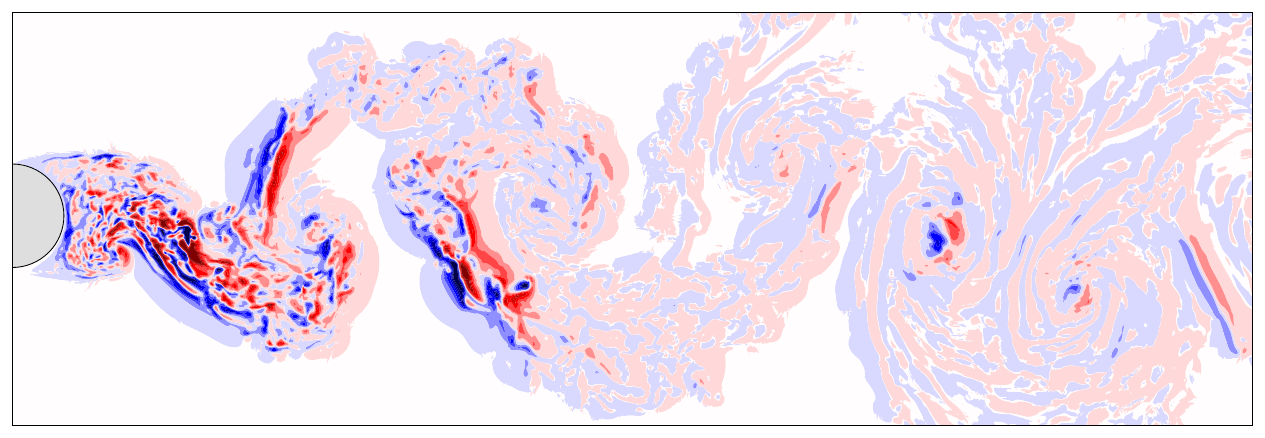}
    \caption{$\mathcal{S}^R_x$} 
\end{subfigure}
\begin{subfigure}{\linewidth}
    \includegraphics[width=\linewidth]{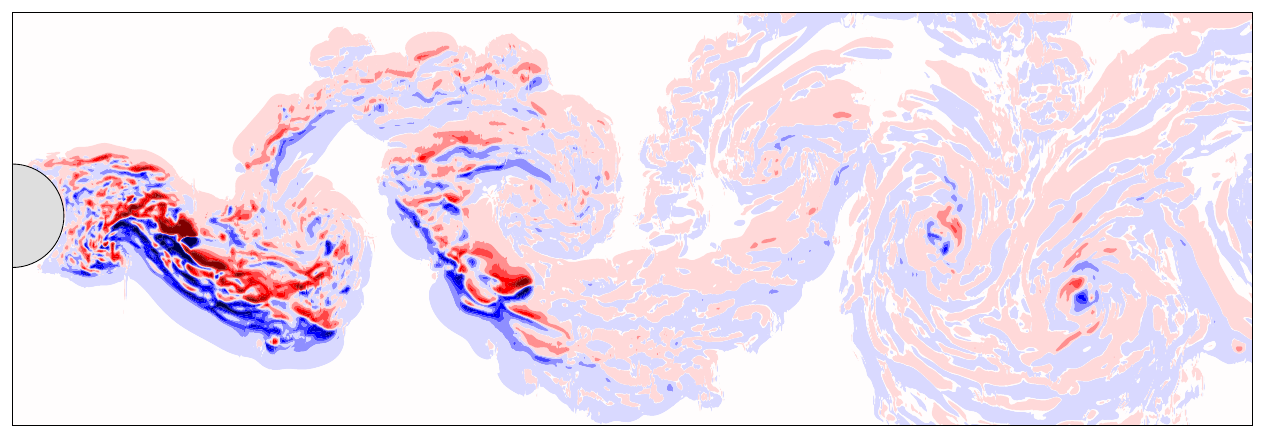}
    \caption{$\mathcal{S}^R_y$} 
\end{subfigure}
\begin{subfigure}{\linewidth}
    \includegraphics[width=\linewidth]{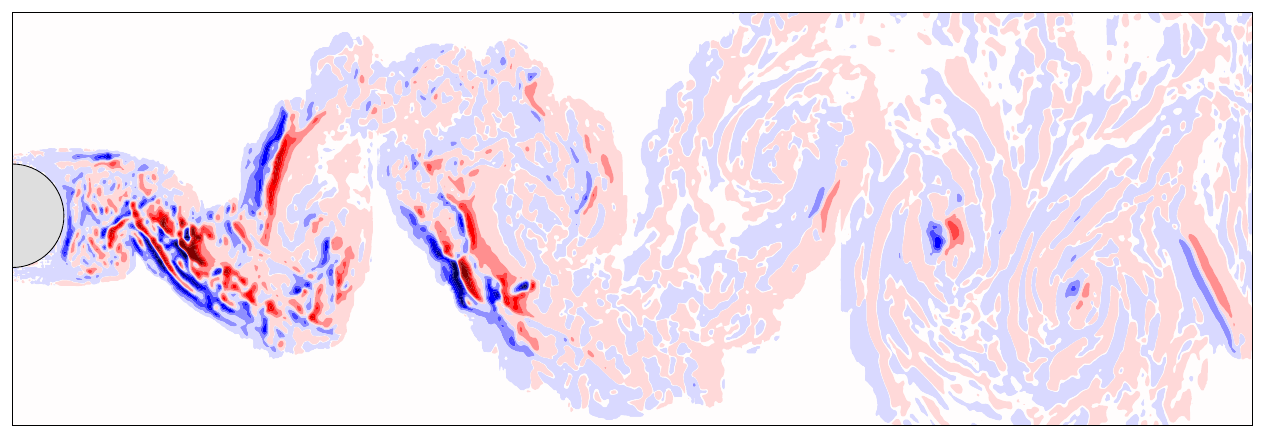}
    \caption{$\mathcal{S}^{R,\,\mathrm{ML}}_x$} 
\end{subfigure}
\begin{subfigure}{\linewidth}
    \includegraphics[width=\linewidth]{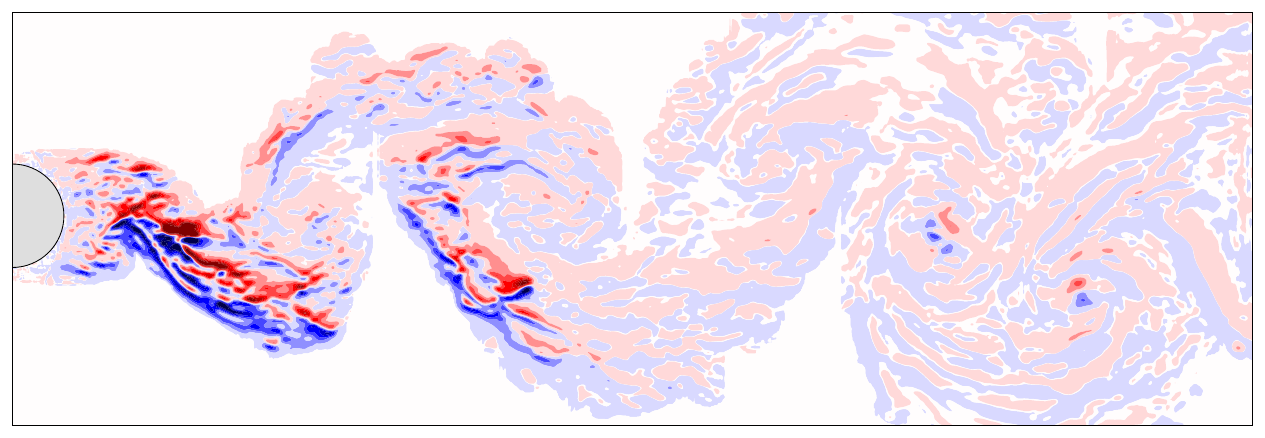}
    \caption{$\mathcal{S}^{R,\,\mathrm{ML}}_y$} 
\end{subfigure}
\end{multicols}
\caption{ML model prediction of the perfect closure components.}
\label{fig:ML_SR_a-priori}
\end{figure*}

Before plugging the ML model prediction of the perfect closure into the 2-D system, an a-priori analysis of the trained model is performed (the training history is displayed in \fref{fig:SR_history}).
High correlation values are found for both perfect closure components, as presented in \tref{tab:ML_ab_a-priori}.
Qualitatively, the ML model prediction of both components is displayed in \fref{fig:ML_SR_a-priori}.
Again, mid- and large-scale structures are correctly captured while small-scale structures present a weaker correlation.

\begin{table}[t]
\centering
\caption{Comparison of different metrics recorded during $\Delta t^*=10$ time units for the 3-D, SANS, and 2-D systems.}
\begin{tabular}{lrrr}
\toprule
 &$\left\langle 3\text{-}\mathrm{D} \right\rangle$ & SANS & $2\text{-}\mathrm{D}$\\
\midrule
$\overline{C}_D$ & 1.17 & 1.14 (-2.6\%)& 1.35 (+15.4\%)\\
$\overline{C}_L$ & 0.46 & 0.51 (+10.9\%) & 0.91 (+97.8\%)\\
$\overline{E}$ & 374.34 & 374.84 (+0.1\%) & 376.44 (+0.6\%) \\
$\overline{Z}$ & 0.0150 & 0.0151 (+0.6\%) & 0.0189 (26.0\%) \\
Cost [h] & 80.4 & 0.43 (-99.5\%) & 0.21 (-99.7\%) \\
\bottomrule
\label{tab:ML_a-posteriori}
\end{tabular}

{\footnotesize The overline indicates a temporal average except for the lift coefficient, for which the r.m.s. value is provided.
Energy $(E)$ and enstrophy $(Z)$ of the 3-D system are calculated using the streamwise and crossflow velocity components alone.
The relative error to the 3-D case is expressed in parenthesis.
The computational cost is expressed in hours and normalised by number of processes and clock speed.
The same constant time step is used in all systems.}
\end{table}

\begin{figure}[!ht]
\centering
\includegraphics[width=0.415\linewidth]{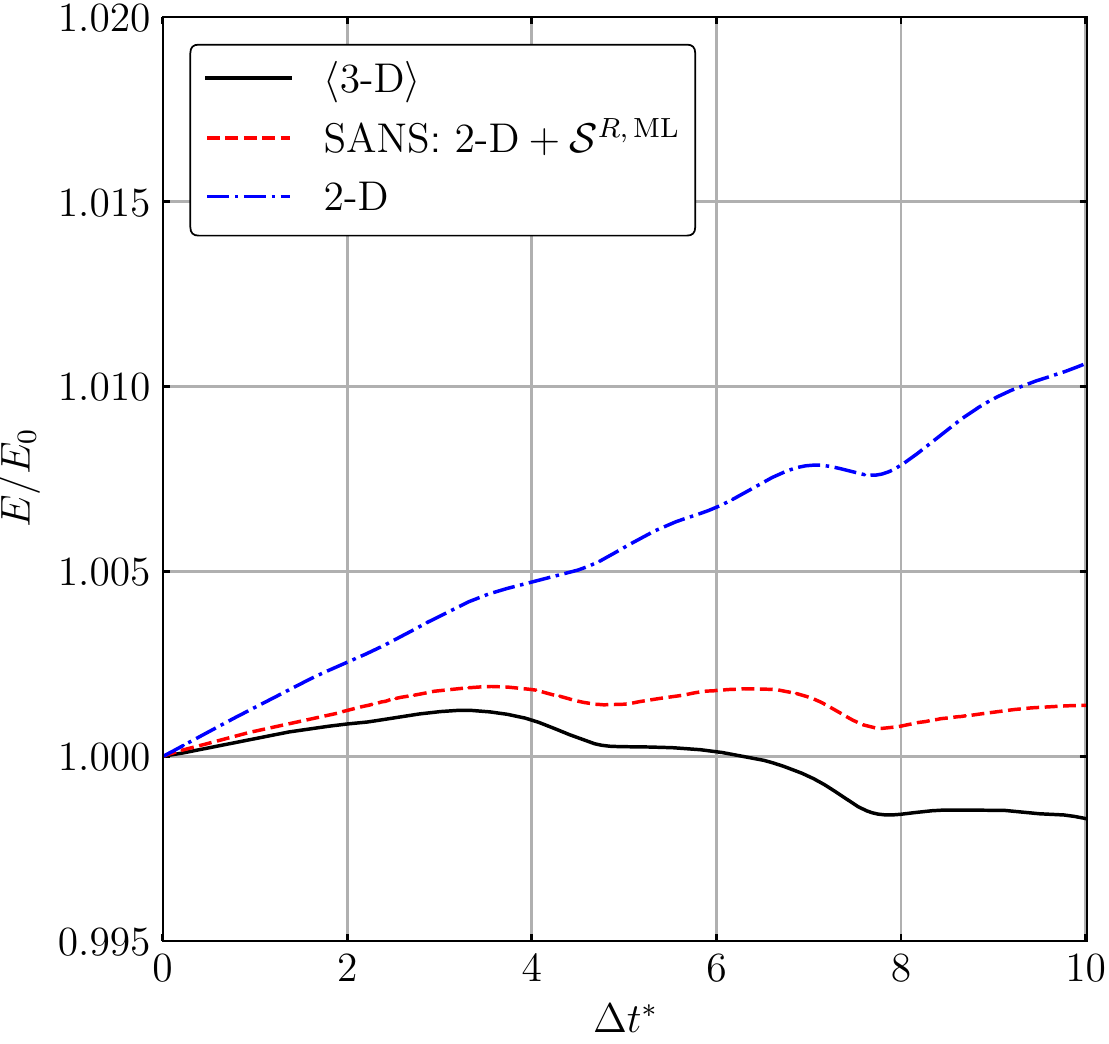}\hspace{0.5cm}
\includegraphics[width=0.4\linewidth]{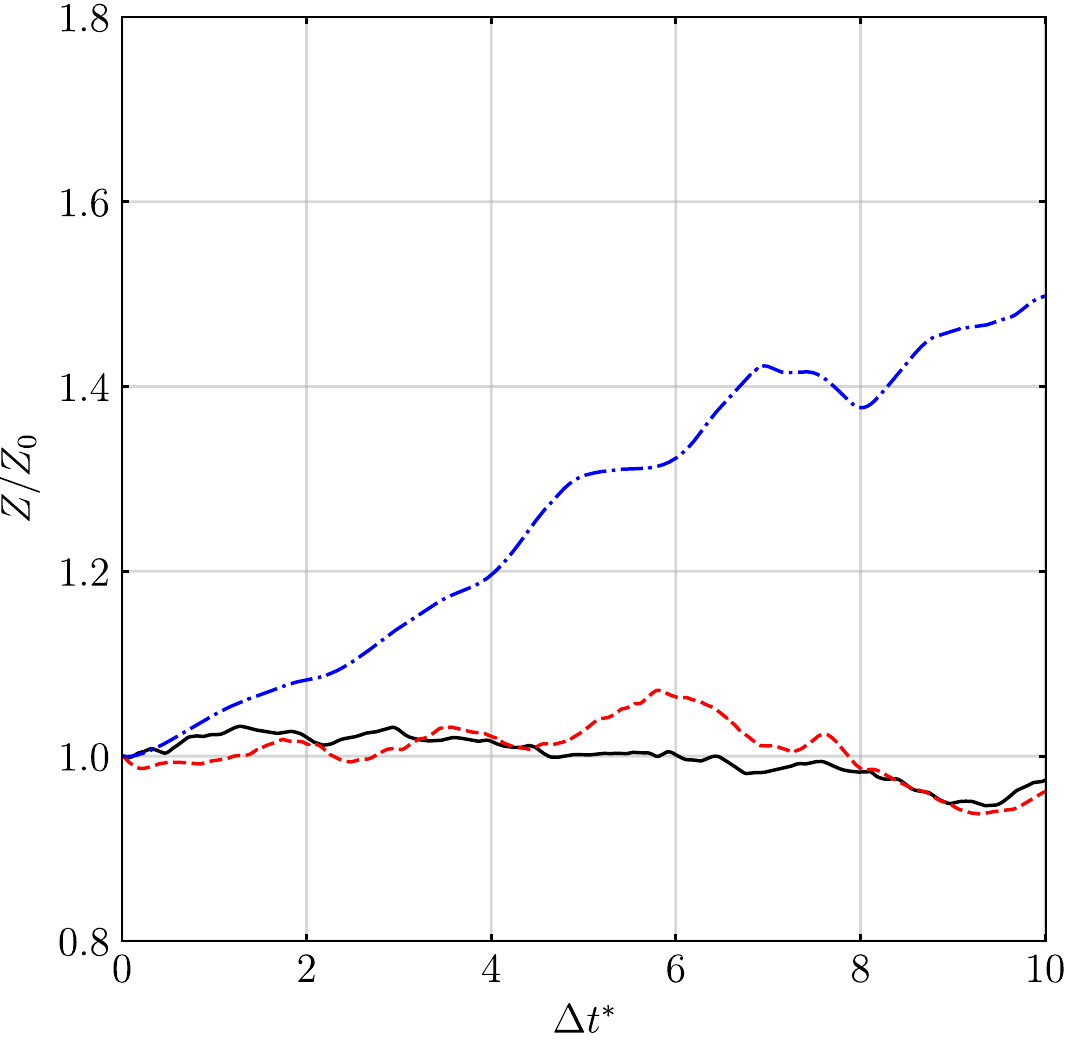}
\caption{Kinetic energy (left) and enstrophy (right) evolution during $\Delta t^*=10$ for the 3-D spanwise-averaged (reference data), SANS, and 2-D systems.
Simulations are started from a 3-D spanwise-averaged snapshot at $t^*_0$, corresponding to the scaling energy $(E_0)$ and enstrophy $(Z_0)$.
Energy and enstrophy are computed as detailed in \fref{fig:t-g_perfect}.}
\label{fig:E_Z_a-posteriori}
\end{figure}

\begin{figure}[!ht]
\centering
\includegraphics[width=0.4\linewidth]{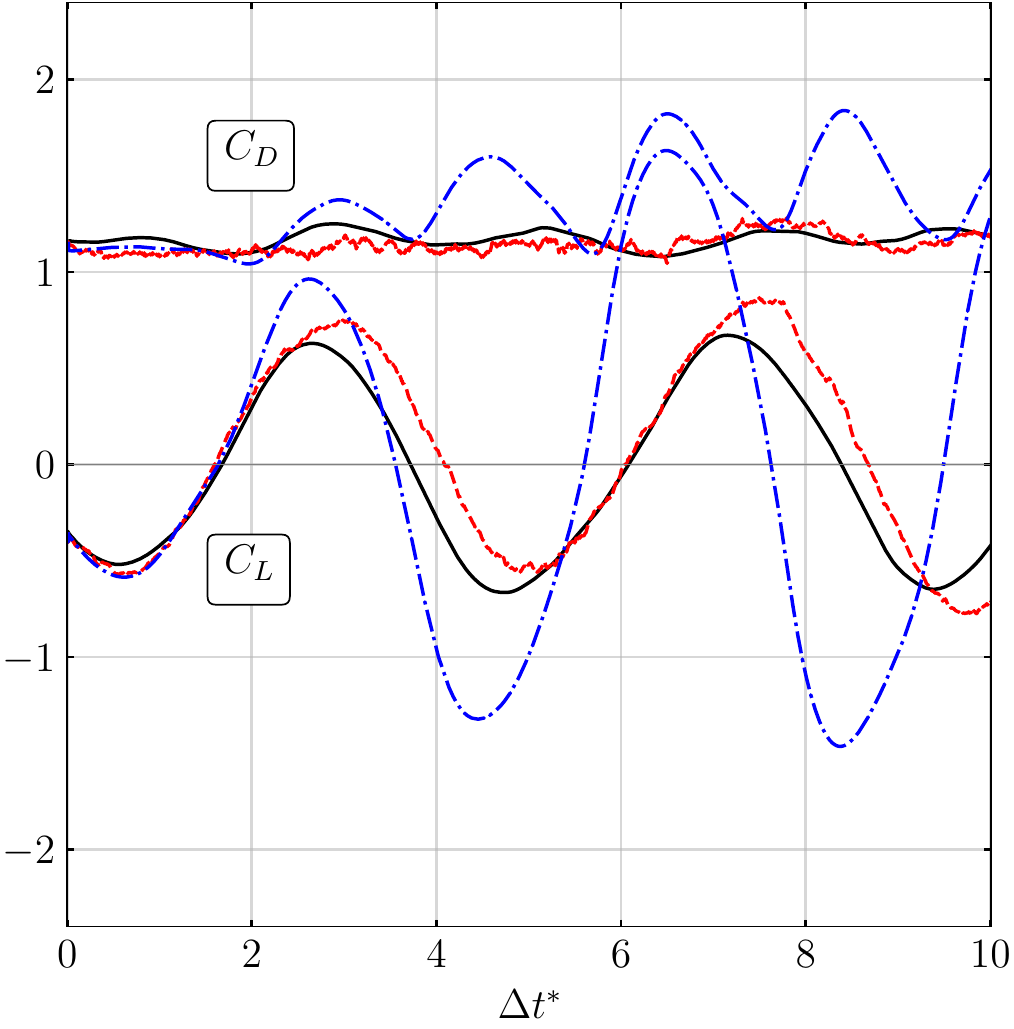}
\caption{Lift (bottom) and drag (top) forces induced to the cylinder (computed as described in \fref{fig:perfect_cc_forces}).
The lines legend is equivalent to \fref{fig:E_Z_a-posteriori}.}
\label{fig:forces_a-posteriori}
\end{figure}

Learning the perfect closure improves the system stability and delays the transition to 2-D dynamics.
Starting from a spanwise-averaged snapshot of the 3-D flow, a-posteriori statistics are recorded as the flow evolves over $\Delta t^*=10$ time units, and these are summarised in \tref{tab:ML_a-posteriori}.
As shown in \fref{fig:E_Z_a-posteriori}, the global enstrophy deviates from the spanwise-averaged reference data obtained in the 3-D system as a result of the ML model error accumulation.
This is also reflected in the kinetic energy evolution, which increases with respect to the reference data.
On the other hand, both metrics still provide a notable improvement compared to the pure 2-D prediction.
The forces induced to the cylinder (\fref{fig:forces_a-posteriori}) also appear to be in better agreement with the reference data when including perfect closure ML predictions.
It becomes clear from \tref{tab:ML_a-posteriori} that the SANS equations yield results closer to a 3-D system while offering the same computational cost order of 2-D simulations.
In this sense, the SANS simulation takes 0.43 hours to compute $\Delta t^*=10$ time units while the 3-D system consumes 80.4 hours, resulting in a reduction of the 3-D simulation computational time of 99.5\%.
On the other hand, the pure 2-D simulation can complete the simulation in 0.21 hours, approximately half of the SANS computational time.
The computational time overhead between the SANS and the 2-D systems is related to the data transfer between the fluid solver and the ML model, plus the ML model prediction time.
Further information on the SANS cost plus software and hardware details can be found in \aref{app:ML_model_details_SWHW}.


\begin{figure}[!ht]
\centering
\begin{subfigure}[t]{0.5\linewidth}
 	\includegraphics[width=\linewidth]{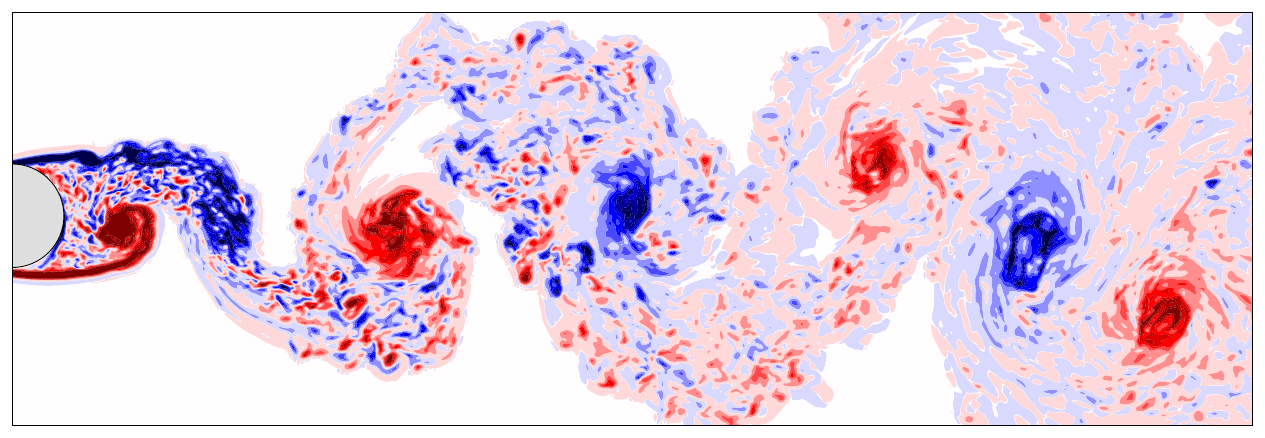}
 	\caption{$\left\langle 3\text{-}\mathrm{D} \right\rangle$}
\end{subfigure}
\begin{subfigure}[t]{0.5\linewidth}
	\includegraphics[width=\linewidth]{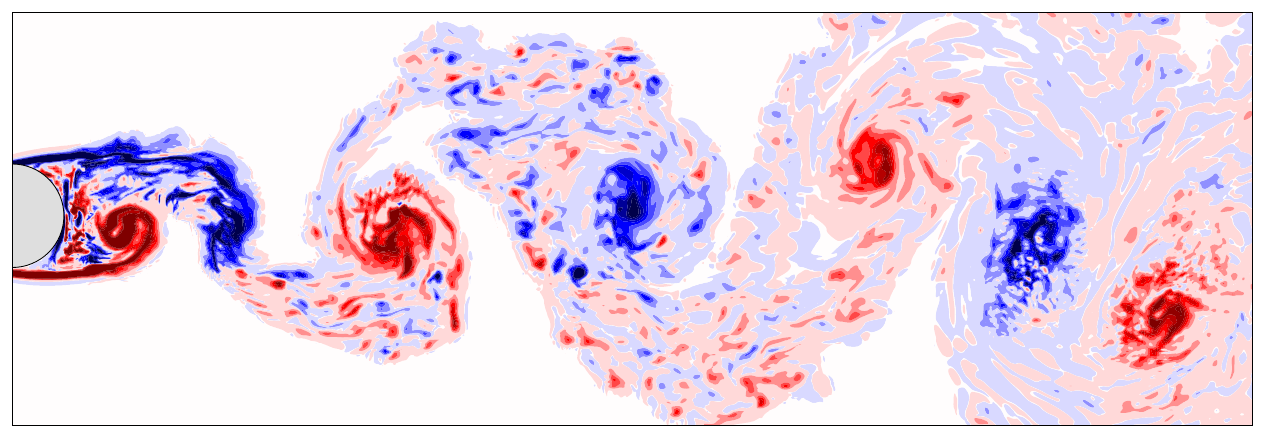}
	\caption{SANS}
\end{subfigure}
\begin{subfigure}[t]{0.5\linewidth}
	\includegraphics[width=\linewidth]{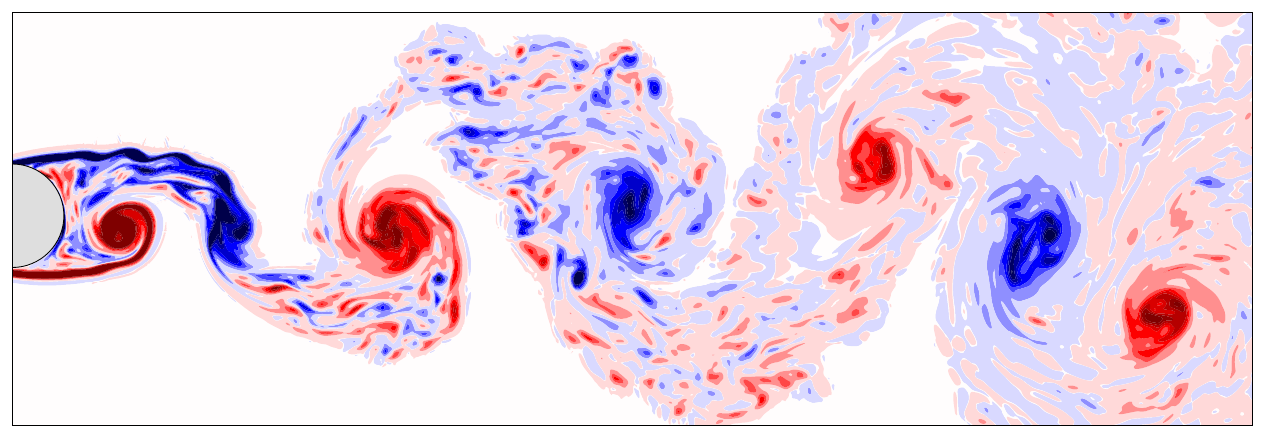}
	\caption{$2\text{-}\mathrm{D}$}
\end{subfigure}
\caption{Spanwise vorticity snapshots for: (a) Spanwise-averaged 3-D flow (reference data).
(b) SANS with $\mathcal{S}^{R,\,\mathrm{ML}}$.
(c) 2-D simulation (no model).
The SANS and the 2-D systems are started from a spanwise-averaged snapshot of the 3-D simulation at $t^*_0$.
The snapshots above correspond to $\Delta t^*=1$.}
\label{fig:ML_a-posteriori}
\end{figure}

Additionally, the instantaneous spanwise vorticity is depicted in \fref{fig:ML_a-posteriori}.
It is worth noting that the shear layer roll-up region of the SANS snapshot is not as coherent as the 2-D system, where structures have already reorganised into large scales.
In particular, the bottom detaching vortex of the SANS simulation is not as energised as the pure 2-D simulation one, and small-scale structures are still found in the SANS snapshot.
The close-wake region dictates the forces induced to the cylinder, explaining the significant difference in lift and drag forces shown in \tref{tab:ML_a-posteriori}.
In general, vortices found in the SANS simulation do not appear as coherent as in the 2-D case, where the inverse energy cascade promoted via vortex-merging and vortex-thinning events rapidly two-dimensionalises the wake structures.
A closer look on the SANS small-scale structures also shows that these are not correlated to the reference data.
This can also be appreciated at the top shear layer roll-up, where the error starts to propagate to the far field.
The ML closure error is responsible for such phenomenon, and ultimately yields the divergence of the system dynamics.

\newpage

\section{Discussion and conclusion}

This paper introduces a ML model for the SANS equations, enabling 3-D turbulent flows with a homogeneous flow direction to be modelled using a 2-D system of equations with an observed speed up of 99.5\%.
The SSR tensor is critical to the success of this description, driving the 2-D flow dynamics to match the 3-D statistics and flow forces.
While EVMs are completely inadequate to model the SSR tensor, we develop a data-driven closure based on a deep-learning network with tested correlation statistics above 90\%.
From a qualitative point of view, the ML model can accurately reconstruct large- and mid-scale structures, while small-scale structures are more difficult to predict.
Additionally, the trained ML model can predict the spanwise stresses of different Reynolds regimes and body shapes to the training case with similar accuracy (except for the shear-layer region).

The success of ML models for turbulent flows in an a-posteriori set-up is still limited, and the error propagation into the resolved quantities remains an open issue \citep{Wang2017, Beck2019}.
Stability has been previously achieved by projecting the ML model output into a known stable closure such as EVMs for time-averaged (RANS) or filtered (LES) systems \citep{Maulik2018, Beck2019, Cruz2019}.
Also, highly-resolved simulations help to mitigate the closure weight resulting in more stable computations \citep{Gamahara2017}.
Unfortunately, these solutions cannot be implemented in SANS because, as previously exposed, the EVMs hypothesis is intrinsically different from the nature of the SSR tensor.

The use of recurrent neural networks (RNN) in SANS is a research direction worth exploring in the future.
RNNs incorporate the system temporal dynamics so they can offer a more stable a-posteriori solution targeting statistically stationary metrics \citep{Vlachas2018, Kim2020b}.
In this sense, the use of long-short term neural networks (a type of RNN) similarly to \citet{Vlachas2018}, together with CNNs could help overcoming the issues exposed in the SANS closure.
Additionally, a physics-informed loss function can also improve the model stability as shown in \citet{Lee2019}.
Finally, while the current model is initialised from a 3-D averaged simulation, a more stable closure should be capable of developing 3-D turbulence dynamics even when initialised with a pure 2-D flow snapshot.

Overall, the new SANS flow description combined with advances in data-driving modelling has the potential to vastly improve the speed and accuracy of turbulence simulations for a prevalent class of engineering and environmental applications.

\section*{Acknowledgments}

The authors acknowledge the support of the Singapore Agency for Science, Technology and Research (A*STAR) Research Attachment Programme award ARAP-2015-1-0035 as a collaboration between the A*STAR Institute of High Performance Computing (IHPC) and the Faculty of Engineering and Physical Sciences of the University of Southampton.
The authors also acknowledge the use of the IRIDIS High Performance Computing Facility, and associated support services at the University of Southampton, in the completion of this work.
G.D.W. acknowledges the support from the Office of Naval Research Global award N62909-18-1-2091.

\appendix

\section{An analogous derivation for the vorticity transport equation}

The spanwise stress residual terms found in Eqs. \ref{eq:z-avg_X} to \ref{eq:z-avg_Z} can be similarly obtained on the vorticity formulation of the Navier--Stokes equations.
The non-dimensional vorticity transport equation can be written in its conservation form as
\begin{equation}
\partial_t\boldsymbol\omega+\nabla \cdot \pars{\vect{u}\otimes\boldsymbol{\omega}}=\nabla \cdot \pars{\boldsymbol{\omega}\otimes\vect{u}}+Re^{-1}\nabla^2\boldsymbol{\omega}.\label{eq:VTE}
\end{equation}
Decomposing and averaging similarly to the momentum equations, the following spanwise-averaged vorticity transport equation can be obtained for the $\vect{e}_z$ component of \eref{eq:VTE}
\begin{gather}
\partial_t {\Omega}_z + U\partial_x{\Omega}_z + V\partial_y{\Omega}_z = Re^{-1}\pars{\partial_{xx}{\Omega}_z + \partial_{yy}{\Omega}_z}-\partial_x\pars{\avg{u\p\omega_z\p}-\avg{\omega_x\p w\p}}-\partial_y\pars{\avg{v\p\omega_z\p}-\avg{\omega_y\p w\p}}+\nonumber \\
+{\Omega}_x\partial_xW+{\Omega}_y\partial_yW+{\Omega}_z\avg{\partial_z w\p}-W\avg{\partial_z\omega_z\p}+Re^{-1}\avg{\partial_{zz}\omega_z\p}, \label{eq:vort_e3_avg}
\end{gather}
where the spanwise-averaged divergence of the velocity and vorticity vector fields, respectively
\begin{gather}
\partial_x{U}+\partial_y{V}=-\avg{\partial_z{w\p}},\\
\partial_x{{\Omega}_x}+\partial_y{{\Omega}_y}=-\avg{\partial_z{\omega_z\p}},
\end{gather}
have been introduced to obtain the non-conservation form.

When spanwise periodicity is assumed, the second line in \eref{eq:vort_e3_avg} vanishes by its last three terms cancelling by definition (\eref{eq:periodic_assumption}), and the first two terms cancelling each other
\begin{gather}
{\Omega}_x\partial_xW+{\Omega}_y\partial_yW=\avg{\partial_y{w}-\partial_z{v}}\partial_x{W}+\avg{\partial_z{u}-\partial_z{w}}\partial_y{W}=\nonumber\\
=\avg{\partial_y{\pars{W+w\p}}-\partial_z{\pars{V+v\p}}}\partial_x{W}+\avg{\partial_z{\pars{U+u\p}}-\partial_z{\pars{W+v\p}}}\partial_y{W}=\nonumber\\
=\pars{\partial_y{W}-\cancelto{0}{\avg{\partial_z{v\p}}}}\partial_x{W}+\pars{\cancelto{0}{\avg{\partial_z{u\p}}}-\partial_x{W}}\partial_y{W}=0.
\end{gather}
The spanwise-averaged vorticity transport equation is therefore simplified to
\begin{equation}
\partial_t {\Omega}_z + \vect{U}\cdot\nabla{\Omega}_z = Re^{-1}\nabla^2{\Omega}_z-\nabla\cdot\boldsymbol\zeta^R,
\end{equation}
where $\vect{U}=\pars{U,V}$ and $\boldsymbol \zeta^R = \pars{\avg{u\p\omega_z\p}-\avg{\omega_x\p w\p}, \avg{v\p\omega_z\p}-\avg{\omega_y\p w\p}}$.

Next, we show how the spanwise stresses in the SANS and the spanwise-averaged VTE frameworks are related.
First, we derive an expression for the vorticity-velocity terms of the spanwise-averaged VTE to write them as velocity-velocity terms.
Considering the following vector identity
\begin{equation}
\frac{1}{2} \nabla \left( \vect{a}\cdot\vect{a} \right) = (\vect{a} \cdot \nabla) \vect{a} + \vect{a} \times (\nabla \times \vect{a}),
\end{equation}
we can rewrite it in terms of fluctuating quantities
\begin{equation}
\frac{1}{2} \nabla \left( \vect{u}\p\cdot\vect{u}\p \right) = (\vect{u}\p \cdot \nabla) \vect{u}\p + \vect{u}\p \times \boldsymbol{\omega}\p,\label{eq:vect_identity_fluct}
\end{equation}
where the definition of the vorticity vector field, $\boldsymbol\omega=\nabla\times\vect{u}$, has been introduced.

Second, the conservation form of the convective forces can be expressed as
\begin{equation}
\nabla\cdot\pars{\vect u \otimes\vect u}=\pars{\vect u \cdot \nabla}\vect u + \vect u\pars{\nabla\cdot\vect u},
\end{equation}
and it can be rewritten for the fluctuating quantities as
\begin{gather}
\nabla\cdot\pars{\vect u\p \otimes\vect u\p}=\pars{\vect u\p \cdot \nabla}\vect u\p + \vect u\p\pars{\nabla\cdot\vect u\p},\\
\pars{\vect u\p \cdot \nabla}\vect u\p = \nabla\cdot\pars{\vect u\p \otimes\vect u\p}-\vect u\p\pars{\nabla\cdot\vect u\p}.\label{eq:conv_term_fluct}
\end{gather}

Third, decomposing the continuity equation for incompressible flows (\eref{eq:cont})
\begin{equation}
\partial_xU+\partial_xu\p+\partial_yV+\partial_yv\p+\partial_zw\p=0,\label{eq:continuity_decomp}
\end{equation}
and subtracting the averaged continuity equation (\eref{eq:cont_decomp_avg}) yields
\begin{equation}
\partial_xu\p+\partial_yv\p+\partial_zw\p-\avg{\partial_z w\p}=0,
\end{equation}
hence
\begin{equation}
\nabla\cdot\vect u\p =\avg{\partial_z w\p}.\label{eq:cont_fluct}
\end{equation}

We now introduce \eref{eq:cont_fluct} into \eref{eq:conv_term_fluct} obtaining
\begin{equation}
\pars{\vect u\p \cdot \nabla}\vect u\p = \nabla\cdot\pars{\vect u\p \otimes\vect u\p}-\vect u\p\avg{\partial_z w\p},\label{eq:conv_term_fluct2}
\end{equation}
and \eref{eq:conv_term_fluct2} is finally introduced into \eref{eq:vect_identity_fluct}
\begin{equation}
\frac{1}{2} \nabla \left( \vect{u}\p\cdot\vect{u}\p \right) = \nabla\cdot\pars{\vect u\p \otimes\vect u\p}-\vect u\p\avg{\partial_z w\p} + \vect{u}\p \times \boldsymbol{\omega}\p,
\end{equation}
which yields
\begin{equation}
\boldsymbol{\omega}\p\times\vect{u}\p = \nabla\cdot\pars{\vect u\p \otimes\vect u\p} - \frac{1}{2} \nabla \left( \vect{u}\p\cdot\vect{u}\p \right) - \vect u\p\avg{\partial_z w\p}.
\label{eq:vect_identity_fluct2}
\end{equation}

With this, a relation between the vorticity-velocity spanwise stresses and the velocity-velocity spanwise stresses is defined in the conservation form.
To derive a more explicit form of the spanwise-averaged stresses, let us average \eref{eq:vect_identity_fluct2} and retrieve its $\vect{e}_x$ and $\vect{e}_y$ components
\begin{gather}
\vect{e}_x: \quad \avg{\omega_y\p w\p}-\avg{\omega_z\p v\p} = \partial_x\avg{u\p u\p}+\partial_y\avg{v\p u\p}+\avg{\partial_z\pars{w\p u\p}}-\partial_x\pars{\avg{u\p u\p}+\avg{v\p v\p}+\avg{w\p w\p}}/2,\label{eq:vec_ident_fluct_e1}\\
\vect{e}_y: \quad \avg{\omega_z\p u\p}-\avg{\omega_x\p w\p} = \partial_x\avg{u\p v\p}+\partial_y\avg{v\p v\p}+\avg{\partial_z\pars{w\p v\p}}-\partial_y\pars{\avg{u\p u\p}+\avg{v\p v\p}+\avg{w\p w\p}}/2\label{eq:vec_ident_fluct_e2}.
\end{gather}
Note that the last term of \eref{eq:vect_identity_fluct2} vanishes because of \eref{eq:rule2}. 

The spanwise stresses of the spanwise-averaged vorticity transport equation (\eref{eq:vort_e3_avg}) can be expanded using \eref{eq:vec_ident_fluct_e1} and \eref{eq:vec_ident_fluct_e2} as follows
\begin{gather}
-\partial_x\pars{\avg{u\p\omega_z\p}-\avg{\omega_x\p w\p}}-\partial_y\pars{\avg{v\p\omega_z\p}-\avg{\omega_y\p w\p}} = \nonumber\\
=-\partial_x\pars{\partial_x\avg{u\p v\p}+\partial_y\avg{v\p v\p}+\avg{\partial_z\pars{w\p v\p}} - \partial_y\pars{\avg{u\p u\p}+\avg{v\p v\p}+\avg{w\p w\p}}/2} +\nonumber\\
+\partial_y\pars{\partial_x\avg{u\p u\p}+\partial_y\avg{v\p u\p}+\avg{\partial_z\pars{w\p u\p}} - \partial_x\pars{\avg{u\p u\p}+\avg{v\p v\p}+\avg{w\p w\p}}/2}=\nonumber\\
=-\partial_{xx}\avg{u\p v\p}-\partial_{xy}\avg{v\p v\p}-\partial_x\avg{\partial_z\pars{w\p v\p}}+\partial_{xy}\pars{\avg{u\p u\p}+\avg{v\p v\p}+\avg{w\p w\p}}/2+\nonumber\\
+\partial_{xy}\avg{u\p u\p}+\partial_{yy}\avg{v\p u\p}+\partial_y\avg{\partial_z\pars{w\p u\p}}-\partial_{xy}\pars{\avg{u\p u\p}+\avg{v\p v\p}+\avg{w\p w\p}}/2=\nonumber\\
=\pars{\partial_{yy}-\partial_{xx}}\avg{u\p v\p}+\partial_{xy}\pars{\avg{u\p u\p}-\avg{v\p v\p}}+\partial_y\avg{\partial_z\pars{w\p u\p}}-\partial_x\avg{\partial_z\pars{w\p v\p}}\label{eq:vte_spanwise_stresses}.
\end{gather}

On the other hand, the SANS equations (Eqs. \ref{eq:z-avg_X} to \ref{eq:z-avg_Z}) have yield the following spanwise stresses
\begin{equation}
-\avg{\nabla\cdot\pars{\vect u\p\otimes\vect u\p}}=
-\begin{pmatrix}
    \partial_x\avg{u\p u\p}+\partial_y\avg{u\p v\p}+\avg{\partial_z\pars{w\p u\p}} \\
    \partial_x\avg{v\p u\p}+\partial_y\avg{v\p v\p}+\avg{\partial_z\pars{w\p v\p}} \\
    \partial_x\avg{w\p u\p}+\partial_y\avg{w\p v\p}+\avg{\partial_z\pars{w\p w\p}}
  \end{pmatrix}.
\end{equation}
Taking the curl of the momentum spanwise stresses
\begin{equation}
-\nabla\times\avg{\nabla\cdot\pars{\vect u\p\otimes\vect u\p}}=-
\begin{pmatrix}
    \vect{e}_x & \partial_x & \partial_x\avg{u\p u\p}+\partial_y\avg{u\p v\p}+\avg{\partial_z\pars{w\p u\p}} \\
    \vect{e}_y & \partial_y & \partial_x\avg{v\p u\p}+\partial_y\avg{v\p v\p}+\avg{\partial_z\pars{w\p v\p}} \\
    \vect{e}_z & \partial_z & \partial_x\avg{w\p u\p}+\partial_y\avg{w\p v\p}+\avg{\partial_z\pars{w\p w\p}}
  \end{pmatrix},
\end{equation}
recovers the VTE spanwise stresses as demonstrated for the $\vect{e}_z$ component
\begin{gather}
-\nabla\times\avg{\nabla\cdot\pars{\vect u\p\otimes\vect u\p}}\vect{e}_z=\nonumber\\
=-\partial_x\pars{\partial_x\avg{v\p u\p}+\partial_y\avg{v\p v\p}+\avg{\partial_z\pars{w\p v\p}}}
+\partial_y\pars{\partial_x\avg{u\p u\p}+\partial_y\avg{u\p v\p}+\avg{\partial_z\pars{w\p u\p}}}=\nonumber\\
=-\partial_{xx}\avg{v\p u\p}-\partial_{xy}\avg{v\p v\p}-\partial_x\avg{\partial_z\pars{w\p v\p}}+\partial_{xy}\avg{u\p u\p}+\partial_{yy}\avg{u\p v\p}+\partial_y\avg{\partial_z\pars{w\p u\p}}=\nonumber\\
=\pars{\partial_{yy}-\partial_{xx}}\avg{u\p v\p}+\partial_{xy}\pars{\avg{u\p u\p}-\avg{v\p v\p}}+\partial_y\avg{\partial_z\pars{w\p u\p}}-\partial_x\avg{\partial_z\pars{w\p v\p}},\label{eq:NS_spanwise_stresses}
\end{gather}
showing that the spanwise stresses in both formulations (Eqs. \ref{eq:vte_spanwise_stresses} and \ref{eq:NS_spanwise_stresses}) are equivalent.

\section{Additional machine-learning model details} \label{app:ML_model_details}

\subsection{Training history}

\begin{figure}[!ht]
\begin{multicols}{2}
\centering
\includegraphics[width=1.02\linewidth]{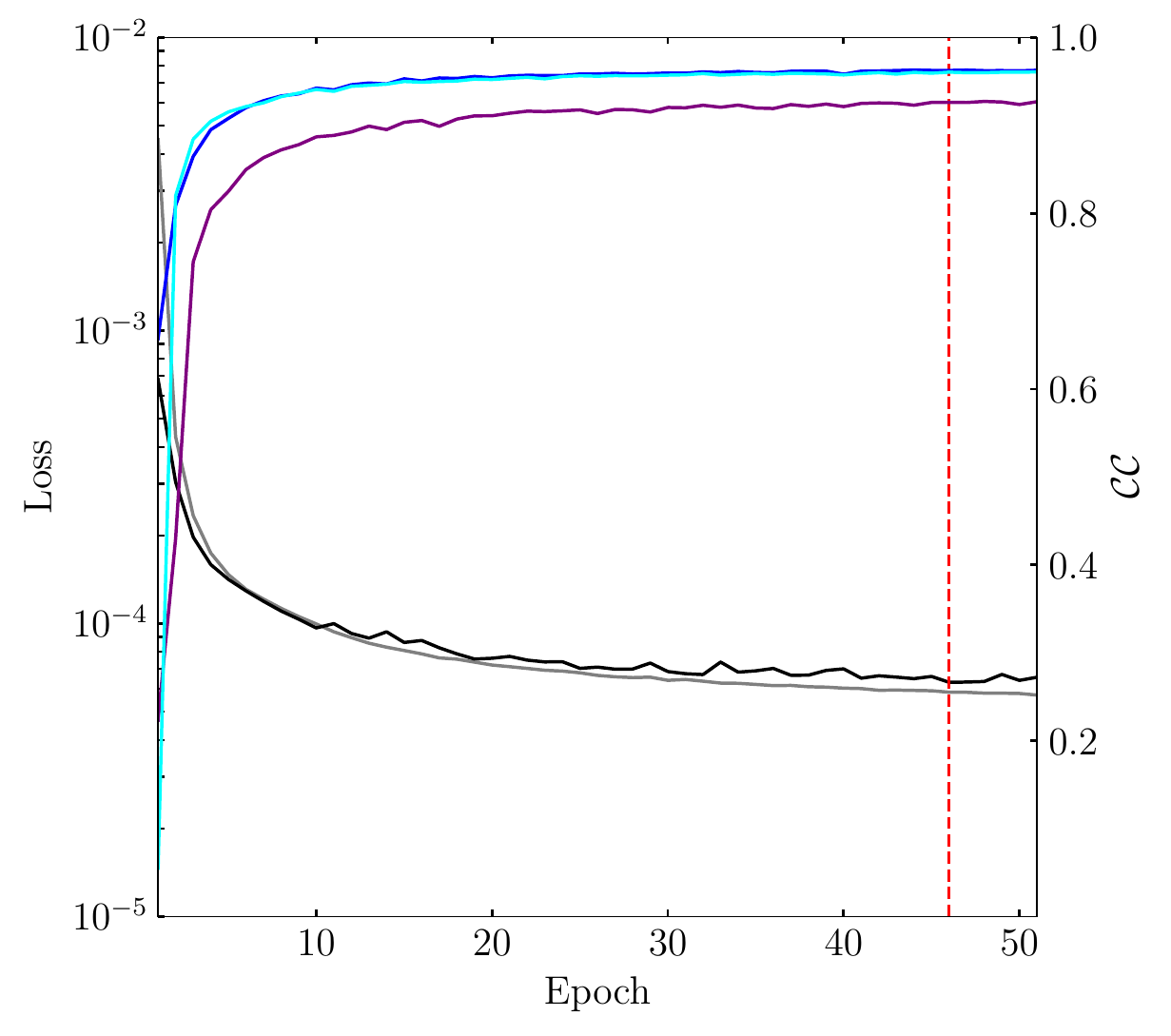}
\caption{Training of the $\mathrm{Y}_n=\big\lbrace\langle u' u'\rangle,\langle u' v'\rangle,\langle v' v'\rangle\big\rbrace$ output set.
The vertical red dashed line indicates the early stop.
Grey: combined training loss.
Black: combined validation loss.
Blue: $\langle u' u'\rangle$ correlation coefficient.
Purple: $\langle u' v'\rangle$ correlation coefficient.
Cyan: $\langle v' v'\rangle$ correlation coefficient.}\label{fig:ab_history}

\includegraphics[width=1.02\linewidth]{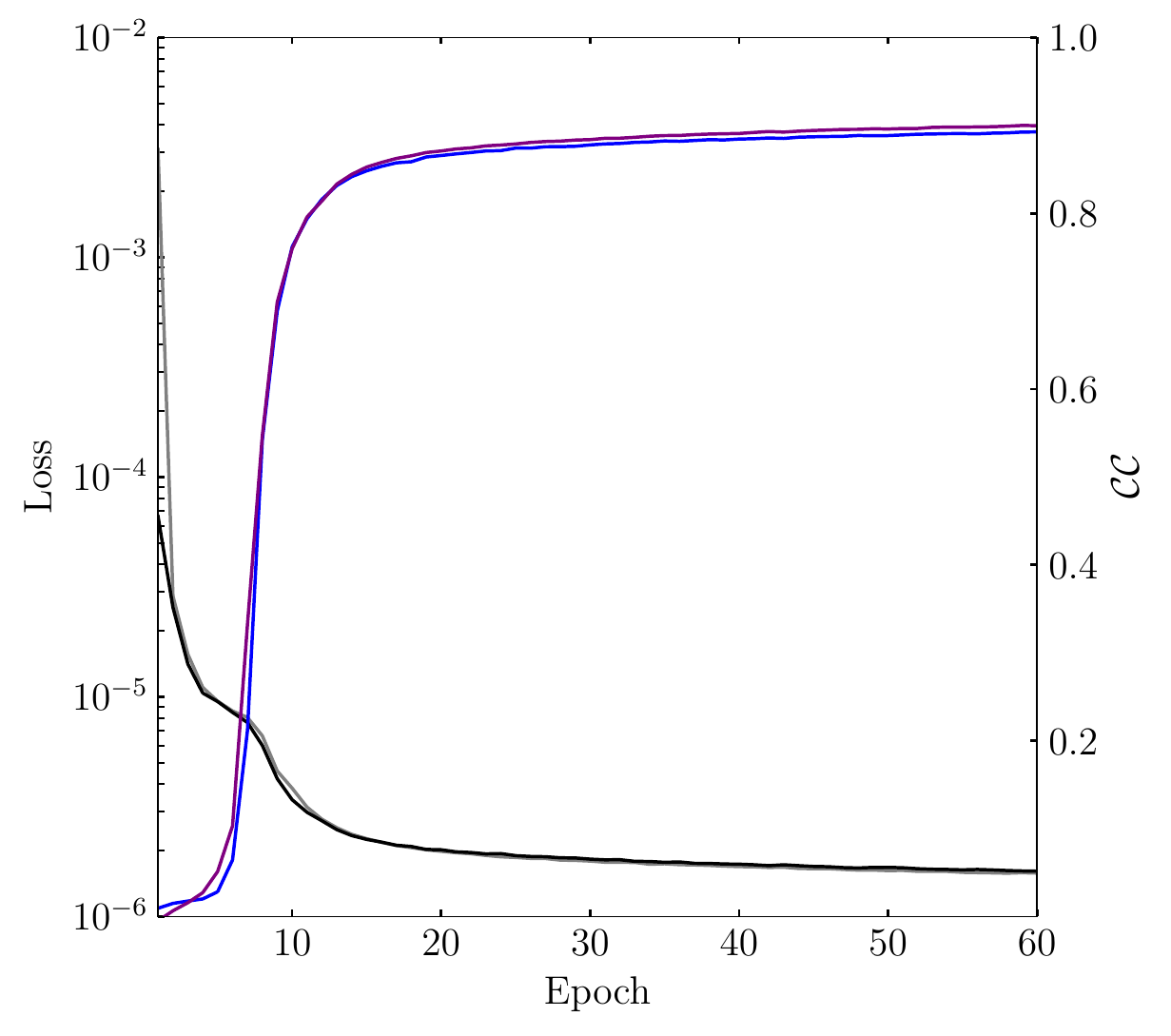}
\caption{Training of the $\mathrm{Y}_n=\big\lbrace \mathcal{S}^R_x, \mathcal{S}^R_y \big\rbrace$ output set.
Grey: combined training loss.
Black: combined validation loss.
Blue: $\mathcal{S}^R_x$ correlation coefficient.
Purple: $\mathcal{S}^R_y$ correlation coefficient.}
\label{fig:SR_history}
\end{multicols}
\end{figure}

The ML model training history of the SSR tensor components (a-priori target) is plotted in \fref{fig:ab_history}, where a rapid convergence cap be appreciated.
We choose a maximum of 60 epochs and allow 5 epochs (full iteration on the training dataset) of patience to stop the simulation if the validation error does not improve (a.k.a early-stopping).
It can be observed that the early-stopping criterion is triggered on the 46th epoch, providing the lowest validation error not improved for the next 5 epochs.
On the other hand, the training history of the perfect closure components (a-posteriori target) is plotted in \fref{fig:SR_history}.
In this case, the early-stopping criterion is not triggered and optimisation is performed during 60 epochs (the maximum defined).

\subsection{Hardware and software} \label{app:ML_model_details_SWHW}

The 3-D simulations carried in this work have been performed in the Iridis4 supercomputer at the University of Southampton.
The fluid solver is run in 128 parallel processes across 8 different nodes.
Each node is equipped with 16x 2.6 GHz Intel\textsuperscript\textregistered $ $ Sandybridge\texttrademark $ $ processors. With this set-up, 205 hours of computational time was spent for the dataset generation.
The ML model is built using Keras \citep{Chollet2015}, a high-level deep learning Python library employing Tensorflow \citep{Tensorflow} as backend, and can be found in \citet{Font2020-sanspy}.
A NVIDIA\textsuperscript\textregistered $ $ Tesla V100 GPU is used for training the ML model in the Iridis5 supercomputer.
It takes approximately 30 minutes to iterate through the training dataset (i.e. 30 min./epoch).

In the a-posteriori analysis, the fluid solver (Fortran) and the ML model (Python) need to run in sync.
At every time step, the resolved quantities data is transferred to the ML model in order to compute the closure terms.
After the prediction is completed, the closure terms are sent back to the fluid solver to compute the solution of the next time step.
This has been accomplished using network sockets, a data transfer protocol between processes living in the same network.
In particular, internet sockets have been employed allowing to transfer data across a network port.
An example of Fortran-Python sockets data transfer can be found in \cite{Font2019-f2py}.
Since the SANS simulation is essentially 2-D, this has been carried in a 4.0 GHz Intel\textsuperscript\textregistered $ $ Core\texttrademark $\,\,$i7-6700K processor using serial computations.
It takes approximately 0.43 hours to complete $\Delta t^*=10$ time units in this a-posteriori configuration.

\bibliographystyle{etc/apalike-refs}
\bibliography{main}
\end{document}